\documentclass[12pt]{article}
\pdfoutput=1

\usepackage{graphicx,psfrag,epsf,color}
\usepackage{xcolor}
\usepackage{amsmath,amssymb,amsfonts}
\usepackage{array}
\usepackage{cite}
\usepackage{rotating}
\usepackage{slashed,mathtools}
\usepackage{xparse}
\usepackage{stmaryrd}
\usepackage{cancel}
\usepackage{todonotes}
\usepackage{tikz}
\usepackage{caption}
\usepackage{subcaption}

\usepackage{centernot}

\definecolor{cBlue}{RGB}{0,110,191}

\definecolor{c}{rgb}{0.90,0.62,0}
\definecolor{cc}{rgb}{0.90,0.62,0}
\definecolor{cb}{rgb}{0.34,0.71,0.91}
\definecolor{s}{rgb}{0,0.62,0.45}
\definecolor{hc}{rgb}{0,0.0,0.9}
\definecolor{sc}{rgb}{0.9,0.42,0.32}

\usepackage[hypertexnames=true]{hyperref}
\hypersetup{bookmarksnumbered,colorlinks,
    linkcolor={black},
    citecolor={cBlue},
    urlcolor={cBlue}}
\numberwithin{equation}{section}

\newcommand{\be}{\begin{equation}}
\newcommand{\ee}{\end{equation}}
\newcommand{\bea}{\begin{eqnarray}}
\newcommand{\eea}{\end{eqnarray}}
\newcommand{\bi}{\begin{itemize}}
\newcommand{\ei}{\end{itemize}}
\newcommand{\ben}{\begin{enumerate}}
\newcommand{\een}{\end{enumerate}}
\newcommand{\bt}{\begin{tabular}}
\newcommand{\et}{\end{tabular}}

\newcommand{\T}{{\bf T}}

\definecolor{darkgreen}{rgb}{0.0,0.6,0.0}
\definecolor{cPurple}{RGB}{93,35,125}

\begin{document}
\allowdisplaybreaks

\begin{titlepage}
		
\begin{flushright}
{\small
IPPP/24/83\\
LTH 1389\\
\today
}
\end{flushright}

\vskip0.1cm
\begin{center}
{\Large \bf The structure of quark mass corrections in the $gg \rightarrow HH$ amplitude at high-energy}
\end{center}
		
\vspace{0.5cm}
\begin{center}
{\sc Sebastian Jaskiewicz$^a$}, 
{\sc Stephen Jones$^b$},
{\sc Robert Szafron$^c$},
{\sc Yannick Ulrich$^{a,d}$}
\\[6mm]

{\it 
$^a$ Albert Einstein Center for Fundamental Physics, Institut f\"ur Theoretische Physik, Universit\"at Bern, Sidlerstrasse 5, CH-3012 Bern, Switzerland
 \\[0.2cm]
$^b$ Institute for Particle Physics Phenomenology, Durham University,\\
South Road, Durham, DH1 3LE, United Kingdom 
 \\[0.2cm]
 $^c$ Department of Physics, Brookhaven National Laboratory, \\Upton, New  York, 11973, U.S.A.
  \\[0.2cm]
 $^d$ Department of Mathematical Sciences, University of Liverpool, \\ Liverpool, L69 3BX, United Kingdom
}
\end{center}
		
\vspace{0.4cm}
\begin{abstract}
\vskip0.2cm\noindent
The leading and next-to-leading order QCD predictions for Higgs boson pair production at hadron colliders suffer from a significant mass renormalisation scheme uncertainty related to the choice of the top-quark mass.
The functional dependence of the result on the value of the intermediate quark mass can be understood in the high-energy limit using the Method of Regions and the tools of Soft-Collinear Effective Theory.
In this work, we study the origin of the sizeable logarithmic mass corrections in the $gg \to HH$ amplitudes at leading and next-to-leading power in the limit $s,|t|,|u| \gg m_t^2 \gg m_H^2$.
We argue that the mass corrections follow a predictable factorised pattern that can be exploited to simplify their computation.
We present results for the leading power leading logarithmic corrections, our analysis leads to a significant reduction in the theoretical uncertainty of the double Higgs production amplitudes at high-energy due to the top-quark mass scheme.
 
\end{abstract}
\end{titlepage}

\tableofcontents

\section{Introduction}

Computing multi-loop scattering amplitudes that retain the complete mass dependence for intermediate and initial/final state particles is exceptionally challenging.
The current state-of-the-art for high multiplicity massive processes includes $2 \rightarrow 3$ results for $W b \overline{b}$~\cite{Badger:2021nhg,Abreu:2021asb,Hartanto:2022qhh}, $H b \overline{b}$~\cite{Badger:2021ega,Badger:2024awe}, 
$H t \overline{t}$~\cite{FebresCordero:2023pww,Agarwal:2024jyq,Devoto:2024nhl}, 
$Wjj$~\cite{Abreu:2021asb}, $W \gamma j$~\cite{Badger:2022ncb}, $Z\gamma \gamma$~\cite{Kermanschah:2024utt},
and 
$W \gamma \gamma$~\cite{Badger:2024sqv} at two-loops.
The relevant master integrals are known for five-point one mass~\cite{Chicherin:2021dyp,Abreu:2023rco} and two-mass processes in the planar limit/leading-colour approximation~\cite{Badger:2022hno, Badger:2024fgb} (partly relying on generalised series solutions~\cite{Hidding:2020ytt, Moriello:2019yhu}).
In many cases, including Higgs boson pair production, fully analytic results have not been obtained for all Feynman integrals appearing beyond one-loop, and numerical~\cite{Smirnov:2015mct, Borowka:2017idc} or semi-numerical~\cite{Hidding:2020ytt, Liu:2022chg, Armadillo:2022ugh} methods are used either exclusively or to supplement partial analytic results.

For Higgs production processes in the limit of large intermediate quark mass ($m_q \rightarrow \infty$), a heavy top effective field theory has been constructed~\cite{Ellis:1975ap, Shifman:1979eb, Kniehl:1995tn} in which massive quark loops are integrated out of the theory and effective couplings are introduced between the Higgs boson and gluons.
The existence of such an effective theory means that higher-order corrections can be computed much more straightforwardly, with the quark mass dependence factorising from the loop integrals.
This effective field theory approximation has been used extensively in Higgs physics and played an essential role in many high-order results in the literature, for example, the N$^3$LO QCD corrections to Higgs boson production~\cite{Anastasiou:2015vya, Anastasiou:2016cez}.
Fixed-order QCD corrections for Higgs boson pair production are known at N$^3$LO in the limit of large top-quark mass~\cite{Chen:2019lzz, Chen:2019fhs, Grazzini:2018bsd}, which are not applicable for production of a Higgs pair with large total invariant mass.

Results for the partial and full electroweak corrections to Higgs pair production have also recently been obtained~\cite{Bi:2023bnq, Heinrich:2024dnz, Zhang:2024rix, Li:2024iio}.
The corrections are also known in the limit of large top-quark mass~\cite{Davies:2023npk} and the high-energy limit for the Yukawa corrections~\cite{Davies:2022ram}.
Retaining the complete top-quark mass dependence, the QCD corrections to Higgs boson pair production are known only at NLO~\cite{Glover:1987nx, Borowka:2016ehy, Borowka:2016ypz, Baglio:2018lrj, Baglio:2020ini, Baglio:2020wgt, Bagnaschi:2023rbx, Campbell:2024tqg}.
The light-fermion $n_f$-contributions have been calculated at NNLO for forward scattering kinematics neglecting the Higgs boson mass~\cite{Davies:2023obx}.
Recently, the one-particle reducible contributions were also computed at NNLO~\cite{Davies:2023obx,Davies:2024znp}.  

In Ref.~\cite{Baglio:2018lrj}, it was noted that the NLO result is sensitive to the precise value chosen for the intermediate top-quark mass and that the reduction in this dependence going from LO to NLO was mild.
This behaviour means that small changes to the top-quark mass can enormously impact the NLO correction. 
Furthermore, the choice of the renormalisation scheme for the top-quark mass can have a sizable effect on the value obtained for the NLO result. 

Taking the difference between results obtained using the on-shell ($\mathrm{OS}$) and $\overline{\mathrm{MS}}$ schemes for the top-quark mass as a genuine theoretical uncertainty leads to the introduction of a \textit{mass scheme} uncertainty that is comparable in size to the usual renormalisation/factorisation scale uncertainties~\cite{Baglio:2020wgt}.
The renormalisation scale uncertainty can be significantly reduced using higher-order computations in the heavy top effective field theory described above; however, the mass scheme uncertainty can not be addressed in this limit, so it becomes the dominant theoretical uncertainty on Higgs pair production.
The same fate is known to be suffered by several other amplitudes in the Higgs sector, including off-shell Higgs boson production~\cite{Mazzitelli:2022scc, Amoroso:2020lgh}, $Z$-Higgs production~\cite{Wang:2021rxu, Chen:2022rua} and $ZZ$ production via an intermediate Higgs boson.

Owing to the complexity of the NLO QCD corrections to Higgs pair production and the fact that they are known only numerically, significant effort has been invested to obtain results in various limits, namely,
\begin{enumerate}
\item High-energy limit\footnote{Note that the term ``high-energy limit'' is often synonymous with the forward scattering -- or Regge -- limit where $-t/s \ll 1$, in this work we do not impose a hierarchy between the kinematic invariants.}: $s, |t|, |u| \gg m_t^2 \gg m_H^2$~\cite{Davies:2018ood,Davies:2018qvx,Davies:2019dfy,Davies:2023vmj},
\item Small Higgs boson mass limit: $s, |t|, |u| \gg m_H^2$~\cite{Xu:2018eos,Wang:2020nnr},
\item Small-$p_T$ and/or small-$t$~\cite{Bonciani:2018omm,Davies:2023vmj},
\item Near to the top-quark threshold: $s \sim 4 m_t^2$~\cite{Grober:2017uho}.
\end{enumerate}
In this work, we focus on the first of these approximations and investigate
to what extent the behaviour of the mass corrections in this limit can be
understood using the Method of Regions (MoR) approach
\cite{Beneke:1997zp, Smirnov:1990rz, Smirnov:1994tg, Smirnov:2002pj}, 
Soft-Collinear Effective Field Theory (SCET) \cite{Bauer:2000yr, Bauer:2001yt,Bauer:2002nz, Beneke:2002ph, Beneke:2002ni} 
and related tools \cite{Jantzen:2011nz, Semenova:2018cwy, Smirnov:2023bxf,Ma:2023hrt, Borinsky:2020rqs,Borinsky:2023jdv, Gardi:2022khw, Gardi:2024axt}. 
Similar analyses have been carried out for the single Higgs boson
production and decay amplitudes 
\cite{Penin:2014msa, Liu:2017axv, Liu:2017vkm, Liu:2018czl, Penin:2018vmt, Wang:2019mym, Liu:2019oav, Liu:2020eqe, Liu:2020wbn, Anastasiou:2020vkr, Liu:2021chn, Liu:2022ajh, Liu:2020tzd}, 
Higgs boson plus jet~\cite{Liu:2024tkc} and power-enhanced QED corrections
for $B_s \to \overline{\ell} \ell$ \cite{Beneke:2017vpq, Beneke:2019slt}.

In Section~\ref{sec:amplitudes}, we describe the general structure of the $gg \rightarrow HH$ amplitudes at fixed order, first by analysing the regions present in the scalar integrals and then discussing how the regions contribute at the amplitude level at one- and two- loop level.
In Section~\ref{sec:scet}, we introduce a SCET-based analysis of mass corrections to $gg \rightarrow HH$ and describe how the leading-power (LP) and next-to-leading power (NLP) mass effects can be resumed.
In Section~\ref{sec:resultsplots}, we discuss to what extent the leading and sub-leading mass effects capture the mass dependence of the amplitude at high-energy and show how resuming the leading mass corrections reduces the mass scheme uncertainty.
Finally, Section~\ref{sec:summary} presents our conclusions and outlook.

\section{Structure of the amplitudes at fixed order}
\label{sec:amplitudes}

\begin{figure}[!ht]
\begin{center}
\includegraphics[width=0.9\textwidth]{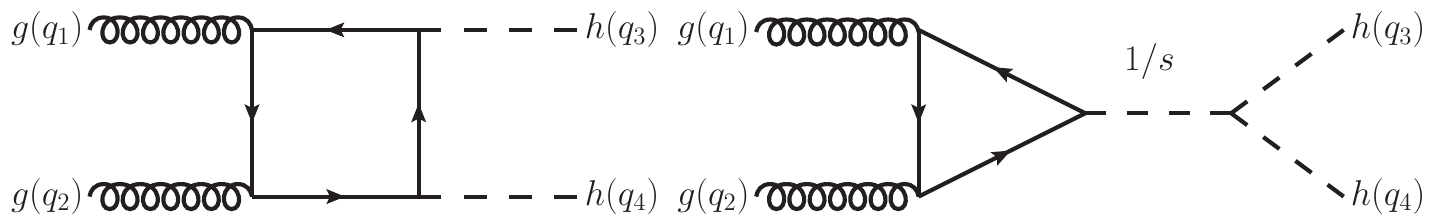}
\caption{\label{fig:gghh_lo} The leading order QCD diagrams contributing to $gg \to HH$. 
The left (box) diagram is proportional to $y_t^2$, while the right (triangle) diagram is proportional to $y_t$.}
\end{center}
\end{figure}

We consider the process $g(q_1) g(q_2) \rightarrow H(-q_3) H(-q_4)$, with all momenta defined to be incoming.
The amplitude can be parametrised in terms of the usual Mandelstam invariants,
\begin{align}
&s=(q_1 + q_2)^2,& &t=(q_1 + q_3)^2,& &u=(q_2 + q_3)^2.&
\end{align}

The $gg \rightarrow HH$ amplitude can be decomposed in terms of two form factors, which can be chosen to be the helicity amplitudes, $A_1 = -\mathcal{M}^{++} = -\mathcal{M}^{--}$, and $A_2 = -\mathcal{M}^{+-} = -\mathcal{M}^{-+}$.
Furthermore, the QCD corrections can be separated into \textit{box} diagrams, $A_{i,y_t^2}$, in which both Higgs bosons couple to a massive quark line and \textit{triangle} diagrams, $A_{i,y_t \lambda_3}$, in which a single off-shell Higgs boson couples to the massive quark line and splits into two Higgs bosons via a trilinear self-coupling (here, for simplicity, we consider only the top quark to be massive).
Explicitly,
\begin{align}
&A_1 = T_F \frac{G_F}{\sqrt{2}} \frac{\alpha_s}{2 \pi} s\  \left[ \frac{3 m_H^2}{s - m_H^2} A_{1,y_t \lambda_3} + A_{1, y_t^2} \right],&
&A_2 = T_F \frac{G_F}{\sqrt{2}} \frac{\alpha_s}{2 \pi} s\  \left[ A_{2, y_t^2} \right].&
\label{eq:Ai}
\end{align}
The form factors can then be further expanded in powers of the strong coupling as
\begin{align}
A_{i,j} = \sum_{k=0} \left(\frac{\alpha_s}{2\pi} \right)^k A_{i,j}^{(k)}, \quad i=1,2,\ j=y_t^2, y_t \lambda_3.
\end{align} 
Representative leading-order diagrams are shown in Fig.~\ref{fig:gghh_lo}.

We will focus on the behaviour of the amplitudes in the high-energy limit,
\begin{equation}
s, |t|, |u| \gg m_t^2 \gg m_H^2.
\end{equation}
Since, from the perspective of QCD,  the dependence on $m_H^2$ is only through the external kinematical invariants and the $s$-channel propagator in the triangle-type contribution, we will focus on the $m_t$ mass dependence and take $m_t^2 \gg m_H^2$, i.e. we first set $m_H=0$ and then perform expansion in the small top-mass, $m_t$. Restoring the kinematical dependence on $m_H$ is straightforward after factorisation of the amplitude at the scale $m_t$. We define our power counting parameter 
as $ \lambda=m_t/Q$, where $Q$ is of the order of large invariants~$s,|t|,|u|$. 

We expand the amplitude around a small Higgs boson mass, $m_H \sim 0$.
To leading power in the $m_H$ expansion, this means that external momenta and Mandelstam invariants obey,
\begin{align}
&q_1^2=q_2^2=q_3^2=q_4^2 = 0,& &s+t+u = 0.&
\end{align}
The contribution of the triangle-type ($y_t \lambda_3$) diagrams is power-suppressed by $m_H^2/s$ relative to the box amplitude  For this case, 
the all-order structure of the mass-logarithms can be obtained from existing results for single Higgs production \cite{Liu:2018czl, Anastasiou:2020vkr, Liu:2021chn, Liu:2022ajh, Schnubel:2023cxu}, and therefore we do not discuss it further.

The starting point of this work is the observation that the result for one- and two-loop box contributions to the  $gg \to HH$ amplitude in the $\mathrm{OS}$ scheme 
 can be written in the high-energy limit as~\cite{Davies:2018qvx, Baglio:2018lrj, Baglio:2020ini},
\begin{align}
\label{eq:1.1}
A^{(0)}_{i,y_t^2} &= y_t^2\, f_i(s,t) + \mathcal{O}(y_t^2 m_t^2)\,, \\ 
A^{(1)}_{i,y_t^2} &= 3 C_F \,A^{(0)}_i\,\log\left[\frac{m_t^2}{s} \right]+ y_t^2 \, g_i(s,t) + \mathcal{O}(y_t^2 m_t^2)\,,
\end{align}
where $g_i$ does not depend on $m_t$.
The functions $f_i$ are the leading order leading power (in $m_t$)  expressions for the form factors, explicitly they are given by~\cite{Davies:2018qvx},
\begin{align}
&f_1 = \frac{8}{s},& 
&f_2 = \frac{2}{s t (s+t)} \left[ -l_{1ts}^2 (s+t)^2 - l_{ts}^2 t^2 - \pi^2 (s^2 + 2 s t +2 t^2) \right],&
\end{align}
and
\begin{align}
&l_{ts} = \log \left( -\frac{t}s{} \right) + i \pi,&
&l_{1ts} = \log \left( 1 + \frac{t}{s} \right) + i \pi.&
\end{align}
The appearance of a leading $C_F$ term at two-loops can be understood as originating from the mass renormalisation counter-term as can seen by converting the top-quark mass to the $\overline{{\rm{MS}}}$ scheme~\cite{Baglio:2020ini}.
Since the leading mass-dependent logarithm comes only from the mass counter-term, we can predict the leading behaviour of $A^{(1)}_{i,y_t^2}$ in the small mass limit just from the leading order $A^{(0)}_{i,y_t^2}$ result and the mass renormalisation counter-term.

Having noted the above, the immediate questions are,
\begin{enumerate}
    \item Why does the box amplitude have such a simple structure?
    \item Can super-leading/power-enhanced $m_t$ terms appear in the integrals at higher loops, and can they spoil this picture? 
\end{enumerate}
In other words, we wish to establish whether or not the simple structure observed at one- and two-loop order persist to all orders in the QCD coupling.

To answer the above questions, we begin by analysing the process using the MoR and write down the modes and regions that contribute at each order in the power expansion in the quark mass.
In the first step, we study individual scalar integrals, then we analyse complete amplitudes.
In Section~\ref{sec:scet}, using the understanding gained from the MoR analysis, we use the tools of SCET to construct a field theory description of the physics and build a framework that allows the small mass expansion to be studied systematically at higher orders.

\subsection{Structure of the scalar integrals}
\label{sec:regions}

Our analysis of the process begins by applying the MoR to individual scalar Feynman integrals.
This procedure was carried out in detail in parameter space for one- and two-loop integrals in Ref.~\cite{Mishima:2018olh} and used to calculate the integrals in an expansion around a small quark mass.
Here, we will repeat this procedure instead focusing on building an interpretation of each region in terms of the scaling of the loop-momentum.

To obtain the regions, we first express the integrals using the Lee-Pomeransky representation~\cite{Lee:2013hzt}.
A scalar integral corresponding to a graph $G$ in $D=4-2\epsilon$ space-time dimensions may be written as,
\begin{align}
\mathcal{I}(G) &= \frac{\Gamma(D/2)}{\Gamma((L+1)D/2-\nu) \prod_{e \in G} \Gamma(\nu_e)} \int_0^\infty \left( \prod_{e \in G} \frac{\mathrm{d} x_e}{x_e}\right) I(\mathbf{x};\mathbf{s}), \\
I(\mathbf{x};\mathbf{s}) &= \left( \prod_{e \in G} x_e^{\nu_e} \right) \cdot \left( \mathcal{P}(\mathbf{x}; \mathbf{s}) \right)^{-D/2}, \quad
\mathcal{P}(\mathbf{x}; \mathbf{s}) \equiv \mathcal{U}(\mathbf{x}) + \mathcal{F}(\mathbf{x};\mathbf{s}),
\end{align}
where $\nu_e$ is the exponent of the denominator associated to the propagator $e$, and $\nu \equiv \sum_{e\in G} \nu_e$. The polynomial $\mathcal{P}(\mathbf{x};\mathbf{s})$ is the Lee-Pomeransky polynomial, for an $N$ propagator integral it depends on $N$ Lee-Pomeransky parameters which we denote by $\mathbf{x}$ and a set of Lorentz invariants (or Mandelstam variables) and masses which we denote collectively by $\mathbf{s}$. The polynomials $\mathcal{U}(\mathbf{x})$ and $\mathcal{F}(\mathbf{x};\mathbf{s})$ are the first and second Symanzik polynomials.

The particular limit we are interested in can be defined by introducing a small parameter $\lambda$ and rescaling the kinematic invariants/masses according to:
\begin{align}
&m_t \rightarrow \lambda m_t,&
&s \rightarrow s,&
&t \rightarrow t,&
&u \rightarrow u.&
\end{align}
In parameter space, each region is defined as a set of scalings (with respect to the small parameter $\lambda$) of each Lee-Pomeransky parameter.
For example, for an integral depending on the parameters $x_1, \ldots, x_N$, a region $(R)$ may be given by the scaling 
\begin{align}
&(R):& &x_1 \rightarrow \lambda^{u_1^R} x_1,&
&x_2 \rightarrow \lambda^{u_2^R} x_2,& &\ldots,&
&x_N \rightarrow \lambda^{u_N^R} x_N,& \label{eq:ux}
\end{align}
along with an overall rescaling of the expansion parameter $\lambda \rightarrow \lambda^{u_{N+1}^R}$. 
The set of scalings can be collected into a region vector $\mathbf{u}^R = (u_1^R, \ldots, u_{N+1}^R )$.
For Euclidean kinematics, the regions that appear in parameter space can be obtained automatically using existing tools~\cite{Jantzen:2012mw, Heinrich:2021dbf}\footnote{It is known that, away from Euclidean kinematics additional work has to be done to reveal hidden/Glauber regions, see for example Refs.~\cite{Jantzen:2012mw,Gardi:2024axt,Becher:2024kmk}.}.
One advantage of using the Lee-Pomeransky representation for the MoR is that the scaling of parameter $x_e$ is identical to the scaling of the corresponding propagator up to an overall shift of the $\mathbf{u}$ vector~\cite{Engel:2022kde, Gardi:2022khw}.

In general, the expanded integrals are no longer regulated in dimensional regularisation.
Instead, we find a new type of singularity that cancels between the different regions.
These rapidity divergences are due to the so-called factorisation (or collinear) anomaly~\cite{BenekeFA, Becher:2010tm}. The physical reason for their appearance has been understood in SCET as a breakdown of the EFT's classical boost (or reparametrisation) symmetry by quantum corrections.
We can regulate this type of singularity by using an analytic regulator, i.e. shifting the powers of the propagators away from 1 to $1\pm\eta$.
Once all regions are added, the procedure involves taking the limit $\eta\to0$ \emph{before} imposing the limit $\epsilon\to0$ to obtain the correct answer.

\subsubsection{One-loop}
\label{ssec:scalar_one_loop}

\begin{figure}[h]
    \centering
    \includegraphics[width=0.3\textwidth]{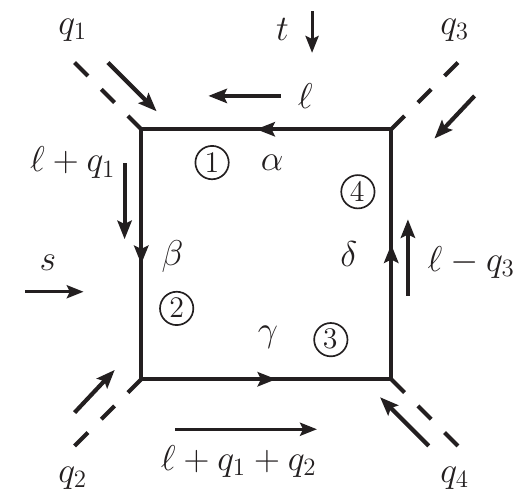}
    \caption{Kinematics of the one-loop scalar box integral. The internal lines are massive, with mass $m_t$, and the external lines are massless. The circled numbers refer to the order in which the propagators appear in the region vectors in \eqref{eq:oneloopregionvectors} and $\alpha,\beta,\gamma,\delta$ refer to the propagator powers. 
    \label{fig:gghh_diag1}}
\end{figure}

Let us consider the one-loop scalar box integral shown in Fig.~\ref{fig:gghh_diag1}, the integral is defined in momentum space as,
\begin{align}\label{eq:loopbox}
\resizebox{.999\hsize}{!}{$\displaystyle
        I = e^{\gamma_E\epsilon} \mu^{2\epsilon} 
    \int
    \frac{d^D \ell}{i\pi^{D/2}}  
    \frac{1}{
        \big[\ell^2 - m_t^2\big]^\alpha
        \big[(\ell+q_{1})^2 - m_t^2\big]^\beta
        \big[(\ell+q_{1}+q_2)^2 - m_t^2\big]^\gamma
        \big[(\ell- q_{3})^2 - m_t^2\big]^\delta
    }\,.
$}
\end{align}
Expanding around small $m_t$ using the MoR, we obtain five regions in parameter space,
\begin{align} \label{eq:oneloopregionvectors}
&\mathbf{u}^{(1)} = (-2, -2, 0, 0),&
&\mathbf{u}^{(2)} = (0, -2, -2, 0),& 
&\mathbf{u}^{(3)} = (-2, 0, 0, -2),& \nonumber\\
&\mathbf{u}^{(4)} = (0, 0, -2, -2),& 
&\mathbf{u}^{(5)} = (0, 0, 0, 0).&
\end{align}
These regions correspond to the various internal propagators having either a hard scaling~$\sim \lambda^0$ or a power suppressed scaling $\sim \lambda^2$.
The regions are identical to those considered in Ref.~\cite{Mishima:2018olh} up to a rescaling required to convert the Lee-Pomeransky parameters to Feynman parameters, see Eq.(2.8) of Ref.~\cite{Gardi:2022khw}.

While it is possible to calculate the integrals directly for each region, we would first like to interpret these regions in terms of the scaling of the loop-momentum $\ell$.
This will help us in Section~\ref{sec:scet}, where we connect our calculation to SCET and derive a factorisation theorem.

Let us begin with the region $\mathbf{u}^{(1)}$.
Based on the region vector, we can infer that the propagators $\ell^2-m_t^2$ and $(\ell+q_1)^2-m_t^2$ must scale as $\lambda^2$ and that $(\ell+q_1+q_2)^2 - m_t^2$ and $(\ell-q_3)^2-m_t^2$ scale as $\lambda^0$.
For energetic particles close to the mass shell, 
it is natural to formulate the analysis in 
terms of components of the momenta
decomposed using light-cone vectors as follows
\begin{align}\label{eq:lc}
  \ell^\mu =
  \underbrace{
  (n_+\cdot \ell)\frac{n_-^\mu}{2}  }_{\ell_-^{\mu}} + \underbrace{  (n_-\cdot \ell)\frac{n_+^\mu}{2} }_{\ell_+^{\mu}}+   \ell^{\mu}_\perp\,=
  \big( n_+\cdot \ell ,  n_-\cdot \ell ,  \ell_{\perp} \big). 
\end{align}
where $n_-^\mu$ and $n_+^\mu$ are the two light-like vectors
satisfying $n_-^2 =n_+^2 =0$, $n_-\cdot n_+=2$, and~$\perp$
indicates the directions perpendicular to both of these vectors.
If we take the $q_1^{\mu}$ momentum to be in the $n_{-}^{\mu}$ 
direction, such that $n_+\cdot q_1$~$\sim \lambda^0$,  
and consider our loop momentum to be collinear to $q_1$, i.e.
\begin{align}
  \ell^\mu = \underbrace{(n_+\cdot \ell)}_{\mathcal{O}(\lambda^0)}\frac{n_-^\mu}2 
  +  \underbrace{(n_-\cdot \ell)}_{\mathcal{O}(\lambda^2)}\frac{n_+^\mu}{2} +  \underbrace{\ell^{\mu}_\perp\,}_{\mathcal{O}(\lambda )},
\end{align}
we obtain the following scalings for the scalar products
\begin{align}
  \ell^2 \sim \lambda^2\,Q^2,\quad
  \ell\cdot q_1 \sim \lambda^2\,Q^2,\quad
  \ell\cdot q_2 \sim \lambda^0\,Q^2,\quad
  \ell\cdot q_3 \sim \lambda^0\,Q^2,
\end{align}
since by construction there is a wide separation between the directions of the four external particles. Inputting these scalings into the propagators, we see that the region described by the region vector~$\mathbf{u}^{(1)}$
is consistently identified as
the one with loop momentum collinear to $q_1$. We denote this region as $c_1$.

To reveal the nature of the second region $\mathbf{u}^{(2)}$, we need to perform a shift in the loop momentum because, according to the region vector, the third propagator, namely $(\ell+q_1+q_2)^2 - m_t^2$ in the original routing, should scale as $\lambda^2$. However, despite the fact that $\ell\cdot q_2\sim\lambda^2$, the propagator scales as $\lambda^0$ since $q_1\cdot q_2\sim \lambda^0$. 
As such, we note a well-known fact: the diagram's momentum routing can obscure a region's physical nature.
In the region where the loop momentum $\ell$ is collinear to $q_2$, the scalar products scale as
\begin{align}
  \ell^2 \sim \lambda^2\,Q^2,\quad
  \ell\cdot q_1 \sim \lambda^0\,Q^2,\quad
  \ell\cdot q_2 \sim \lambda^2\,Q^2,\quad
  \ell\cdot q_3 \sim \lambda^0\,Q^2,
\end{align}
and we find that the correct scaling for the propagators is obtained after shifting $\ell\to \ell-q_1$ in Fig.~\ref{fig:gghh_diag1}.
With the correct routing, we can now identify the region described by region vector $\mathbf{u}^{(2)}$ as collinear to $q_2$, i.e. $c_2$.

Finally, to reveal $\mathbf{u}^{(3)}$ and $\mathbf{u}^{(4)}$ we must consider the frame spanned by $q_3$ and $q_4$.
It should not be surprising that our ability to identify regions depends on which of the four-momenta we eliminate or, in other words, which frame we choose.
One can easily show that using the shifts $\ell\to \ell+q_3$ and $\ell \to \ell-q_1 - q_2$ we can identify the momentum-space representation for regions defined by vectors $\mathbf{u}^{(3)}$ and $\mathbf{u}^{(4)}$ respectively.
The fifth region, $\mathbf{u}^{(5)}$, is easily identified as the hard region, which always manifests as $u_i^{R} = 0$.
As all components of the loop momentum are large, it does not matter which frame or routing we consider.

\begin{table}
    \centering
    \begin{tabular}{l|l|l|l}
        \bf $\mathbf{u}^R$      & \bf order & \bf interpretation  & \bf routing\\\hline

        (-2, -2, 0, 0)   & $4-2(\epsilon + \alpha+\beta)$ & $c_1 $   & $\ell$\\
        (0, -2, -2, 0) & $4-2(\epsilon + \beta+\gamma)$ & $c_2 $   & $\ell-q_1$\\
        (-2, 0, 0, -2)   & $4-2(\epsilon + \alpha+\delta)$ & $c_3 $   & $\ell+q_3$\\ 
        (0, 0, -2, -2) & $4-2(\epsilon + \gamma+\delta)$ & $c_4 $   & $\ell - q_1 - q_2$\\
        (0, 0, 0, 0)   & 0 & $h$                       & n/a
    \end{tabular}
    \caption{The regions of the one-loop diagram and their interpretation in terms of hard and collinear modes. The routing is identified by specifying the momentum flowing through the propagator labelled $\alpha$ in Fig.~\ref{fig:gghh_diag1}.
    Note that for the scalar ``corner'' integral $\alpha=\beta=\gamma=\delta=1$ and all regions enter at leading power.
    }
    \label{tab:regions1L}
\end{table}

\begin{figure}
    \centering
    \includegraphics[width=0.9\textwidth]{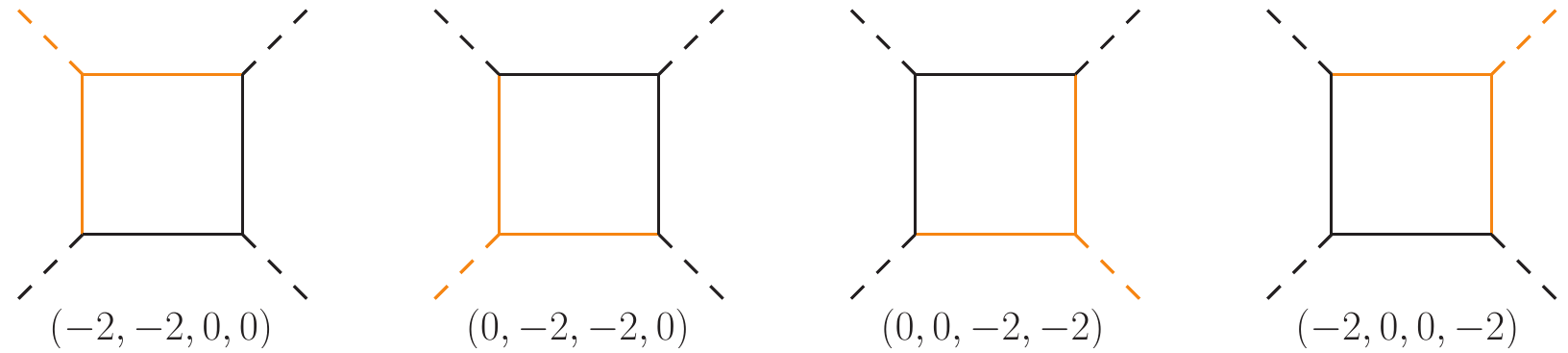}
    \caption{Regions for the one-loop diagram in Fig.~\ref{fig:gghh_lo}. The \textbf{\textcolor{c}{orange}} propagators are collinear to one of the external momenta (also coloured), while the \textbf{black} propagators have a hard scaling. Purely hard region with all propagators \textbf{black} is not depicted.}
    \label{fig:oneloopregions}
\end{figure}

We summarise our findings in Table~\ref{tab:regions1L}.
The order in $\lambda$ at which each region begins contributing is shown in the second column.
At the level of the scalar integral, in the limit $\epsilon \rightarrow 0$ and $\alpha,\beta,\gamma,\delta \rightarrow 1$, all regions enter and contribute at leading power $\sim \lambda^0$.
In Fig.~\ref{fig:oneloopregions}, we depict each region and their physical interpretation, propagators shown in \textbf{\textcolor{c}{orange}} are collinear to external momentum, while \textbf{black} propagators are hard modes.

Our choice of rapidity regulator makes the soft region scaleless at one-loop, but this is not true for a different choice of regulator.
For example, by setting the analytic regulators $\alpha=\beta=\gamma=\delta=1$ and instead introducing an additional propagator
$1/(2\ell \cdot (q_2 - q_1))^\eta$ to regulate the integral, we obtain two additional regions,
\begin{align}
&\mathbf{u}^{(6)} = (-2, -1, 0, -1, -1),&
&\mathbf{u}^{(7)} = (0, -1, -2, -1, -1).& 
\end{align}
The new regions enter  at leading power in $\lambda$ for the scalar integrals  and can be interpreted as soft regions.
The loop momentum has the following scaling in the newly revealed soft regions,
\begin{align}
  \ell^\mu = \underbrace{(n_+\cdot \ell)}_{\mathcal{O}(\lambda)}\frac{n_-^\mu}2 
  +  \underbrace{(n_-\cdot \ell)}_{\mathcal{O}(\lambda)}\frac{n_+^\mu}{2} +  \underbrace{\ell^{\mu}_\perp\,}_{\mathcal{O}(\lambda )},
\end{align}
yielding the following scaling for the scalar 
products
\begin{align}
&\ell^2 \sim \lambda^2\,Q^2,&
&\ell \cdot q_1 \sim \lambda\,Q^2,&
&\ell \cdot q_2 \sim \lambda\,Q^2.&
\end{align}
The particular choice of the regulator adopted here is consistent with Ref.~\cite{Liu:2019oav} for on-shell kinematics and preserves the analytic properties of the integral. 

Once the regions and routings which reveal their nature in 
momentum space have been identified, we can perform  calculation of the integrals directly in momentum space. 
The integral required in the hard region corresponds to the integral in \eqref{eq:loopbox} with $m_t=0$. It well-defined 
without analytic regulators, so $\alpha=\beta=\gamma=\delta=1$, and it is known in literature. 
However, the collinear regions require a new integral to be considered, which we perform with arbitrary powers $\alpha$, $\beta$, $\gamma$, and $\delta$.
As an example, we will discuss the $c_1$ region, which, in momentum space, takes the form
\begin{align}
    I = e^{\gamma_E\epsilon} \mu^{2\epsilon} 
    \int
    \frac{d^D \ell}{i\pi^{D/2}}  
    \frac{1}{
        \big[\ell^2 - m_t^2\big]^\alpha
        \big[(\ell+q_{1-})^2 - m_t^2\big]^\beta
        \big[2(\ell+q_{1-})\cdot q_{2+}\big]^\gamma
        \big[-2 \ell\cdot q_{3+}\big]^\delta
    }\,.
\end{align}
From this we can infer that $I\propto (n_-\cdot q_3)^{-\delta} (n_-\cdot q_2)^{-\gamma}$.
Since $q_{1-}^{\mu}$ is the only external vector in the $n_-^{\mu}$ direction, we can deduce 
\begin{align}\label{eq:scaling1l}
    I =
      \Big(-\underbrace{(n_+\cdot q_1)(n_-\cdot q_3)}_{t} \Big)^{-\delta}
      \Big(-\underbrace{(n_+\cdot q_1)(n_-\cdot q_2)}_{s} \Big)^{-\gamma}
      \Big( m_t^2 \Big)^{D/2-\alpha-\beta}
      I_0(\alpha,\beta,\gamma,\delta; D)\,,
\end{align}
where we have found the dependence on $m_t$ from dimensional analysis.
The remaining integral $I_0$ is now a function of the powers of propagators
and the space-time dimension.
Explicit calculation shows that
\begin{align}
    I_0 = \Big(  e^{\gamma_E\epsilon} \mu^{2\epsilon}  \Big)
    \, (-1)^{\alpha+\beta+\gamma+\delta}
    \frac{\Gamma [-2+\epsilon+\alpha+\beta]}{\Gamma[\beta]\Gamma[\alpha] }
    \frac{\Gamma[\alpha-\gamma] \Gamma [\beta-\delta]}{\Gamma[\alpha+\beta-\delta-\gamma]}\,.
\end{align}
This result concludes our discussion of the one-loop box, all other box integrals appearing in the amplitude can be obtained from crossing the diagram in Fig.~\ref{fig:gghh_diag1}.

\subsubsection{Two-loop}

\begin{figure}[h!]
    \centering
    \input{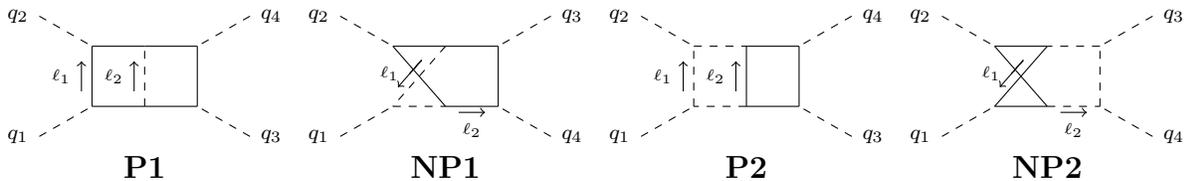}
    \renewcommand{\secondroutinginfo}[1]{#1}
    \begin{tikzpicture}[y=0.4cm,x=0.7cm]
        \begin{scope}
            \diagPT{}{}{}{}{}{}{}{}{}{}{}
            \node at (0,-3) {$\mathbf{P1}$};
        \end{scope}
        \begin{scope}[xshift=4cm]
            \diagNT{}{}{}{}{}{}{}{}{}{}{}
            \node at (0,-3) {$\mathbf{NP1}$};
        \end{scope}
        \begin{scope}[xshift=8cm]
            \diagPG{}{}{}{}{}{}{}{}{}{}{}
            \node at (0,-3) {$\mathbf{P2}$};
        \end{scope}
        \begin{scope}[xshift=12cm]
            \diagNG{}{}{}{}{}{}{}{}{}{}{}
            \node at (0,-3) {$\mathbf{NP2}$};
        \end{scope}
    \end{tikzpicture}
    \caption{
    The four top-level two-loop topologies. Massless particles (Higgs bosons and gluons) are depicted with dashed lines, massive top quarks are shown as solid lines.
    We indicate the lines that carry the loop momenta for our default routing.
    }
    \label{fig:gghh_diag2l}
\end{figure}

At the two-loop order, there are four top-level (7-propagator) topologies to consider, they are shown in Fig.~\ref{fig:gghh_diag2l}: a planar box and a non-planar box that each contain either predominantly top quarks or gluon propagators.
As the number of possible routings, frame choices, and regions proliferates beyond one-loop, we have automated the routing-finding algorithm.
Our code considers the scaling of each scalar product for each possible region and frame and then applies it to all possible shifts.
To limit the number of shifts, we only consider those that result in two propagators being directly identical to the two-loop momenta.
We observe that this procedure is sufficient for the problem of identifying soft and collinear regions at hand, but it is possible to construct regions that cannot be straightforwardly revealed this way, for example Glauber, threshold and potential regions.
This code is currently being integrated as part of pySecDec
\cite{JUunpublished}.

{
\newcommand{\extroutinginfo}[1]{#1}
\newcommand{\routinginfo}[1]{}
\newcommand{\secondroutinginfo}[1]{}
\newcommand\diagextP[4]{
    \draw[#1,dashed] (-2,-2) \extroutinginfo{ node[left ] {\scriptsize $q_1$}}-- (-1,-1) ;
    \draw[#2,dashed] (-2, 2) \extroutinginfo{ node[left ] {\scriptsize $q_2$}}-- (-1, 1) ;
    \draw[#3,dashed] ( 2,-2) \extroutinginfo{ node[right] {\scriptsize $q_3$}}-- ( 1,-1) ;
    \draw[#4,dashed] ( 2, 2) \extroutinginfo{ node[right] {\scriptsize $q_4$}}-- ( 1, 1) ;
}
\newcommand\diagextNP[4]{
    \draw[#1,dashed] (-2,-2) \extroutinginfo{ node[left ] {\scriptsize $q_1$}}-- (-1,-1) ;
    \draw[#2,dashed] (-2, 2) \extroutinginfo{ node[left ] {\scriptsize $q_2$}}-- (-1, 1) ;
    \draw[#3,dashed] ( 2,-2) \extroutinginfo{ node[right] {\scriptsize $q_4$}}-- ( 1,-1) ;
    \draw[#4,dashed] ( 2, 2) \extroutinginfo{ node[right] {\scriptsize $q_3$}}-- ( 1, 1) ;
}
\newcommand\diagP[7]{
    \draw[#1] (-1, 1) -- ( 0, 1) \routinginfo{node [midway] {\scriptsize $\ell_1+q_2           $}};
    \draw[#2] (-1,-1) -- (-1,+1) \routinginfo{node [midway] {\scriptsize $\ell_1               $}};
    \draw[#3] ( 0,-1) -- (-1,-1) \routinginfo{node [midway] {\scriptsize $\ell_1-q_1           $}};
    \draw[#4] ( 1,-1) -- ( 0,-1) \routinginfo{node [midway] {\scriptsize $\ell_1+\ell_2-q_1    $}};
    \draw[#5] ( 1, 1) -- ( 1,-1) \routinginfo{node [midway] {\scriptsize $\ell_1+\ell_2+q_2+q_4$}};
    \draw[#6] ( 0, 1) -- ( 1, 1) \routinginfo{node [midway] {\scriptsize $\ell_1+\ell_2+q_2    $}};
    \draw[#7] ( 0,-1) -- (0,1)   \routinginfo{node [midway] {\scriptsize $\ell_2               $}};

   \secondroutinginfo{
        \draw [->] (-1.2,-0.5) -- (-1.2,+0.5) node [midway,anchor=east] {\tiny$\ell_1$};
        \draw [->] (-0.2,-0.5) -- (-0.2,+0.5) node [midway,anchor=east] {\tiny$\ell_2$};
   }
}
\newcommand\diagN[7]{
    \draw[#1] ( 0, 1) -- (-1,-1) \routinginfo{node [near start, anchor=west] {\scriptsize $\ell_1                    $}  };
    \draw[#2] (-1,-1) -- ( 0,-1) \routinginfo{node [midway, anchor=north] {\scriptsize $\ell_1+q_1                $}  };
    \draw[#3] ( 0,-1) -- ( 1,-1) \routinginfo{node [midway, anchor=north] {\scriptsize $\ell_2                    $}  };
    \draw[#4] ( 1,-1) -- ( 1, 1) \routinginfo{node [midway, anchor=west ] {\scriptsize $\ell_2+q_4                $}  };
    \draw[#5] ( 1, 1) -- ( 0, 1) \routinginfo{node [midway, anchor=south] {\scriptsize $\ell_2+q_3+q_4            $}  };
    \draw[#6] ( 0, 1) -- (-1, 1) \routinginfo{node [midway, anchor=south] {\scriptsize $\ell_2-\ell_1+q_3+q_4    $}  };
    \draw[#7] (-1, 1) -- ( 0,-1) \routinginfo{node [near start, anchor=east] {\scriptsize $-\ell_1+\ell_2-q_1$}  };

   \secondroutinginfo{
        \draw [->] (-0.5,0) ++ ({90+atan(2)}:0.2) --+ ({atan(2)}:0.5) --+ ({180+atan(2)}:0.5) node [midway,anchor=east] {\tiny$\ell_1$};
        \draw [->] ( .25,-1.2) -- ( .75,-1.2) node [midway,anchor=north] {\tiny$\ell_2$};
   }
}

\newcommand\diagPT[7]{\diagP{#1}{#2}{#3}{#4}{#5}{#6}{#7,dashed}\diagextP}
\newcommand\diagPG[7]{\diagP{#1,dashed}{#2,dashed}{#3,dashed}{#4}{#5}{#6}{#7}\diagextP}
\newcommand\diagNT[7]{\diagN{#1,dashed}{#2,dashed}{#3}{#4}{#5}{#6}{#7}\diagextNP}
\newcommand\diagNG[7]{\diagN{#1}{#2}{#3,dashed}{#4,dashed}{#5,dashed}{#6}{#7}\diagextNP}
\begin{figure}[h!]
    \centering
    \scalebox{0.8}{\begin{tabular}{cccc}
    \begin{tikzpicture}[y=0.4cm,x=0.8cm]\diagPT{c}{c}{}{cb}{cb}{}{}{}{c}{cb}{}\end{tikzpicture}&
\begin{tikzpicture}[y=0.4cm,x=0.8cm]\diagPT{c}{c}{}{}{cb}{cb}{}{}{c}{}{cb}\end{tikzpicture}&
\begin{tikzpicture}[y=0.4cm,x=0.8cm]\diagPT{c}{c}{}{}{}{cc}{cc}{}{c}{}{}\end{tikzpicture}&
\begin{tikzpicture}[y=0.4cm,x=0.8cm]\diagPT{c}{c}{}{}{}{}{}{}{c}{}{}\end{tikzpicture}\\

$(-2,-2,0,-2,-2,0,0)$&
$(-2,-2,0,0,-2,-2,0)$&
$(-2,-2,0,0,0,-2,-2)$&
$(-2,-2,0,0,0,0,0)$\\

\begin{tikzpicture}[y=0.4cm,x=0.8cm]\diagPT{cc}{}{}{}{c}{c}{cc}{}{}{}{c}\end{tikzpicture}&
\begin{tikzpicture}[y=0.4cm,x=0.8cm]\diagPT{}{c}{c}{cb}{cb}{}{}{c}{}{cb}{}\end{tikzpicture}&
\begin{tikzpicture}[y=0.4cm,x=0.8cm]\diagPT{}{c}{c}{cc}{}{}{cc}{c}{}{}{}\end{tikzpicture}&
\begin{tikzpicture}[y=0.4cm,x=0.8cm]\diagPT{}{c}{c}{}{cb}{cb}{}{c}{}{}{cb}\end{tikzpicture}\\

$(-2,0,0,0,-2,-2,-2)$&
$(0,-2,-2,-2,-2,0,0)$&
$(0,-2,-2,-2,0,0,-2)$&
$(0,-2,-2,0,-2,-2,0)$\\

\begin{tikzpicture}[y=0.4cm,x=0.8cm]\diagPT{}{c}{c}{}{}{}{}{c}{}{}{}\end{tikzpicture}&
\begin{tikzpicture}[y=0.4cm,x=0.8cm]\diagPT{}{}{cc}{c}{c}{}{cc}{}{}{c}{}\end{tikzpicture}&
\begin{tikzpicture}[y=0.4cm,x=0.8cm]\diagPT{}{}{}{c}{c}{}{}{}{}{c}{}\end{tikzpicture}&
\begin{tikzpicture}[y=0.4cm,x=0.8cm]\diagPT{}{}{}{}{c}{c}{}{}{}{}{c}\end{tikzpicture}\\

$(0,-2,-2,0,0,0,0)$&
$(0,0,-2,-2,-2,0,-2)$&
$(0,0,0,-2,-2,0,0)$&
$(0,0,0,0,-2,-2,0)$
    \end{tabular}}
    \caption{Regions for the two-loop diagram $\textbf{P1}$ in Fig.~\ref{fig:gghh_diag2l}. 
    Propagators and external lines are coloured
    \textbf{\textcolor{c}{orange}} for the first collinear mode, \textbf{\textcolor{cb}{blue}} for the second collinear mode,  and \textbf{black} for the hard modes. Purely hard region is not depicted. }
    \label{fig:regions2LP1}
\end{figure}
}

We begin by considering the diagram $\mathbf{P1}$, shown in Fig.~\ref{fig:gghh_diag2l}.
We first use standard techniques to obtain the region vectors of expansion around small $m_t$ in parameter space.
In total, we obtain 13 regions for this diagram.
In the limit $\epsilon \rightarrow 0$ we find that, for this planar scalar integral, all regions enter at leading power.
Next, using the routing algorithm, we find a suitable routing that enables us to interpret the regions in terms of the scaling of the loop momenta.
In Fig.~\ref{fig:regions2LP1} we display each region vector, using colours to represent the physical interpretation of each region, with \textbf{\textcolor{c}{orange}} for the first collinear mode, \textbf{\textcolor{cb}{blue}} for the second collinear mode and \textbf{black} for hard modes.
We note that all regions can be described purely in terms of the loop momenta becoming collinear to some external momenta. 
The soft mode is absent from the expansion of this diagram.

Next, we focus on the non-planar two-loop diagram $\mathbf{NP1}$.
Repeating the above procedure, we obtain 14 regions for this diagram.
Here, we observe a new feature that was not present in the one-loop or two-loop planar diagrams: there is a mode that causes the propagators to scale as $\sim \lambda^1$.
This new mode can be understood by allowing the loop momentum to have a soft scaling.

The interpretation of each region is again shown graphically in Fig.~\ref{fig:regions2LNP1} with \textbf{\textcolor{c}{orange}} for the first collinear mode, \textbf{\textcolor{cb}{blue}} for the second collinear mode, \textbf{\textcolor{s}{green}} for soft modes and \textbf{black} for hard modes.
However, the presence of the soft region constrains the possible valid momentum routing further as it fixes the routing completely (up to permutations of $\ell_1$ and $\ell_2$).
We also present the interpretations, along with shifts in routing needed to reveal the regions, in Table~\ref{tab:regions2LNP1} for the convenience of the reader.

{

\begin{figure}
    \centering
    \scalebox{0.8}{\begin{tabular}{cccc}
    \begin{tikzpicture}[y=0.4cm,x=0.8cm]\diagNT{c}{c}{cc}{}{}{}{cc}{c}{}{}{}\end{tikzpicture}&
\begin{tikzpicture}[y=0.4cm,x=0.8cm]\diagNT{c}{c}{}{}{cc}{cc}{}{c}{}{}{}\end{tikzpicture}&
\begin{tikzpicture}[y=0.4cm,x=0.8cm]\diagNT{s}{}{}{}{s}{s}{}{}{}{}{}\end{tikzpicture}&
\begin{tikzpicture}[y=0.4cm,x=0.8cm]\diagNT{cc}{}{}{c}{c}{cc}{}{}{}{}{c}\end{tikzpicture}\\

$(-2,-2,-2,0,0,0,-2)$&
$(-2,-2,0,0,-2,-2,0)$&
$(-2,-1,0,-1,-2,-2,-1)$&
$(-2,0,0,-2,-2,-2,0)$\\

\begin{tikzpicture}[y=0.4cm,x=0.8cm]\diagNT{cc}{}{}{}{cc}{c}{c}{}{c}{}{}\end{tikzpicture}&
\begin{tikzpicture}[y=0.4cm,x=0.8cm]\diagNT{}{s}{s}{}{}{}{s}{}{}{}{}\end{tikzpicture}&
\begin{tikzpicture}[y=0.4cm,x=0.8cm]\diagNT{}{cc}{c}{c}{}{}{cc}{}{}{c}{}\end{tikzpicture}&
\begin{tikzpicture}[y=0.4cm,x=0.8cm]\diagNT{}{cc}{cc}{}{}{c}{c}{}{c}{}{}\end{tikzpicture}\\

$(-2,0,0,0,-2,-2,-2)$&
$(-1,-2,-2,-1,0,-1,-2)$&
$(0,-2,-2,-2,0,0,-2)$&
$(0,-2,-2,0,0,-2,-2)$\\

\begin{tikzpicture}[y=0.4cm,x=0.8cm]\diagNT{}{}{c}{c}{}{cb}{cb}{}{cb}{c}{}\end{tikzpicture}&
\begin{tikzpicture}[y=0.4cm,x=0.8cm]\diagNT{}{}{c}{c}{}{}{}{}{}{c}{}\end{tikzpicture}&
\begin{tikzpicture}[y=0.4cm,x=0.8cm]\diagNT{}{}{}{c}{c}{cb}{cb}{}{cb}{}{c}\end{tikzpicture}&
\begin{tikzpicture}[y=0.4cm,x=0.8cm]\diagNT{}{}{}{c}{c}{}{}{}{}{}{c}\end{tikzpicture}\\

$(0,0,-2,-2,0,-2,-2)$&
$(0,0,-2,-2,0,0,0)$&
$(0,0,0,-2,-2,-2,-2)$&
$(0,0,0,-2,-2,0,0)$\\

\begin{tikzpicture}[y=0.4cm,x=0.8cm]\diagNT{}{}{}{}{}{c}{c}{}{c}{}{}\end{tikzpicture}\\

$(0,0,0,0,0,-2,-2)$
    \end{tabular}}
    \caption{Regions for the two-loop diagram $\mathbf{NP1}$ in Fig.~\ref{fig:gghh_diag2l}.  Propagators and external lines are coloured
    \textbf{\textcolor{c}{orange}} for the first collinear mode, \textbf{\textcolor{cb}{blue}} for the second collinear mode, \textbf{\textcolor{s}{green}} for the soft modes,  and \textbf{black} for the hard modes. Purely hard region is not depicted. }
    \label{fig:regions2LNP1}
\end{figure}

}

\begin{table}
    \small
    \centering
    \begin{tabular}{l|l|l|l|l}
        \bf $\mathbf{u}^R$   & \bf order & \bf interpretation & \bf routing\\\hline                    
        (-2, -2, -2, 0, 0, 0, -2)  & $-4 \epsilon$  & $c_1c_1$      & $\ell_1$, $\ell_2$\\                              
        (-2, -2, 0, 0, -2, -2, 0)  & $-4 \epsilon$ & $c_1c_1$      & $\ell_1$, $\ell_2 - q_3 - q_4$\\                  
   (-2, -1, 0, -1, -2, -2, -1)& $-1-4 \epsilon$ & $ss$          & $\ell_1$, $\ell_2 - q_3 - q_4$\\                  
        (-2, 0, 0, -2, -2, -2, 0)  & $-4 \epsilon$ & $c_3c_3$      & $\ell_1$, $\ell_2 - q_4$\\                        
        (-2, 0, 0, 0, -2, -2, -2)  & $-4 \epsilon$ & $c_2c_2$      & $\ell_1$, $\ell_2 - q_3 - q_4$\\                  
        (-1, -2, -2, -1, 0, -1, -2)& $-1-4 \epsilon$ & $ss$          & $\ell_1 - q_1$, $\ell_2$\\                        
        (0, -2, -2, -2, 0, 0, -2)  & $-4 \epsilon$ & $c_4c_4$      & $\ell_1 - q_1$, $\ell_2$\\                        
        (0, -2, -2, 0, 0, -2, -2)  & $-4 \epsilon$ & $c_2c_2$      & $\ell_1 - q_1$, $\ell_2$\\                        
        (0, 0, -2, -2, 0, -2, -2)  & $-4 \epsilon$ & $c_4\bar c_2$ & $\ell_1 - \ell_2 + q_3 + q_4$, $\ell_1$\\         
        (0, 0, -2, -2, 0, 0, 0)    & $-2 \epsilon$ & $c_4h$        & $\ell_1 - \ell_2 + q_3 + q_4$, $\ell_1$\\         
     (0, 0, 0, -2, -2, -2, -2)  & $-4 \epsilon$ & $c_3\bar c_2$ & $\ell_1 - \ell_2 + q_3$, $\ell_1 - q_4$\\         
        (0, 0, 0, -2, -2, 0, 0)    & $-2 \epsilon$ & $c_3h$        & $\ell_1 - \ell_2 + q_3$, $\ell_1 - q_4$\\         
        (0, 0, 0, 0, 0, -2, -2)    & $-2 \epsilon$ & $hc_2$        & $\ell_1$, $\ell_1 + \ell_2 - q_3 - q_4$\\         
        (0, 0, 0, 0, 0, 0, 0)         & 0    & $hh$           & n/a                                               
    \end{tabular}
    \caption{The regions of the two-loop diagram $\mathbf{NP1}$ and their interpretation in terms of hard, collinear, and soft modes. 
    The routing shifts are relative to one given for the $\mathbf{NP1}$ diagram in Fig.~\ref{fig:gghh_diag2l}.
    The order given is valid when all propagators are raised to power 1 and the dependence on additional analytic regulators is suppressed.}
    \label{tab:regions2LNP1}
\end{table}

The region analysis of diagram $\mathbf{P2}$ and $\mathbf{NP2}$ proceeds similarly to $\mathbf{P1}$ and $\mathbf{NP1}$, respectively.
In all four topologies, all regions obtained can be interpreted as loop momenta becoming soft or collinear to an external momentum, see Appendix~\ref{app:two_loop_regions} for a depiction of each region and their interpretation in momentum space according to the colour coding described above.

In summary, we find that at the one- and two-loop level the following loop momenta modes are sufficient to interpret all regions appearing in the small-mass expansion,
\begin{align}
&\mathrm{hard}:& &\ell^\mu_H \sim Q\, (1,1,1),& \\
&\mathrm{collinear}-q_i:& &\ell^\mu_{C_i} \sim Q\, (1,\lambda^2,\lambda),& \\
&\mathrm{soft}:& &\ell^\mu_S \sim Q\, (\lambda,\lambda,\lambda).&
\end{align} 
For the collinear-$q_i$ mode we use light-cone vectors $n_+^{\mu}$ and $n_-^{\mu}$ defined above generalised for each of the collinear directions. See beginning of Section~\ref{sec:scet} for explicit construction.

\subsubsection{Three-loop and beyond}

In the previous sections, we have categorised the loop momenta scaling relevant for understanding the region expansion up to two-loops.
The complete knowledge of these regions allows us to compute the one-loop and two-loop amplitudes not only at leading power but also at any order in the power expansion.

\begin{figure}[h]
\centering
    \begin{tikzpicture}[y=0.6cm,x=1.2cm]
    \draw[c,dashed ] (-2,-2) node[left ] {\scriptsize $q_1$}-- (-1,-1) ;
    \draw[hc,dashed] (-2,+2) node[left ] {\scriptsize $q_2$}-- (-1,+1) ;
    \draw[dashed   ] (+2,+2) node[right] {\scriptsize $q_4$}-- (+1,+1) ;
    \draw[dashed   ] (+2,-2) node[right] {\scriptsize $q_3$}-- (+1,-1) ;

    \draw[         ]( 1  ,-1) -- ( 1  , 1);
    \draw[s        ](+0.5,-1) -- ( 1  ,-1);
    \draw           (+0.5, 1) -- ( 1  , 1);
    \draw[s        ](-1  ,-1) -- (+0.5,-1);
    \draw[c        ](+0.5,+1) -- (-1,-1);
    \draw[hc,dashed](-1  , 1) -- (-0.5, 0);
    \draw[hc,dashed](-1  ,1 ) -- (-0.5, 1);
    \draw[s, dashed](+0.5,-1) -- (-0.5, 0);
    \draw[hc,dashed](-0.5, 1) -- (+0.5, 1);
    \draw[hc,dashed](-0.5, 1) -- (-0.5, 0);

    \draw [->] (-0.8,-1.2) -- (-0.2,-1.2) node [midway,anchor=north] {\tiny$\ell_1$};
    \draw [->] (0,-0.5) ++ (45:0.2) ++ (-45:-0.3) --+ (-45:0.6) node [midway,anchor=south west] {\tiny$\ell_2$};
    \draw [->] (-0.9,1.2) -- (-0.6,1.2) node [midway,anchor=south] {\tiny$\ell_3$};
\end{tikzpicture}
    \quad
    \begin{tikzpicture}[y=0.6cm,x=1.2cm]
    \draw[dashed,c] (-2,-2) node[left ] {\scriptsize $q_1$}-- (-1,-1) ;
    \draw[dashed  ] (-2,+2) node[left ] {\scriptsize $q_2$}-- (-1,+1) ;
    \draw[dashed  ] (+2,+2) node[right] {\scriptsize $q_4$}-- (+1,+1) ;
    \draw[dashed  ] (+2,-2) node[right] {\scriptsize $q_3$}-- (+1,-1) ;

    \draw[c](-1, 1) -- (-1, 0); 
    \draw [->] (-1.2,0.8) -- (-1.2,0.2) node [midway,anchor=east] {\tiny$\ell_1$};
    \draw[c](-1, 0) -- (-1,-1);
    \draw[s]( 1, 1) -- ( 1, 0);
    \draw [->] (1.2,0.8) -- (1.2,0.2) node [midway,anchor=west] {\tiny$\ell_2$};
    \draw[s]( 1, 0) -- ( 1,-1);
    
    \draw(1, -1) -- ( 0,-1);
    \draw[c](0, -1) -- (-1,-1);
    \draw(1,  1) -- ( 0, 1);
    \draw(0,  1) -- (-1, 1);

    \draw[dashed](0,-1) -- (0,+1);
    \draw[sc,dashed](-1,0) -- (+1,0); 
    \draw [->] (-0.8,0.2) -- (-0.2,0.2) node [midway,anchor=south] {\tiny$\ell_3$};
\end{tikzpicture}
    \caption{Example three-loop diagrams containing modes beyond soft, collinear, and hard. (left) A region with a hard-collinear-$q_2$ loop momentum, shown in \textbf{\textcolor{hc}{blue}}. (right) A region containing a soft-collinear-$q_1$ loop momentum, shown in \textbf{\textcolor{sc}{pink}}. Both diagrams also contain soft in \textbf{\textcolor{s}{green}} and collinear in \textbf{\textcolor{c}{orange}} loop momenta. }
    \label{fig:three_loop}
\end{figure}
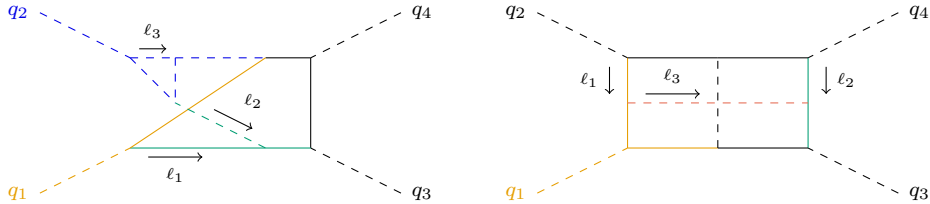

If we wish to understand the all-order structure of the QCD corrections to $HH$ production at any order in the power expansion, we must ask if additional loop momenta scalings appear in the region expansion beyond two-loops.
We generate all diagrams relevant to Higgs pair production at three-loops and obtain a list of all regions appearing in the MoR analysis for each diagram.
Using the tool described in the previous section, we find that considering only the collinear and soft loop momenta modes is insufficient to capture the regions appearing at three-loops, see Figure~\ref{fig:three_loop} for example diagrams that contain additional loop momenta modes.
However, adding the following three-loop momenta modes allows us to describe the regions appearing in diagrams containing up to three-loops,
\begin{align}
    &\mathrm{hard-collinear}-q_i:& &\ell^\mu_{HC_i} \sim Q\, (1,\lambda, {\lambda}^{1/2}),& \\
    &\mathrm{soft-collinear}-q_i:& &\ell^\mu_{SC_i} \sim Q\, (\lambda,\lambda^2, \lambda^{3/2}),& \\
        &\mathrm{ultra-soft}:& &\ell^\mu_{US} \sim Q\, (\lambda^2,\lambda^2,\lambda^2).&
\end{align}
The appearance of new loop momentum scalings at each loop order in the small mass expansion is consistent with other results in the literature, see for example Ref.~\cite{Ma:2023hrt}, which discusses the modes appearing in the small mass expansion of heavy-to-light decays.
In the present analysis, we expect further modes to enter at four-loop and for each additional loop order.

Suppose we instead only wish to understand the all-order structure at leading power.
In that case, it is important to consider not only the regions relevant to the expansion of the scalar integrals but also the order in which they are relevant to the power expansion of the complete amplitude.
In particular, regions can be power-suppressed (or enhanced) by the non-trivial numerator structure of gauge amplitudes.
In Sections~\ref{sec:structure_one_loop_amplitude} and~\ref{sec:structure_two_loop_amplitude} we will analyse the power suppression due to the numerator of the amplitude at one- and two-loops, respectively.
In Section~\ref{sec:scet}, we will discuss the general all-order structure of the amplitude. 

\subsection{Structure of the one-loop amplitude}
\label{sec:structure_one_loop_amplitude}

The hard region can be computed by rescaling parameters according to Eq.~\eqref{eq:ux} with region vector $\mathbf{u}^{(h)}$ and then expanding about small $\lambda$. This procedure is equivalent to \emph{directly} expanding the original integrand in $m_t$. Expanding to next-to-leading power, the hard region result for the first form factor is given by,
\begin{align}
A^{(h)}_{1,y_t^2} =& \frac{4 y_t^2}{s} 
\left\{ 2  
- 2 m_t^2 \left[
-\frac{2}{\epsilon^2 s}
-\frac{1}{\epsilon}\left( \frac{s^2+2 t u }{s t u}\, l_s + \frac{l_t}{u} + \frac{l_u}{t} \right)
\right. \right. \nonumber \\
& \left. \left. 
+
\frac{-l_s^2 + 2 l_t^2 + 2 l_u^2}{s} 
+ \frac{l_s\, l_t}{t} + \frac{l_s\, l_u}{u} + \frac{(t - u)^2\, l_t\, l_u}{s t u}
\right. \right. \nonumber \\
& \left. \left.
-\left( \frac{2}{s} + \frac{1}{t} + \frac{1}{u} \right) l_s - \frac{t\, l_t}{s u} - \frac{u\, l_u}{s t}
+ \frac{60 + 13 \pi^2}{6 s}
\right]
+ \mathcal{O}(m_t^4) \right\},
\end{align}
where $l_x = \ln(-\mu^2/x)$.
The hard region can be evaluated without an additional analytic regulator and does not generate logarithms involving the top-quark mass.
The leading power term in the $m_t$ expansion is finite in the $\epsilon \rightarrow 0$ limit and corresponds to the massless, $m_t = 0$, amplitude.
The terms at next-to-leading power and beyond contain poles in $\epsilon$ that must be cancelled by the poles appearing in the collinear regions.

The collinear regions can be computed similarly after rescaling parameters according to the region vectors $\mathbf{u}^{(c_1)},\ldots,\mathbf{u}^{(c_4)}$.
The singularities in the collinear regions are not fully regulated by $\epsilon$ alone. We introduce an additional analytic regulator $\eta$ with $\alpha=\beta=1+\eta$ and $\gamma=\delta=1-\eta$.
The singular dependence on the additional regulator will cancel among the collinear regions, and we can ultimately take the limit $\eta \rightarrow 0$ after summing the results for the collinear regions.
Computing each region of the first form factor to next-to-leading power we obtain:
\begin{eqnarray}
   A^{(c_1)}_{1,y_t^2} = &&  \frac{-4 y_t^2 m_t^2 \,  \mu ^{2 \epsilon }  e^{\gamma_E  \epsilon }
     \Gamma (2 \eta )^2 \Gamma (\epsilon +2 \eta ) }{s (s t u)   (1-2 \eta -\epsilon ) \Gamma (4 \eta ) \Gamma (1+\eta )^2 }
     \left(\frac{1}{m_t^2}\right)^{2 \eta +\epsilon}
\nonumber \\ &&  \times
     \bigg\{ (-t)^{\eta } (-u)^{\eta } \Big[s^2 \left(1+2 \epsilon ^2-(2+\eta ) \epsilon \right)-2 \eta \, t u\Big]
     \nonumber \\ && 
     +(-s)^{\eta } (-t)^{\eta } \Big[s u \left(1+2 \epsilon ^2-(2+\eta ) \epsilon \right)-t u (1-3 \eta -(2-5 \eta ) \epsilon)\Big]
     \nonumber \\ && 
     +(-s)^{\eta } (-u)^{\eta } \Big[s t \left(1+2 \epsilon ^2-(2+\eta ) \epsilon \right)-t u (1-3 \eta -(2 -5 \eta ) \epsilon)\Big]\bigg\}\,,
\end{eqnarray}

\begin{eqnarray}
   A^{(c_2)}_{1,y_t^2} = &&    \frac{-4y_t^2 m_t^2    e^{\gamma_E  \epsilon } \Gamma (-2 \eta )}{s(s t u) \Gamma (-\eta )^2} \left(\frac{\mu ^{2 }}{m_t^2}\right)^{\epsilon } 
    \nonumber \\ && 
   \bigg\{\left({m_t^2}\right)^{2 \eta } (-t)^{-\eta } (-u)^{-\eta } \frac{ \left(s^2 \left(1 + 2 \epsilon ^2-(2 - \eta ) \epsilon \right)+2 \eta \, t u\right)  \Gamma (\epsilon -2 \eta ) \Gamma (-2 \eta ) }{\eta ^2  (1+ 2 \eta -\epsilon  )\Gamma (-4 \eta )}
      \nonumber \\ &&
      + \Big[(-s)^{-\eta } (-u)^{\eta } + (-s)^{-\eta } (-t)^{\eta } \Big]  \frac{4\, t u \, \Gamma (2 \eta ) \Gamma (-\eta ) \Gamma (1+ \epsilon )}{ \epsilon\,  \Gamma (1+ \eta )}
        \bigg\}\,,
\end{eqnarray}

\begin{eqnarray}
   A^{(c_3)}_{1,y_t^2} = &&   \frac{-2 y_t^2 m_t^2  \, e^{\gamma_E  \epsilon } \Gamma (-2 \eta )  }{s(stu)\,\eta ^2  \Gamma (-\eta )^2}\left(\frac{\mu ^{2  }}{m_t^2}\right)^{\epsilon}
   \bigg\{
   \left( {m_t^2}\right)^{2 \eta } (-s)^{-\eta } (-u)^{-\eta }  
   \frac{ t^2    \Gamma (-2 \eta ) \Gamma (\epsilon -2 \eta )
   }{s   (1 + 2 \eta -\epsilon ) (1-4 \eta )  \Gamma (-4 \eta )}
    \nonumber \\ &&  \hspace{0.7cm} \times 
     \Big[ -2 s (1-4 \eta )  \left(1 -(2-\eta ) \epsilon + 2 \epsilon ^2\right)
    +u \left(4 -5 \eta -(4-8 \eta ) \epsilon +4 \eta ^2\right)\Big] 
   \nonumber \\ &&  \hspace{-0.7cm} 
   +(-t)^{-\eta }(-u)^{\eta } 
   \frac{16  tu  }{\Gamma (\eta )} 
   \Gamma (1-\eta ) \Gamma (1+ 2 \eta )    \Gamma (\epsilon )
   +(-s)^{\eta } (-t)^{-\eta }
   \frac{4 tu  \Gamma (1-\eta ) \Gamma (2 \eta )  \Gamma (\epsilon )}{s (1 -\epsilon ) \Gamma (\eta )} 
     \nonumber \\ && \hspace{+1.2cm}  \times
   \Big[s \left(2-9 \eta +( 2+ 8 \eta ) \epsilon -4 \epsilon ^2\right)
   +t (4-9 \eta -(4-8 \eta ) \epsilon )\Big]
   \bigg\}\,,
\end{eqnarray}

\begin{eqnarray}
   A^{(c_4)}_{1,y_t^2} = &&    \frac{-2y_t^2m_t^2\,e^{\gamma_E  \epsilon } \Gamma (-2 \eta )   }{s(stu)\,\eta ^2  \Gamma (-\eta )^2} \left(\frac{\mu ^{2 }}{m_t^2}\right)^{\epsilon }
   \bigg\{ \left({m_t^2}\right)^{2 \eta }(-s)^{-\eta } (-t)^{-\eta }   
   \frac{  u^2\, \Gamma (-2 \eta )  \Gamma (\epsilon -2 \eta )}{s  (1+ 2 \eta -\epsilon )(1 - 4 \eta )  \Gamma (-4 \eta ) }
     \nonumber \\ && \hspace{+0.7cm}  \times
   \Big[(- 2s  (1 - 4 \eta )  \left(1 - (2 -\eta ) \epsilon + 2 \epsilon ^2\right)  
     +t  \left(4 -5 \eta -(4 -8 \eta ) \epsilon + 4 \eta ^2 \right)\Big]
    \nonumber \\ && \hspace{-0.7cm} 
  +(-t)^{\eta } (-u)^{-\eta } 
   \frac{16 tu }{ \Gamma (\eta )} \Gamma (1-\eta ) \Gamma (1+ 2 \eta )   \Gamma (\epsilon )
   +(-s)^{\eta } (-u)^{-\eta } \frac{4 t u  \Gamma (1-\eta ) \Gamma (2 \eta )\Gamma (\epsilon )}{s (1 - \epsilon ) \Gamma (\eta )} 
     \nonumber \\ && \hspace{+1.2cm}  \times
   \Big[ s \left(2-9 \eta +(2+8 \eta ) \epsilon-4 \epsilon ^2 \right)+u (4-9 \eta -(4-8 \eta ) \epsilon )\Big] 
   \bigg\}\,.
\end{eqnarray}
Two features are immediately apparent:
firstly, the leading power $y_t^2 m_t^0$ term is not present for the collinear regions, and so their contribution to the amplitude starts only at the next-to-leading power,
secondly, the collinear regions can generate logarithms of $m_t$ from $(1/m_t^2)^{2\eta + \epsilon}$ hitting poles in either $\eta$ or $\epsilon$.
The second feature means that the logarithmic dependence on $m_t$ beyond leading power is significantly more complicated and necessitates the study of the collinear regions at next-to-leading power.

Here, we have presented results only for the first form factor. The second form factor can be computed straightforwardly using the same techniques; the expressions are significantly longer, but, the collinear regions are again suppressed relative to the hard region.

As will become apparent in what follows, the collinear regions are suppressed relative to the hard region at the level of the amplitude due to helicity conservation.
We note that, as discussed in Section~\ref{ssec:scalar_one_loop}, all regions contribute at leading power in the limit $\epsilon, \eta \rightarrow 0$ for the scalar integrals. 
However, when considering a collinear region, at amplitude level we can see that we obtain an extra power of $m_t^2$ from the Dirac algebra.
We will discuss the reason behind this power suppression of soft/collinear regions in Section~\ref{sec:scet}.

\subsection{Structure of the two-loop amplitude}
\label{sec:structure_two_loop_amplitude}
In Section~\ref{sec:structure_one_loop_amplitude}, it was shown by explicit calculation that at one-loop, the collinear regions are suppressed by two powers of $m_t$ relative to the hard region and thus do not contribute at leading power.
We stated that the origin of this suppression at the amplitude level is helicity conservation.
We now examine to what extent the collinear and soft regions are suppressed relative to the hard region beyond one-loop.
As a starting point, let us consider the top-level two-loop topologies shown in Fig.~\ref{fig:gghh_diag2l}.
In principle, we could compute each region's amplitude at the two-loop level. However, it is possible to show that the contributions must be suppressed by considering the behaviour of the numerator of the amplitude without computing the scalar integrals.

The analysis of the non-planar topologies is more involved than the planar topologies due to the presence of the soft regions.
Here, we will present the analysis of the regions of diagram $\mathbf{NP1}$ in detail; the procedure for the remaining diagrams follows identically.
Considering the first region given in Table~\ref{tab:regions2LNP1}, both loop momenta, $\ell_1, \ell_2$, are collinear to the external momentum $p_1$.
Taking just the numerator of the diagram, we can shift the loop momenta according to the ``routing'' column of Table~\ref{tab:regions2LNP1}.
Next, we can insert the following scalings of the loop momenta, 
\begin{align}
&\ell_1^2 \sim \lambda^2\,Q^2,
& &\ell_2^2 \sim \lambda^2\,Q^2,
& &\ell_1 \cdot \ell_2 \sim \lambda^2\,Q^2,
& &\ell_1 \cdot q_1 \sim \lambda^2\,Q^2,
& &\ell_2 \cdot q_1 \sim \lambda^2\,Q^2,&
\end{align}
which follows from the fact that the loop momenta are both collinear to $q_1$.
Furthermore, to leading power, if the loop momenta are collinear to an external momentum, then we can additionally use that
\begin{align}
\ell_1 \propto q_1,\qquad\ell_2 \propto q_1.
\end{align} 
Inserting these relations, we find that the numerator of the amplitude, after applying projectors and computing traces, scales as $\lambda^2$ and is suppressed by two powers of $m_t$ relative to the hard region.
Since the scalar integral contributes at $m_t^0$ for this region, we can conclude that the region is suppressed by at least two powers of $m_t$.

Let us now consider the third region of $\mathbf{NP1}$ given in Table~\ref{tab:regions2LNP1}.
In this region both loop momenta are instead soft and therefore scale as,
\begin{align}
&\ell_1^2 \sim \lambda^2\,Q^2,& 
&\ell_2^2 \sim \lambda^2\,Q^2,&
&\ell_1\cdot \ell_2 \sim \lambda^2\,Q^2,& \nonumber \\
&\ell_1\cdot q_2 \sim \lambda\,Q^2,& 
&\ell_1\cdot q_3 \sim \lambda\,Q^2,& 
&\ell_1\cdot q_4 \sim \lambda\,Q^2,& \\
&\ell_2\cdot q_2 \sim \lambda\,Q^2,& 
&\ell_2\cdot q_3 \sim \lambda\,Q^2,& 
&\ell_2\cdot q_4 \sim \lambda\,Q^2.& \nonumber
\end{align}
Inserting these relations into the numerator of the amplitude, we again find that it scales as $\lambda^2$.
However, unlike the collinear and hard regions, the underlying scalar integral itself is \emph{enhanced} by a power of $m_t^{-1}$ as can be seen in the 3rd and 6th rows of the ``order'' column of Table~\ref{tab:regions2LNP1}.
Therefore, from this analysis, we conclude that, at the level of individual diagram topologies, the soft-soft region is suppressed by at least one power of $m_t$ relative to the hard region.

This exercise may be repeated for each diagram and region entering the two-loop amplitude.
We have performed this procedure for each of the top-level two-loop diagrams and observe that the soft and collinear regions are indeed suppressed by at least one power of $m_t$ relative to the hard. This is consistent with the result of Ref.~\cite{Davies:2018qvx} in which the leading term of the two-loop amplitude appears to originate solely from the hard region.
The discussion above was focussed only on topologies that include one closed fermion loop. We have verified that the outcome does not change when considering the full set of diagrams. This is also true beyond the two-loop level.

We remark that the above analysis can also be carried out at one-loop, which makes it possible to establish that the collinear regions are suppressed without having to compute any integrals as we did in Section~\ref{sec:structure_one_loop_amplitude}.

Finally, we note that the hard, collinear, and soft regions can be computed separately at two-loops.
We have argued that only the hard region is required to describe the amplitude at leading power, but all modes will eventually contribute beyond leading power.
In principle, the amplitude computation in a given region is simpler than computing the entire amplitude and then expanding.
This is because the amplitude will depend non-trivially on fewer scales after applying the method of regions, which simplifies the evaluation of the master integrals and, naively, results in a simpler IBP reduction. 
However, the presence of the analytic regulator complicates matters.
Some IBP codes, such as Kira~\cite{Klappert:2020nbg}, support analytic regulators, but the regulator still increases the complexity, acting like an additional scale or dimensional regulator.
Luckily, one may use the factorising properties of the amplitude (see~Section~\ref{sec:scet}) and only calculate the regions that are required to obtain the two-loop contribution to a jet function.
Crucially, this implies that we can ignore the mixed hard and collinear regions that arise from a one-loop times one-loop contribution, which usually have a more complicated kinematic dependence.
For integrals where, as in Eq.~\eqref{eq:scaling1l}, the scaling factorises, the IBP can be carried out without any scales, greatly simplifying the calculation.
Finally, by first expanding we also avoid having to calculate integrals to a higher power in $m_t$ than required.

\section{All-order factorisation in Effective Field Theory}
\label{sec:scet}
The high-energy structure of the amplitude can be captured using subleading power SCET formalism suitable for processes with multiple collinear directions. Building on the fixed-order MoR investigations in the preceding sections, we construct an effective field theory description, which formalises and elucidates our findings to all orders in perturbation theory. We explore the structure of factorisation to next-to-leading power in the high-energy expansion where hard, collinear, and soft regions all contribute to the amplitude, and discuss the underlying reasons for the straightforward leading power result where only the hard region is present.   In this section, we follow conventions introduced in \cite{Beneke:2017ztn, Beneke:2018rbh, Beneke:2019kgv}, see also \cite{Feige:2017zci, Moult:2017rpl}.

To set up the framework, we generalise the notation introduced
in the previous section and introduce a set of light-cone reference
vectors $n_{i-}^{\mu}$ and $n_{i+}^{\nu}$ for each collinear direction
$i,j,\ldots$, such that 
$n_{i-}\cdot  n_{j-}\sim n_{i+}\cdot  n_{j+} \sim \mathcal{O}(1)$ 
for all $i\neq j \in \{1,2,3,4\}$, 
and $n^2_{i+}=n^2_{i-}=0$, $n_{i-}\cdot n_{i+}=2$. 
In a generic reference frame, the on-shell external momenta
for massless scattering can  then be written as
\begin{align}\label{eq:icollinearmomenta}
    q^\mu_i  = (n_{i+}\cdot  q_i) \frac{n^\mu_{i-}}{2}\,.
\end{align}
The metric tensor is decomposed  as
\begin{eqnarray}\label{eqn:metricDecompGeneraliDirection}
g^{\mu \nu} = n_{i+}^{\mu}\frac{n_{i-}^{\nu}}{2}+n_{i-}^{\mu}
\frac{n_{i+}^{\nu}}{2}    +g^{\mu \nu}_{\perp_i}\,,
\end{eqnarray}
which implicitly defines $g^{\mu \nu}_{\perp_i}$, 
and for an arbitrary momentum $k$ the decomposition along the $i-$th light-cone direction is given by 
\begin{align}
    k^\mu  = (n_{i+}\cdot  k) \frac{n^\mu_{i-}}{2}  + (n_{i-}\cdot k) \frac{n^\mu_{i+}}{2}+ k_{\perp_i}^{\mu}\,.
\end{align}
The label $\perp_i$ denotes  directions perpendicular  to $n_{i-}^{\mu}$ and
$ n_{i+}^{\mu}$  such that  $n_{i-}\cdot k_{\perp_i}=n_{i+}\cdot k_{\perp_i}=0$,
but not necessarily with respect to another direction $j$:
$n_{i\pm}\cdot k_{\perp_j}\neq 0$. 
For the 2-to-2 kinematics in the problem at hand, it is often
convenient to use the partonic centre-of-mass frame, 
where $n_{1\pm}^{\mu}=n_{2\mp}^{\mu}$ and $n_{3\pm}^{\mu}=n_{4\mp}^{\mu}$.

\begin{figure}
    \centering
    \includegraphics[width=0.5\textwidth]{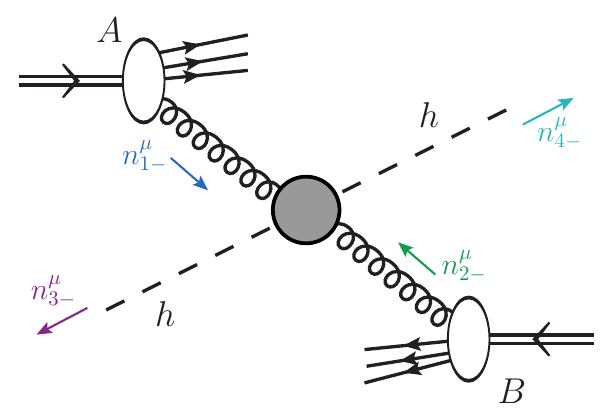}
    \caption{The light-cone directions for $gg\to HH$. In the partonic center-of-mass frame, the gluons and Higgs bosons are back-to-back; hence $n_{2\pm}$ can be eliminated in favour of $n_{1\mp}^{\mu}$, $n_{2\pm}^{\mu}=n_{1\mp}^{\mu}$  and similarly for $n_{4\pm}^{\mu}=n_{3\mp}^{\mu}$.  }
    \label{fig:comgghhV2}
\end{figure}

\subsection{Hard scale \texorpdfstring{$\mu^2\sim s$}{musq ~ s}}

The large kinematic invariants $s,t,u$ for the wide-angle scattering amplitude set the hard scale for the process, at which the Standard Model has to be matched onto the effective field theory. The power-counting arguments given below are valid to all orders in the strong coupling constant $\alpha_s$, and we work only to the leading order in the electroweak couplings. Nonetheless, we provide the relevant expression for the collinear Higgs Lagrangian below that can be used to extend this work to all orders in the electroweak expansion.

First, let us briefly review the SCET formalism for multi-jet processes. The SCET operators are constructed from collinear-gauge invariant
building blocks given by \cite{Hill:2002vw},
\begin{equation}
\label{eq:gaugeInvBuildBlocksCi}
{\psi_{c_i}}(x)  \in
\quad
 \left\{ \begin{array}{ll}
\displaystyle 
\chi_{c_i}(x)=W_{c_i}^{\dagger}(x)\xi_{c_i}(x)  & {\quad{\text{${i}$-collinear quark,}}} \\[0.3cm]
\displaystyle 
\mathcal{A}^{\mu}_{{c_i}\perp_i}(x)= 
W_{c_i}^{\dagger}(x)\big[iD^{\mu}_{{c_i}\perp_i}W_{c_i}(x)\big]
 & {\quad{\text{${i}$-collinear gluon,}}} 
\end{array}\right.
\end{equation}
where $	\xi_{c_i}(x)$ is the $i-$collinear quark field and the corresponding collinear covariant derivative is
$i D_{c_i}^{\mu}(x)= i\partial^{\mu} +g_s A_{c_i}^{\mu}(x)$.
The $i$-collinear Wilson line is defined as,
\begin{equation}\label{collinearWilsonline}
W_{c_i}(x)= \textbf{P}\exp\bigg[ig_s\int_{-\infty}^{0} ds\,
 n_{i+}\cdot A_{c_i}(x+sn_{i+})   \bigg]\,,
\end{equation}
where $\textbf{P}$ denotes the path ordering operator. 
A generic $N$-jet operator, up to relative $\mathcal{O}(\lambda^2)$ power-corrections, has the following structure \cite{Beneke:2017ztn},
\begin{eqnarray}\label{eq:NjetOperator}
J= \int \bigg[\prod_{ik} dt_{i_k}\bigg]
C(\{t_{i_k}\})\, \prod_{i=1}^NJ_{c_i}(t_{i_1},t_{i_2}...) \,,
\end{eqnarray}  
where $C(\{t_{i_k}\})$ is a generalised Wilson coefficient, depending on the complete set $\{t_{i_k}\} = \{t_{i_k}:i\in\{1,\ldots, N\} \wedge k \in \{ 1,\ldots, N_i \}  \}$ of light-cone positions, that captures the hard modes and $J_{c_i}$ is a product of $N_i$  collinear building blocks associated
with a specific collinear direction~$n_{i+}^{\mu}$,
\begin{eqnarray}\label{eq:ColliOperator}
J_{c_i}(t_{i_1},t_{i_2}...)=
\prod_{k=1}^{N_{i}\leq 3}\psi_{{c_i}_k}(t_{i_k} n_{i+}).
\end{eqnarray}
Since the collinear particles have large momenta components of the order of the hard scale, the operators are non-local along the light-cone directions as indicated by the position variable $t_{i_k}$. 

In SCET, every field has a unique scaling in the expansion parameter $\lambda$. Thus, analysing operators and their $\lambda$ suppression is straightforward. A single building block is present at leading power in each collinear direction. 
There are three ways to extend the basis of operators to subleading powers \cite{Beneke:2017ztn, Beneke:2019slt}.  
First, we can introduce $\partial_{\perp}^{\mu}$ derivatives, which act on the building blocks present already in the leading power configuration, bringing an $\mathcal{O}(\lambda)$ suppression. Second, more building blocks can be added within each collinear direction. Each additional insertion of $\psi_{c_i}$ induces an $\mathcal{O}(\lambda)$ suppression. Third, power suppression is induced by direct insertions of factors of $m_t$, which is treated as a building block to maintain homogeneous power-counting; see also \cite{Beneke:2019slt}. The basis of subleading power currents includes time-ordered products of  subleading power Lagrangian terms with lower power currents.

\subsubsection{Leading power matching}
\label{sec:lpmatching}

In the case under consideration, the number of collinear directions is set to $N=4$ and up to the two possible helicity projections\footnote{We could construct the alternative version of the basis using helicity building blocks as in \cite{Feige:2017zci, Moult:2017rpl}. These two methods are equivalent except for the definition of evanescent operators.}, there exists a unique field structure of leading power operators constructed from collinear gluon and Higgs fields,
\begin{align}\label{eq:lp-col-op}
J^{[i]}_{{\rm{LP}}}(t_1,t_2,t_3,t_4) =y_t^2  \mathcal{P}^{\mu\nu}_{[i]}
\mathcal{A}_{{\rm c}_1\perp_1\;\mu}(t_1 n_{1+})
\mathcal{A}_{{\rm c}_2 \perp_2\;\nu}(t_2 n_{2+}) 
h_{{\rm c}_3}(t_3 n_{3+})h_{{\rm c}_4}(t_4 n_{4+})  \;,
\end{align}
with $i=1,2$ and the tensors  $\mathcal{P}^{\mu\nu}_{[i]}$ are,
\begin{eqnarray}\label{eq:projectors1}
\mathcal{P}^{\mu\nu}_{[1]}  &=&
g^{\mu\nu} - \frac{n_{1-}^{\nu} n_{2-}^{\mu}}{n_{1-}\cdot n_{2-}}, \\
\mathcal{P}^{\mu\nu}_{[2]}  &=&  g^{\mu\nu} + 
\frac{n_{1-}\cdot n_{2-}\; n_{3-}^{\nu} n_{3-}^{\mu}}{n_{1-}\cdot n_{3-}\;n_{2-}\cdot n_{3-}}  
    - \frac{n_{1-}^{\nu} n_{3-}^{\mu}}{n_{1-}\cdot n_{3-}} 
    - \frac{n_{3-}^{\nu} n_{2-}^{\mu}}{n_{3-}\cdot n_{2-}}\,.
\end{eqnarray}
\begin{figure}
    \centering
    \includegraphics[width=0.93\textwidth]{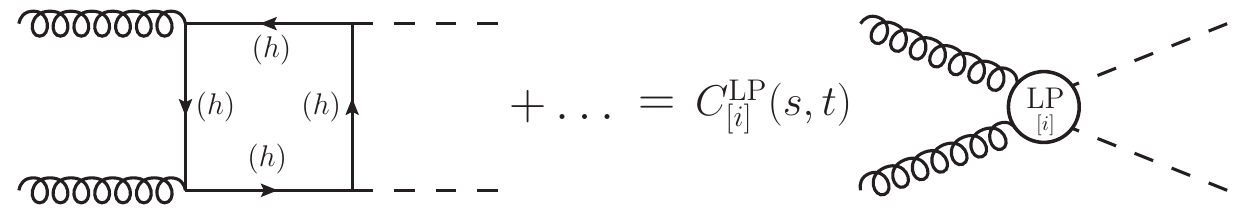}
    \caption{Graphical representation of the leading order matching to SCET at leading power. The label
    $(h)$ denotes the hard scaling of the propagator. The ellipses indicate the remaining five box diagrams, namely the $t$- and $u$-type topologies, in addition to the depicted $s$-type topology, as well as the corresponding diagrams with reversed fermion flow of the top quark loop. }
    \label{fig:lp_matching_graph}
\end{figure}
To perform the matching as depicted in Fig.~\ref{fig:lp_matching_graph}, we calculate in six-flavour QCD the following matrix element of the operators defined above,
\begin{align}\label{eq:lpmatching}
\hspace{-0.36cm}
\left .   
 \left< H(q_3)H(q_4)\right| \int d{\mathbf{t}} \frac{1}{g_s^2}\, {C}_{[i]}^{{\rm{LP}}}(t_1,t_2,t_3,t_4) J^{[i]}_{{\rm{LP}}}(t_1,t_2,t_3,t_4) \left| 
    g(q_1) g(q_2)\right> \equiv  A_{i,y_t^2} (s,t)
\right|_{\rm hard}
\end{align}
where $d\mathbf{t}=dt_1 dt_2 dt_3 dt_4$. 
The form factor $A_{i,y_t^2}$ was defined in~\eqref{eq:Ai}. Here all particles are massless, i.e. only the hard region contributes to this matching as discussed in Section~\ref{sec:regions}. We could, in principle, perform this matching step in the unbroken phase of the Standard Model, but it does not offer any advantage when focusing only on the QCD corrections. 

It is customary to define  the Fourier-transformed Wilson coefficients,
\begin{eqnarray}
    C_{[i]}^{{\rm{LP}}}(n_{1+}\cdot q_1, n_{2+}\cdot q_2 ,n_{3+}\cdot q_3 ,n_{4+}\cdot q_4 ) 
   & =& \int  dt_1 dt_2 dt_3 dt_4\, {C}^{{\rm{LP}}}_{[i]}(t_1,t_2,t_3,t_4) 
    \\ && \hspace{-0.4cm} \times \nonumber   e^{-i  \,( n_{1+}\cdot  q_1 )\, t_{1} }
e^{-i  \,(n_{2+}\cdot  q_{2})\, {t}_{2} }
 e^{-i  \,( n_{3+}\cdot  q_3 )\, t_{3}} 
e^{-i  \,(n_{4+}\cdot  q_{4})\, {t}_{4}}
\,,
\end{eqnarray}
which depend on the large momentum components of all the collinear fields. Their respective arguments represent quantities in coordinate space and their Fourier-transformed counterparts. To shorten the notation, using reparametrisation invariance, we can trade the large momentum components for Mandelstam variables in the massless kinematics
\begin{align}
 C_{[i]}^{{\rm{LP}}}(n_{1+}\cdot q_1, n_{2+}\cdot q_2 ,n_{3+}\cdot q_3 ,n_{4+}\cdot q_4)  \equiv  C_{[i]}^{{\rm{LP}}}(s,t)\,,
\end{align}
with 
$s=2 q_1 \cdot q_2= \frac{n_{1-}\cdot n_{2-}}{2}(n_{1+}\cdot q_1)(n_{2+}\cdot q_2)$, and $t=2q_1 \cdot  q_3= \frac{n_{1-}\cdot n_{3-}}{2}(n_{1+}\cdot q_1)(n_{3+}\cdot q_3)$. 

In addition to the leading power 
current given in \eqref{eq:lp-col-op}, 
we will also require LP type currents
with a single (anti)quark field building
block in one or two of the collinear
directions, $\chi_{c_i}(x)$
in \eqref{eq:gaugeInvBuildBlocksCi}.  
These currents can enter the description
of the $gg\to HH$ amplitude through 
time-ordered products with subleading power
Lagrangian terms, see the construction 
in~\cite{Beneke:2017ztn}.
In particular, through interactions with the 
$\mathcal{L}_{\xi q}$ Lagrangian an incoming collinear
gluon  can be turned into a collinear-quark
soft-antiquark pair, the first of which can then 
participate in the hard scattering described
by a current of the form: 
\begin{align}\label{eq:lp-col-op-chi}
J^{}_{{\rm{LP}}}(t_1,t_2,t_3,t_4) =y_t^2  \,
\bar{\chi}_{{\rm c}_1}(t_1 n_{1+})\,
\chi_{{\rm c}_2 }(t_2 n_{2+}) \,
h_{{\rm c}_3}(t_3 n_{3+})h_{{\rm c}_4}(t_4 n_{4+})  \;.
\end{align}
We discuss the role of these types of
currents further in 
Section~\ref{sec:attopmassscale}. 

\subsubsection{Next-to-leading power matching}
\label{sec:nlpmatching}
We are now ready to discuss the leading power $m_t$ dependence of the amplitude. Still, before we proceed, it is instructive to sketch the factorisation at next-to-leading power in $m_t$ expansion to explain the origin of power suppression at NLO and NNLO for collinear regions observed in Section~\ref{sec:amplitudes} and generalise it to all orders in $\alpha_s$.

At next-to-leading power, several power-suppressed operators contribute. As outlined below Eq.~\eqref{eq:ColliOperator}, the basis is constructed from the basic building blocks considered in \cite{Beneke:2017ztn} and explicit insertions of $m_t$, in analogy with \cite{Beneke:2019slt}. This section gives examples of two operators that appear already at leading order in $\alpha_s$ and are relevant for the leading logarithmic resummation at NLP.  

\begin{figure}
    \centering
    \includegraphics[width=0.92\textwidth]{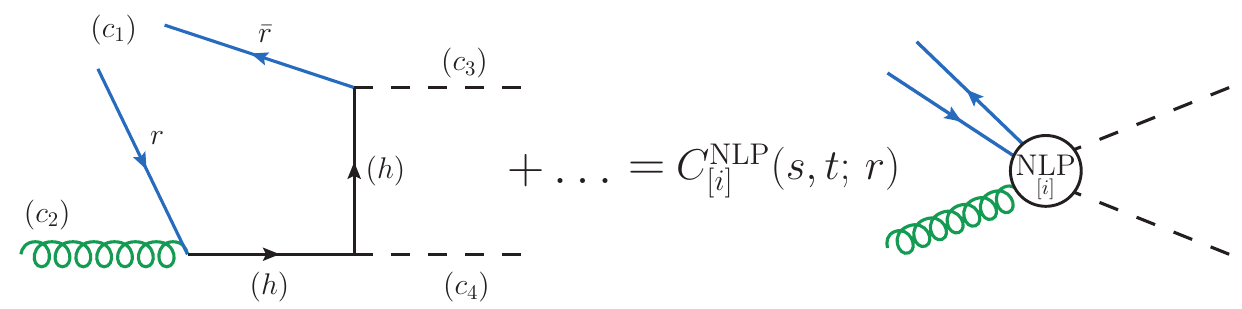}
    \caption{Matching to SCET at next-to-leading power. The $(h),(c_i)$ labels depict the scaling of each line. The fraction of collinear momentum carried by the (anti)quark in the $(c_1)$ sector is denoted by $r (\bar{r})$, where $\bar{r}=(1-r)$. The ellipses contain the remaining five permutations of the possible attachments.  }
    \label{fig:matching}
\end{figure}

The matching of the $gg\to HH$ amplitude to a subleading power SCET operator is presented graphically in Fig.~\ref{fig:matching}. Besides trivial operators with explicit mass suppression, which are directly related to the leading power operator  $J^{[i]}_{{\rm{NLP}}}(t_1,t_2,t_3,t_4)= m_t^2 J^{[i]}_{{\rm{LP}}}(t_1,t_2,t_3,t_4)$, we need to consider operators with two collinear quark fields in a single collinear direction. 

We begin the analysis with the expression for the tree-level QCD process $t( q) +\bar{t}( q')+ g(q_2) \to h(q_3) +h(q_4)$, with $q+q' = q_1$, depicted on the left-hand side. The full QCD expression for the drawn diagram reads, 
\begin{equation}\label{eqn:matchingexpr}
  \mathcal{M}^{\rm{QCD}}_{\rm{Fig}\ref{fig:matching}}=  
 \bar{v}(  q')
   \bigg[\frac{-iy_t}{\sqrt{2}}\bigg]^2
   \frac{i (  -\slashed{{q}}' - \slashed{q}_3   + m_t)}{(q'+q_3)^2-m_t^2}
   \frac{i ( \slashed{q} + \slashed{q}_2 + m_t)}{(q+q_2)^2-m_t^2}
   ig_s\T^B\gamma_{\nu} u(q) \varepsilon^{\nu }(q_2)\,.
\end{equation}
The matching is performed on-shell, i.e. the external gluon is chosen to be transverse, and spinors obey the Dirac equation $ \slashed{q}u(q)=m_t u(q)$, $\bar{v}({q'})\slashed{{q'}}=-\bar{v}({q'})m_t$. 
Next, we expand the amplitude in $m_t$. The QCD spinor can be expanded according to~\eqref{eq:spin_exp}, and all the momenta respect their collinear scaling. We also introduce momentum fraction $r$, such that $n_{1+}\cdot q = rn_{1+}\cdot q_1 $ and $n_{1+}\cdot q' = \bar{r}n_{1+}\cdot q_1 $, where $\bar{r}=(1-r)$.
The leading term in $m_t$ is 
{\small\begin{equation}\label{eqn:matchingexpr10}
\mathcal{M}^{\rm{QCD}}_{\rm{Fig}\ref{fig:matching}-{{\rm{LP}}}} = - ig_s\T^B
    \frac{ y_t^2 }{2}  
   \frac{\varepsilon^{\nu }_{\perp_2}(q_2)  \bar{v}_{c_1}(\bar{r}q_1)   \slashed{n}_{3-} }{   \bar{r}(n_{1+}\cdot q_1)  n_{1-} \cdot n_{3-} }
   \frac{ \big[   2 r(n_{1+}\cdot q_1)  n_{1-\nu}  + (n_{2+}\cdot q_2)  \slashed{n}_{2-} \gamma_{\nu} \big]}{ r(n_{1+}\cdot q_1)(n_{2+}\cdot q_2) n_{1-}\cdot n_{2-}  }   u_{c_1}(rq_1) \,.   
\end{equation}}
For the matching calculation, it is convenient to choose a frame where $q_1$ and $q_2$ are back-to-back, such that $n^{\mu}_{1\pm}=n^{\mu}_{2\mp}$ and $\perp_1=\perp_2$ is perpendicular to the directions $n^{\mu}_{1\pm}$. With this choice,   $\slashed{n}_{3-}\to n_{3-}\cdot \gamma_{\perp_1} $  since the contributions proportional to  $\slashed{n}_{1\pm}$ vanish when acting on the collinear spinors. Thus, the results read 
\begin{eqnarray}\label{eq:treematchingres}
  \mathcal{M}^{\rm{QCD}}_{\rm{Fig}\ref{fig:matching}-{{\rm{LP}}}} &=& - ig_s\T^B
   \frac{ y_t^2 }{2} 
   \frac{\bar{v}_{c_1}(\bar{r}q_1)   n_{3-}\cdot \gamma_{\perp_1} }{   \bar{r}(n_{1+}\cdot q_1)  n_{1-} \cdot n_{3-} }
   \frac{     \slashed{n}_{1+} \gamma_{\perp_1\nu}  u_{c_1}(rq_1) }{2  r(n_{1+}\cdot q_1)    }   \varepsilon^{\nu }_{\perp_1}(q_2)\,.
\end{eqnarray}
We can further use the identity 
\begin{align}
    \gamma_\perp^\mu \gamma_\perp^\nu  \slashed n_{\pm} = \left(g_\perp^{\mu \nu} \mp  i \epsilon_\perp^{\mu \nu} \gamma_5 \right) \slashed n_{\pm}, \quad {\rm where } \quad 
    \epsilon^{\mu\nu}_{\perp}=\frac{1}{2} \epsilon^{\mu\nu\alpha\beta}n_{+\alpha}n_{-\beta},
\end{align}
with the convention $\epsilon^{0123}=+1$. Eq.~\eqref{eq:treematchingres} then reduces to,
\begin{eqnarray}\label{eq:LOmatching}
  \mathcal{M}^{\rm{QCD}}_{\rm{Fig}\ref{fig:matching}-{{\rm{LP}}}} &=& \frac{ig_s\T^B}{ n_{1-} \cdot n_{3-}   } 
    \frac{ y_t^2 }{2 (n_{1+}\cdot q_1)^2}  
   \frac{1 }{  r \bar{r}  }  \bigg(  \bar{v}_{c_1}(\bar{r}q_1)    \frac{\slashed{n}_{1+}}{2}  u_{c_1}(rq_1) n_{3-\nu} \varepsilon^{\nu}_{\perp_1}(q_2)
  \nonumber  \\ && \hspace{3.6cm} +  \bar{v}_{c_1}(\bar{r}q_1) \frac{\slashed{n}_{1+}}{2} \gamma_5 u_{c_1}(rq_1)  n_{3-}^{\mu}  i \epsilon^{\perp_1}_{\mu \nu}    \varepsilon_{\perp_1}^{\nu}(q_2)  \bigg)\,,
\end{eqnarray}
where the result transforms as a scalar and pseudoscalar under rotations in the transverse plane in the first and second lines, respectively.

\begin{figure}
    \centering
    \includegraphics[width=0.35\textwidth]{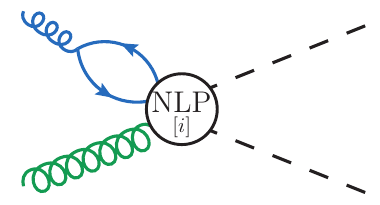}
    \hspace{1.5cm}
    \includegraphics[width=0.35\textwidth]{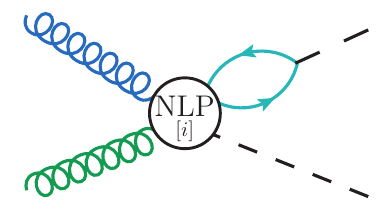}
    \caption{Mixing of the subleading power collinear quark-antiquark operators with a collinear gluon (left) and Higgs (right). The mixing is forbidden for leading power scalar and pseudoscalar operators in the gluon case and vector operators in the Higgs case, to all orders in perturbation theory due to helicity conservation in the massless limit.}
    \label{fig:NLPfactorization}
\end{figure}

The most general basis for an operator built of two collinear fermions consists of scalar $\frac{\slashed n_{i+}}{2}$, pseudoscalar  $\frac{\slashed n_{i+}}{2}\gamma_5$, and vector $\frac{\slashed n_{i+}}{2}\gamma_{\perp_i}^\mu$ 
Dirac structures\footnote{Scalar, pseudoscalar, and vector refer to transformation properties under rotations in the transverse plane.}.
Concretely, the relevant operators are given by 
\begin{align}\label{eq:nlpscalar}
    J_{S_i}( t_{i_1}, t_{i_2} ) &= 
  \bar{ \chi}_{{\rm c}_i }(t_{i_2} n_{i+})
   \frac{\slashed{n}_{i+}}{2}\,\chi_{{\rm c}_i }(t_{i_1} n_{i+}) , \\
   J_{P_i}( t_{i_1}, t_{i_2} )  &= 
    \bar{ \chi}_{{\rm c}_i }(t_{i_2} n_{i+})
   \frac{\slashed{n}_{i+}}{2}\,\gamma_5\,\chi_{{\rm c}_i }(t_{i_1} n_{i+}) ,  \\ \label{eq:nlpvector}
   J^{\nu\; A }_{V_i}( t_{i_1}, t_{i_2} )  &= 
     \bar{ \chi}_{{\rm c}_i }(t_{i_2} n_{i+})
   \frac{\slashed{n}_{i+}}{2}\,\gamma^{\nu}_{\perp_i} \T^A\,\chi_{{\rm c}_i }(t_{i_1} n_{i+}) \,.
\end{align}
Here, we write only colour singlet operators for scalar and pseudoscalars and colour octet for vector operators since only these operators contribute to the matrix element we consider. 
The Fourier transformation with respect to the positions $t_{i_k}$ is defined by 
\begin{eqnarray}
\label{eq:nlpFT}
    J_{X_i}(n_{i+}\cdot q_i,r,\bar{r}) = (n_{i+}\cdot q_i)^2 \int dt_{i_1}dt_{i_2}
    e^{i( t_{i_1}r +  t_{i_2} \bar{r}\,  )n_{i+}\cdot q_i} J_{X_i}( t_{i_1}, t_{i_2} )
\end{eqnarray}
for $X=S,P,V$.
Only the vector operator has a non-zero overlap with the gluon matrix element, while the scalar and pseudo-scalar operators do not contribute
\begin{align}
    \left< 0 \right| J_{S,P}(r) \left| g(q_1)\right> =0 \;.
\end{align}
Since only scalar and pseudoscalar structures are
present in the result in Eq.~\eqref{eq:LOmatching},
the leading term in $m_t$ expansion does not contribute
to vector operators. Therefore, it does not mix with the
gluon; see Fig.~\ref{fig:NLPfactorization}. 
\emph{This result holds to all orders in $\alpha_s$, since helicity is conserved in the limit $m_t \to 0$.}
Moreover, summing all leading order diagrams, denoted by
ellipses in Fig.~\ref{fig:matching}, yields a vanishing
result.

To get a non-zero contribution, we consider the subleading term in $m_t$ in Eq.~\eqref{eqn:matchingexpr}. The result reads
\begin{eqnarray}\label{eq:nlpmatrixelement}
\mathcal{M}^{\rm{QCD}}_{\rm{Fig}
\ref{fig:matching}{{\rm{NLP}}}}    
&& = ig_s\T^B
\frac{y_t^2 }{2(n_{1+}\cdot q_1)^2} 
\frac{1}{\bar{r}^2r^2 }
\bar{v}_{c_1}(q')  
\frac{\left(2\,m_t\right)}{n_{1+}\cdot q_1 }
\frac{ n_{4-}^{\mu}}{ n_{1-} \cdot n_{4-}  }
\frac{n_{3-}^{\eta}}{ n_{1-}\cdot n_{3-}}
\frac{\slashed{n}_{1+}}{2}
\nonumber \\ && \hspace{-0cm}  \times 
\Big[g_{\perp_1 \mu\nu}\gamma_{\perp_1 \eta}
- g_{\perp_1 \mu\eta}\gamma_{\perp_1\nu}
+ g_{\perp_1 \nu\eta}\gamma_{\perp_1\mu} \Big] 
u_{c_1}(q) \varepsilon^{\nu }_{\perp_1 }(q_2)\,
\end{eqnarray}
This time, we obtain non-zero matching onto the vector operator. Consequently, this structure can contribute to the amplitude at all orders starting from subleading power in the $\lambda$ expansion, as explicitly demonstrated in Section~\ref{sec:amplitudes}. 

On the effective field theory side, the structure encountered in the above calculation is naturally reproduced by the next-to-leading power operator of the following form 
\begin{align}\label{eq:c1gluonnlpOP}
J^{[1]}_{{\rm{NLP}}}( t_{1_1}, t_{1_2} , t_2,t_3,t_4) &=y_t^2 
 m_t\,J^{\nu\; A}_{V_1}( t_{1_1}, t_{1_2} )\,
\mathcal{A}_{\nu}^{{\rm c}_2 \perp_2\; A}(t_2 n_{2+})
h_{{\rm c}_3}(t_3 n_{3+})h_{{\rm c}_4}(t_4 n_{4+}) \,,
\end{align}
where an additional collinear field generates power suppression in one collinear direction and the $m_t$ insertion. To avoid cumbersome notation in the example, we have dropped the projection operators introduced in Eq.~\eqref{eq:projectors1}. However, these can be reinstated to obtain the two scalar amplitude structures defined in \eqref{eq:Ai}. The matching coefficient $C^{{\rm{NLP}}}_{[i]}(s,t\,;r)$ is extracted by considering a matrix element of the operator similarly to the leading power case in Eq.~\eqref{eq:lpmatching}, now with an incoming collinear quark-antiquark pair instead of one of the gluons, and comparing with results in Eq.~\eqref{eq:nlpmatrixelement}. For the first projection operator, which ultimately corresponds to the  $g_{\perp}^{\mu\nu}$ structure, we find 
\begin{eqnarray}
C^{{\rm{NLP}}}_{[1]}(s,t;r) &&= -  
\frac{1}{2(n_{1+}\cdot q_1)^2} \frac{1}{\bar{r}^2r^2 }
\frac{2g_{\perp_1 \mu\nu}}{n_{1+}\cdot q_1 }
\frac{ n_{4-}^{\mu}}{ n_{1-} \cdot n_{4-}  }
\frac{n_{3-}^{\nu}}{ n_{1-}\cdot n_{3-}}\,.
\end{eqnarray}
We point out that the endpoint divergences present in the matching coefficient as $\bar{r},r\to 0$ must be dealt with before consistent NLP resummation can be achieved. Single power endpoint divergences have previously been observed in \cite{Beneke:2020ibj, Beneke:2022obx, Liu:2019oav,Bell:2022ott}. Here, the additional momentum fractions arise from the power-suppressed parts of the collinear spinors and hence are constrained by reparametrisation invariance \cite{Beneke:2019kgv}. 

\begin{figure}
    \centering
    \includegraphics[width=0.92\textwidth]{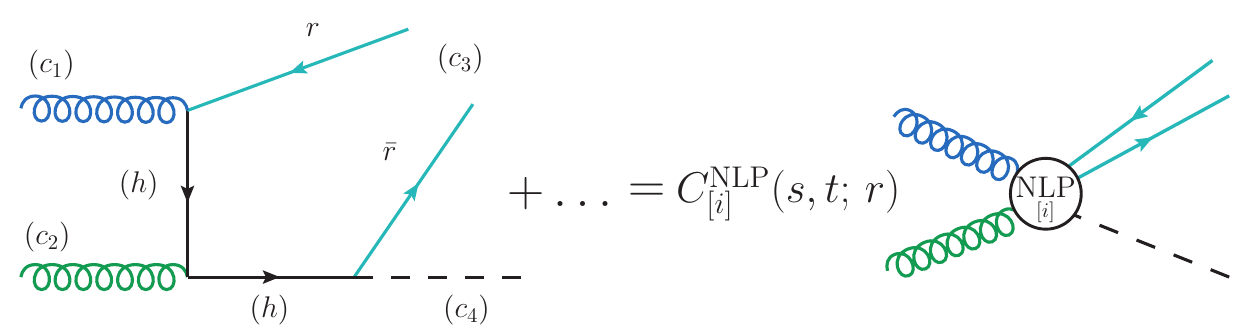}
    \caption{Matching to SCET at next-to-leading power. The $(h),(c_i)$ labels depict the scaling of each line. The fraction of collinear momentum carried by the (anti)quark in the $(c_3)$ sector is denoted by $r (\bar{r})$, where $\bar{r}=(1-r)$. The ellipses contain the remaining five permutations of the possible attachments.  }
    \label{fig:higgsmatching}
\end{figure}
Identical considerations follow for the calculation of the matching coefficient of the next-to-leading power operator contributing in the sector collinear to the second gluon, which is given by
\begin{align}\label{eq:c2gluonnlpOP}
J^{[2]}_{{\rm{NLP}}}(t_1, t_{2_1}, t_{2_2} ,t_3,t_4) &=y_t^2 m_t\,
\mathcal{A}_{\nu}^{{\rm c}_1 \perp_1\; A}(t_1 n_{1+})
J^{\nu\; A}_{V_2}( t_{2_1}, t_{2_2} )\,
h_{{\rm c}_3}(t_3 n_{3+})h_{{\rm c}_4}(t_4 n_{4+}) \,.
\end{align}
However, as our second concrete example, we focus on the structure of contributions in the collinear direction defined by one of the final state Higgs bosons. The graphical matching equation analogous to one appearing in Fig.~\ref{fig:matching} is depicted in Fig.~\ref{fig:higgsmatching}. Starting from the full QCD amplitude for the diagram present on the left-hand side, we perform the expansion in $m_t$ and find the following leading power result
\begin{eqnarray}
\mathcal{M}^{\rm{QCD}}_{\rm{Fig}\ref{fig:higgsmatching}-{{\rm{LP}}}} &=&
ig_s^2\T^B\,\T^A \frac{y_t}{\sqrt{2}}
\bar{v}_{c_3}(q')\frac{1}{\bar{r}(n_{3+}q_3)}
\bigg[\frac{2n_{3-\mu }}{(n_{1+}q_1)n_{3-}\cdot n_{1-}}
\frac{\slashed{n}_{3+}}{2}\gamma_{\nu \perp_3}u_{c_3}(q) 
\nonumber    \\ &&  \hspace{-2.5cm}     +  
\frac{1}{r (n_{3+}q_3) n_{3-}\cdot n_{1-}  } \Big(  n_{1-}\cdot n_{3-} n_{3+\nu}
\frac{\slashed{n}_{3+}}{2}    \gamma_{\mu \perp_3}  u_{c_3}(q)
- n_{3+\nu} n_{3-\mu}  \frac{\slashed{n}_{3+}}{2} 
n_{1-}\cdot \gamma_{\perp_3}  u_{c_3}(q)
\nonumber    \\ &&  \hspace{-2.5cm}  
+ n_{1-}\cdot n_{3+}n_{3-\mu}  \frac{\slashed{n}_{3+}}{2}
\gamma_{\nu \perp_3}  u_{c_3}(q)
+  \frac{\slashed{n}_{3+}}{2}  \gamma_{\nu \perp_3}  
n_{1-}\cdot \gamma_{\perp_3}\gamma_{\mu \perp_3}u_{c_3}(q)\Big)\bigg]
\varepsilon^{\nu }_{\perp_2}(q_2)\,\,\varepsilon^{\mu }_{\perp_1}(q_1)
\end{eqnarray}
We note that the leading power result here exhibits the vector structure in contrast to the leading power result found in the $c_1$ sector in Eq.~\eqref{eq:LOmatching}. Nonetheless, this structure cannot contribute to the amplitude since here, the mixing has to occur with the Higgs boson, which has a vanishing overlap
\begin{align}
    \left< h(q_3) \right| J^{\mu}_{V}(r) \left| 0\right> =0 \;.
\end{align}
To arrive at a non-vanishing contribution, we consider the subleading term in $m_t$ to the QCD diagram in Fig.~\ref{fig:higgsmatching}. The resulting expression is a few lines long and it is not enlightening to express here. It suffices to note that the structure in the transverse spin plane is scalar and pseudoscalar, as expected. 
Therefore, 
the operator which takes part in this matching is given by 
\begin{align}\label{eq:c3higgsnlpOP}
    J^{[3]}_{{\rm{NLP}}}(t_1,t_2, t_{3_1}, t_{3_2} ,t_4) &=y_t 
 m_t\,
 \mathcal{A}_{\nu}^{{\rm c}_1 \perp_1\; A}(t_1 n_{1+})
   \mathcal{A}^{\nu\;A}_{{\rm c}_2 \perp_2\; }(t_2 n_{2+})
 J^{ }_{S_3}( t_{3_1}, t_{3_2} )\,
 h_{{\rm c}_4}(t_4 n_{4+}) \,,
\end{align}
where the $m_t$ power suppression is generated by the two
collinear quark fields in the $c_3$ direction. It is also
noteworthy that this operator contains one less explicit 
power of $y_t$ than the leading power operator since there is
one less Higgs field present. This factor is compensated at
the level of the NLP matrix element by the collinear
quark-antiquark-Higgs interaction vertex since it is required
to have an external Higgs present, see the explicit result
in Eq.~\eqref{eq:NLPhiggsMatrixElement}. The analogous
operator is needed for the case where power suppression is
generated by the collinear quark and antiquark fields in the $c_4$ sector
\begin{align}\label{eq:c4higgsnlpOP}
J^{[4]}_{{\rm{NLP}}}(t_1,t_2,t_3, t_{4_1}, t_{4_2} ) &=y_t m_t\,
\mathcal{A}_{\nu}^{{\rm c}_1 \perp_1\; A}(t_1 n_{1+})
\mathcal{A}^{\nu\;A}_{{\rm c}_2 \perp_2\; }(t_2 n_{2+})
h_{{\rm c}_3}(t_3 n_{3+}) J^{ }_{S_4}( t_{4_1}, t_{4_2} )\,.
\end{align}

\subsection{Scale \texorpdfstring{$\mu^2\sim m_t^2$}{musq ~ mt2}}\label{sec:attopmassscale}

At the scale parametrically of the order of  $m_t$, we evaluate matrix elements of the operators in equations \eqref{eq:lp-col-op}, \eqref{eq:c1gluonnlpOP}, \eqref{eq:c2gluonnlpOP}, \eqref{eq:c3higgsnlpOP}, and \eqref{eq:c4higgsnlpOP} in SCET$_{{\rm{II}}}$. 
Here, the top-quark mass is considered heavy, and we must consistently perform matching on a theory with the relevant low-energy degrees of freedom, which is carried out in Section~\ref{sec:mumtmatchingLP} and Section~\ref{sec:mumtmatchingNLP}.  
First, we recall the leading power collinear Lagrangian for massive fermions \cite{Beneke:2002ni, Leibovich:2003jd}
\begin{eqnarray}\label{eqn:SCETLagrangianLP}
\mathcal{L}^{(0)}_{c_i}=  
\bar{\xi}_{c_i}\bigg[ i n_{i-}\cdot D_{c_i} 
+ \Big( i  \slashed{D}_{{c_i} \perp_i} -m_q \Big)
\frac{1}{i  n_{i+}\cdot D_{c_i}}\Big( i \slashed{D}_{{c_i} \perp_i} + m_q \Big)  
\bigg]\frac{\slashed{n}_{i+}}{2} \xi_{c_i} \,,
\end{eqnarray}
where, unlike in the SCET$_{{\rm{I}}}$ Lagrangian, the covariant derivative does not contain a soft gauge field
\begin{eqnarray}
\label{eq:covDc}
i D_{c_i}^{\mu}(x)&=& i \partial^{\mu} +g_s A_{c_i}^{\mu}(x)\,.
\end{eqnarray}
In addition, we need the collinear Higgs Lagrangian. Since we work to the leading order in the electroweak parameters, only a single insertion of this Lagrangian is required for the NLP matrix elements. 
The  collinear Higgs fields scale as $h_{{\rm c}_i}\sim \lambda$, $i=3,4$, and the Lagrangian reads
\begin{align}
    \mathcal{L}_h^{(0)} &= \sum_{i=3,4}\Bigg[\frac{1}{2}\left(n_{i+}\cdot\partial h_{{\rm c}_{i}} n_{i-}\cdot\partial h_{{\rm c}_{i}}  + \partial^{\perp_i}_\mu h_{{\rm c}_{i}}\partial^\mu_{\perp_i}h_{{\rm c}_{i}} - m^2_h h_{{\rm c}_{i}}^2 \right) - \frac{g}{2m_W}\frac{m_h^2}{2}h^3_{{\rm c}_i} \nonumber \\
    & - \left(\frac{g}{2m_W}\right)^2\frac{m_h^2}{8}h^4_{{\rm c}_i} -\left( \sum_{q}^{n_f}\frac{y_q}{\sqrt{2}} \overline{q}_{{\rm c}_i} \frac{i \slashed D_{{\rm c}_i\perp_i} -m_q}{n_{i+}\cdot D_{{\rm c}_i}} \frac{\slashed n_{i+}}{2} q_{{\rm c}_i} h_{{\rm c}_i}  +h.c. \right)\Bigg]\,.
\end{align}
The parameter renormalisation and decoupling constants, which are required in the following section, are listed in Appendix~\ref{app:renormalisationconstants}.

\subsubsection{Leading power amplitude}
\label{sec:mumtmatchingLP}

We integrate out the collinear scale with the virtuality of the order of
$m_t^2$ and match the collinear fields onto the PDF-collinear fields 
\cite{Beneke:2018gvs, Beneke:2019oqx,Beneke:2019mua,Broggio:2023pbu} at
leading power in the ${\Lambda^2_{\rm QCD}}/{s}$
expansion, i.e. at leading twist.
The relevant operator consists of two PDF-collinear gluon fields
\begin{align}\label{eq:lp-pdfc-op}
    \mathcal{J}^{[i]}_{{\rm{LP}}}(t_1,t_2) = y_t^2 \,\mathcal{P}_{[i]}^{\mu\nu} \mathcal{A}^{{\rm PDF-c}_1\perp_1}_\mu(t_1 n_{1+})\mathcal{A}^{{\rm PDF-c}_2\perp_2}_\nu(t_2 n_{2+}) ,
\end{align}
where the projectors $\mathcal{P}_{[i]}^{\mu\nu}$
are defined above in \eqref{eq:projectors1}. We
now work with $n_f=5$ since the top quark is no
longer an active degree of freedom below the scale
$m_t^2$ and here $y_t^2$ is fixed at that scale.
The matching at leading power reads
\begin{align}
\label{eq:mumtmatchingLP}
   \left< H(q_3) H(q_4)  \right| J^{[i]}_{{\rm{LP}}}(0,0,0,0) \left| g(q_1) g(q_2) \right> = \mathcal{C} \left<0  \right|    \mathcal{J}^{[i]}_{{\rm{LP}}}(0,0) \left| g(q_1) g(q_2) \right>  ,
\end{align}
where, as will be shown, the non-zero contribution to $\mathcal{C}$ due
to radiative corrections starts at the two-loop order, and consequently, for our loop induced process $gg\to HH$, it enters at the three-loop level. 
This term compensates precisely for the change in the cusp anomalous
dimension when matching the theory with $n_f=6$
on $n_f=5$ at $\mu^2\sim m_t^2$. 
The non-trivial structure of this term contradicts
the claim in \cite{Liu:2022ajh}
that the matching is exact to all orders of
perturbation theory. In fact, it is an
interesting case of the IR matching (massification) procedure~\cite{Penin:2005eh, Mitov:2006xs, Becher:2007cu, Liu:2017axv, Engel:2018fsb, Wang:2023qbf, Wang:2024pmv}: 
at scales $\mu^2 \sim s$ the quarks are massless,
but below $m_t^2$ the heavy top quarks can enter
through soft loops, even though there are no 
external-state top quarks.

We obtain the form of $\mathcal{C}$ at
$\mathcal{O}(\alpha_s^2)$ starting from the
bare matrix element of the operator for the
theory on the left-hand side of our matching
equation in~\eqref{eq:mumtmatchingLP} with one
heavy and five light quarks provided 
in~\cite{Mitov:2006xs, Wang:2023qbf}. 
The IR divergences of this matrix element
correspond as usual to the UV renormalisation
factors in the EFT below the scale $m_t$.  
We remove these divergences from the amplitudes
on both sides of \eqref{eq:mumtmatchingLP} and
rearrange for the matching coefficient.
In our conventions, the operator $J^{[i]}_{{\rm{LP}}}$
is renormalised multiplicatively with $Z_{(n_f=6)}$
and the bare $\mathcal{C}$ with $Z^{-1}_{(n_f=5)}$.
We find that the renormalised $\mathcal{C}$
is therefore given by
\begin{eqnarray} \label{eq:matchAtmt}
    \mathcal{C} 
    &&=Z^{-1}_{(n_f=5)} Z_{(n_f=6)}  \mathcal{Z}_g^{(m|0)}(m_t) \, \mathcal{S} (\{p\},m_t )
   \nonumber   \\  && \hspace{-0.0cm}
    = 1 + \left(\frac{\alpha_s}{2\pi}\right)^2 \frac{C_A}{4} T_F \bigg(\frac{112}{27}  \ln \left(-\frac{m_t^2}{s}\right)-\frac{28}{9}  \zeta_3-\frac{5}{27} \pi ^2   
    +\frac{262}{27}  \bigg)
    + \mathcal{O}(\alpha_s^3),
\end{eqnarray}
where the anomalous dimensions required to construct
renormalisation factors $Z_{(n_f=6)}$ and $Z_{(n_f=5)}$
are listed in Appendix~A of \cite{Becher:2009qa} to
sufficient accuracy. The cusp anomalous dimensions are
known to four-loop order \cite{Henn:2019swt}. 
The $\mathcal{S}(\{p\},m_t )$ term is taken
from Eq.~(2.3) of  \cite{Wang:2023qbf}.
Similarly, $\mathcal{Z}_g^{(m|0)}(m_t)$
is constructed from terms in (2.6), (2.7),
and~(2.8), according to Eq.~(2.2) of~\cite{Wang:2023qbf}. 
Moreover, to arrive at the result in \eqref{eq:matchAtmt},
we have used the decoupling of the strong coupling
constant given in \eqref{eq:decoupling}
to write $\alpha_s^{(n_f=6)}$ appearing in the last
three terms of \eqref{eq:matchAtmt} in terms of
$\alpha_s^{(n_f=5)}$ for consistency with the
strong coupling constant appearing in $Z^{-1}_{(n_f=5)}$. 
We note that this matching involves decoupling of the
heavy top quark also from the $\alpha_s$, which is 
explicitly present in operators given 
in~\eqref{eq:lp-col-op} and \eqref{eq:lp-pdfc-op}.

We note that large logarithm $\ln( - {m_t^2}/s)$ appearing in \eqref{eq:matchAtmt} does not have its origin due to top-quark mass renormalisation but rather stems from the collinear anomaly. This tower of logarithms is universal and not specific to the $gg\to HH$ amplitude and can be obtained using rapidity renormalisation group techniques  \cite{Chiu:2012ir}.  
To perform the resummation of these logarithms, we note that our matching coefficient $\mathcal{C}$ is, in fact, a composite object obeying rapidity type factorisation into jet and soft functions,
\begin{eqnarray}
\mathcal{C}  = J_{n_{1-}}\big(m_t^2;\mu^2,\nu^2/s \big)\, J_{n_{2-}}\big(m_t^2;\mu^2,\nu^2/s \big)\, S\big(m_t^2,\mu^2,\nu^2/m_t^2 \big)  \,,
\end{eqnarray}
where the soft and jet functions are given by the matrix element of Wilson lines and
collinear fields 
\begin{eqnarray}
    S\big(m_t^2,\mu^2,\nu^2/m_t^2 \big)  && = \langle 0 | S^\dagger S |0 \rangle
\nonumber \\ J_{n_{1-}}\big(m_t^2;\mu^2,\nu^2/s \big) && =\langle 0 | \mathcal{A}_{c_1\perp_1} |g_{c_1} \rangle
\nonumber \\ J_{n_{2-}}\big(m_t^2;\mu^2,\nu^2/s \big) && =\langle 0 | \mathcal{A}_{c_2\perp_2} |g_{c_2} \rangle\,,
\label{eq:SJJ}
\end{eqnarray}
where $S$ is the adjoint soft Wilson line 
${\cal Y}_{i+}$ 
defined as 
\begin{eqnarray}
\label{adj-soft-Wilson-lines}
    {\cal Y}^{AB}_{i+}(x) = {\bf P} \exp \bigg\{g_s \int _{-\infty}^0 ds 
\, f^{ABC}\, n_{i-} A^C_s(x + s n_{\mp}) \bigg\},
\end{eqnarray} 
and the collinear building blocks are given
in Eq.~\eqref{eq:gaugeInvBuildBlocksCi}. 

\subsubsection{Next-to-leading power amplitude}
\label{sec:mumtmatchingNLP}
Matching of subleading power operators at the scale $\mu^2\sim m_t^2$
proceeds analogously to the analysis performed for the 
leading power operators in the section above. As before, the
relevant operators below $m_t^2$
are constructed from PDF-collinear fields. 

We recall that the next-to-leading power contributions 
are generated in each collinear sector, either through suppressed 
vector-type currents in the sectors collinear to the incoming 
gluons, or scalar type for the Higgses. 
To avoid repetition, we focus on the sector collinear to one of the initial state gluons, and the rest follows analogously. 
We first need the matrix element of the relevant subleading 
power operator with an external collinear gluon. Using the SCET Feynman rule for collinear gluon interaction with massive collinear quarks and a physical polarisation for the gluon, we find
\begin{align}\label{eq:3.41}
    \left< 0 \right| J^{\nu A}_{V}(r) \left| g^B(q_1)\right> = - g_s \delta^{AB} m_t 
    \frac{1}{4\epsilon}\Gamma[1+\epsilon] e^{\epsilon \gamma_E} \left(\frac{\mu^2}{m_t^2} \right)^{\epsilon}
    \varepsilon^{\nu}_{\perp_1}(q_1) \,,
\end{align}
where $J^{\nu A}_{V}(r)$ is the Fourier transform, see \eqref{eq:nlpFT}, 
of the vector-type subleading 
power operator given in \eqref{eq:nlpvector} with momentum fraction $r$
carried by the collinear quark in the NLP operator. Similarly, for the 
subleading power operator relevant in the Higgs sector,
see \eqref{eq:nlpscalar}, the matrix element 
calculation yields 
\begin{align}\label{eq:NLPhiggsMatrixElement}
    \left< 0 \right| J_{S}(r) \left| H(q_3)\right> = - \frac{y_t m_t \bar{r}}{2\sqrt{2}}  
    \frac{1}{\epsilon}\Gamma[1+\epsilon] e^{\epsilon \gamma_E} \left(\frac{\mu^2}{m_t^2} \right)^{\epsilon}
    \,.
\end{align}
Since at next-to-leading power the whole power suppression is generated in one of the collinear sectors, we now combine the NLP operator from above with leading power operators in the remaining three sectors. 
Analogously to \eqref{eq:mumtmatchingLP}, the complete NLP operator, which includes power suppression generated in each of the collinear sectors, has to be 
matched to leading-power (or leading twist) PDF-collinear  operator.
This can be obtained in a straightforward way using Eq.~\eqref{eq:SJJ} for the leading power components of the operators and \eqref{eq:3.41} as well as \eqref{eq:NLPhiggsMatrixElement} for the power-suppressed sectors. We note that colour conservation and multipole-expansion of the soft-fields guarantees that the soft function appearing with NLP hard operator is identical to the one accompanying  the leading power hard operator. 

\begin{figure}
    \centering
    \includegraphics[width=0.4\textwidth]{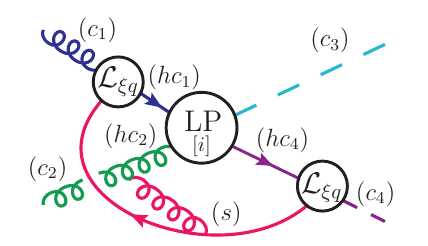}\hspace{1cm}
    \includegraphics[width=0.4\textwidth]{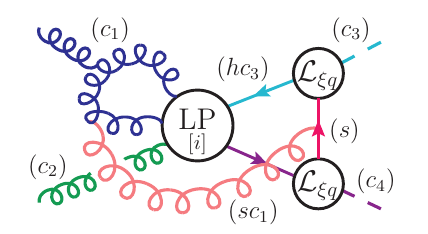}
    \caption{Effective field theory diagrams with scaling of internal lines indicated by labels $(c_i), (hc_i), (s)$, and $(sc_i)$ corresponding to a collinear$-q_i$, $\mathrm{hard-collinear}-q_i$,  soft, and $\mathrm{soft-collinear}-q_i$ scaling, respectively. (left) Power suppressed contributions to the amplitude with time-ordered product insertions of subleading power Lagrangian terms and hard-collinear fields in the operator. This type of contribution reproduces the $(\ell_1,\ell_2)=(s,s)$ diagrams in Fig.~\ref{fig:regions2LNP1}. (right) Sample three-loop diagram (corresponding to one on the right-hand side of Fig.~\ref{fig:three_loop}) power suppressed diagram with a new dynamical soft-collinear mode appearing at three-loop order, see cascading modes in \cite{Ma:2023hrt}. In this process we expect that this contribution enters only at NNLP due to decoupling of leading power soft-collinear interactions.}
    \label{fig:scetLqxi}
\end{figure}

As remarked at the end of Section~\ref{sec:lpmatching},
there also exist contributions to the amplitude 
due to double insertions of 
$\mathcal{L}_{\xi q}$ Lagrangian terms through time-ordered
products with lower-power currents \cite{Beneke:2017ztn}.
An example where this
type of contribution appears is the two-loop diagram
considered in the MoR analysis in Section~\ref{sec:regions} 
where both of the loop momenta are soft. 
On the left-hand side of Fig.~\ref{fig:scetLqxi} 
we have drawn a corresponding EFT
diagram to depict how this contribution is 
reproduced in SCET. Since each  component of the soft 
momentum is scaling as $\lambda$, the interaction
with a collinear mode results in a hard-collinear 
momentum scaling. Therefore, in the basis of operators 
we must also include leading power type operators 
with one hard-collinear field present in each direction.
As noted above, with the regulator used in \cite{Liu:2019oav}
these contributions start at one-loop order. 
Moreover, as the MoR analysis in the preceding 
section uncovered, at higher loop orders we see a cascade of
modes in accordance with our expectations following the 
analysis of contributing modes with massive particles present
\cite{Ma:2023hrt}. Expressing the corresponding diagrams 
in the SCET set up, it becomes evident from the power counting
that these contributions can only start to enter at NLP, 
and in fact, due to decoupling, we expect these
contributions are further power suppressed. 
For example, the SCET representation of the
three-loop graph with an ultra-collinear mode is 
shown on the right-hand side of Fig.~\ref{fig:scetLqxi}.
Using power counting for the objects entering the diagram,
at first sight, it appears that this type of contribution
can contribute at NLP. Indeed, diagram by diagram
these diagrams can produce non-vanishing contributions
at this order.
However, focusing on the attachment of the
soft-collinear gluon to the collinear loop, 
we notice a similarity with the case of leading power soft interaction coupling to
a collinear loop,  which has been studied
in \cite{Beneke:2019oqx}.
In that work, it has been shown that indeed the decoupling of 
soft and collinear effects in the leading power SCET Lagrangian is ultimately responsible for the explicit 
cancellation of such type of diagrams, since effectively, due to the decoupling, there is no external scale that can be
associated with the collinear loop.
For this reason, we expect that this type of contribution
can only enter the power expansion of the Higgs pair production amplitude starting from NNLP. However, this can only be
verified via an explicit calculation. In case there exists a mechanism which prevents
the cancellation of such diagrams, it would be interesting to
study in its own right. However, 
the lowest possible contribution is nonetheless an NLP effect 
and does not threaten our leading power analysis of the 
structure of the amplitude, which is the central focus of this
work. This mechanism could have more interesting consequences in
the case of $gg\to H$ amplitude studied in \cite{Liu:2022ajh}
since this process begins at NLP. Therefore, 
in case of non-cancellation of this type contributions
from cascading modes appearing at higher orders, the process
would receive corrections at the first order
in the power counting.   
In general, for the derivation of the NLP factorisation 
formula for our process under consideration, all possible 
operators and structures need to be included.  
The explicit derivation of the NLP factorisation for $gg\to HH$
amplitude and resummation of the NLP corrections in $m_t$ is 
left for future work. 

\subsection{Resummation}
Having understood the structure of the amplitude,
we are now ready to discuss the leading power resummation of the
large logarithms of $m_t$ contributing to the $gg\to HH$
amplitude in the high-energy limit. We discuss their numerical
impact in Section~\ref{sec:resultsplots}.
Resummation of the large logarithms is performed
using renormalisation group equations (RGEs). 
At leading power, the RGE structure 
is remarkably simple due to the fact that the
full leading power amplitude is equivalent to 
the hard-matching Wilson coefficient  $C_{[i]}^{{\rm{LP}}}(s,t,\mu)$.
From the anomalous dimension of the leading power SCET  
operator given in \eqref{eq:lp-col-op}
it follows that $C_{[i]}^{{\rm{LP}}}(s,t,\mu)$
obeys the RGE
\begin{equation}
\label{eq:RGE_CA0}
\frac{d}{d\ln\mu}C_{[i]}^{{\rm{LP}}}(s,t,\mu)  
=\left(\Gamma_{\text{cusp}}\ln\frac{s}{\mu^2}+ \gamma\right)
C_{[i]}^{{\rm{LP}}}(s,t,\mu) \,.
\end{equation}
The relevant anomalous dimensions are given by 
\begin{equation}
\Gamma_{\text{cusp}}=\frac{\alpha_s}{2\pi}\, 2 C_{A}+\mathcal{O}(\alpha_s^2), 
\qquad
\gamma=0+2\gamma_m+\mathcal{O}(\alpha_s^2)\,,
\label{eq:adhard}
\end{equation}
where $\alpha_s$ is the $\overline{\rm MS}$ QCD coupling at the 
scale $\mu$, the $0$ in $\gamma$ is the $\mathcal{O}(\alpha_s)$
contribution to the gluon collinear anomalous dimension,
and $\gamma_m = \frac{3\alpha_s}{2\pi}C_F+\mathcal{O}(\alpha_s^2)$
is the anomalous dimension governing 
the running of the top-quark mass. 
Higher orders for $\gamma_m$ 
can be found in \cite{Vermaseren:1997fq}. 
The general solution to \eqref{eq:RGE_CA0} reads
\begin{equation}
C_{[i]}^{{\rm{LP}}}(s,t,\mu) = 
\exp\Big[2 S(\mu_h,\mu) -  a_{\gamma}(\mu_h,\mu) \Big]
\left(\frac{s}{\mu_h^2} \right)^{- a_\Gamma(\mu_h,\mu)}\!
C_{[i]}^{{\rm{LP}}}(s,t,\mu_h) \,,
\label{eq:hardRGE}
\end{equation}
where the auxiliary functions are given by \cite{Neubert:2004dd}
\begin{eqnarray}
S(\nu,\mu) &=& - \int\limits_{\alpha_s(\nu)}^{\alpha_s(\mu)}\!
    d\alpha\,\frac{\Gamma_{\rm cusp}(\alpha)}{\beta(\alpha)}
    \int\limits_{\alpha_s(\nu)}^\alpha
    \frac{d\alpha'}{\beta(\alpha')}, 
\label{eq:Sdef}
\\ 
a_\Gamma(\nu,\mu) 
&=& - \int\limits_{\alpha_s(\nu)}^{\alpha_s(\mu)}\!
    d\alpha\,
    \frac{\Gamma_{\rm cusp}(\alpha)}{\beta(\alpha)} \,, \qquad  
a_\gamma(\nu,\mu) 
   = - \int\limits_{\alpha_s(\nu)}^{\alpha_s(\mu)}\!
    d\alpha\,\frac{\gamma(\alpha)}{\beta(\alpha)}.
\label{eq:adef}
\end{eqnarray}
At this point it is worthwhile to discuss the fate
of the different types of logarithmic corrections 
captured by our solution. Firstly, the function $S(\mu_h,\mu)$ 
contains logarithms of the type $\ln^2\big({\mu}/{\mu_h}\big)$
with a $C_A$ colour prefactor.
This type of logarithms does not appear in the
$\mathcal{O}(\alpha_s^2)$ amplitude results of \cite{Davies:2018qvx}
due to the fact that they are of IR origin and are cancelled 
in the process of removing IR poles using Catani's scheme
\cite{Catani:1998bh}. This is a scheme choice for defining
finite reminder of the amplitude. The SCET approach enables more
natural implementation in the $\overline{\rm MS}$ scheme which
generalises easily to all orders. Indeed, as described above, 
we perform matching at the scale $\mu^2 \sim m_t^2$ to the
relevant operator with PDF-collinear fields given in
\eqref{eq:lp-pdfc-op}. 
Therefore, the structure of these logarithms is reflected in the
running also below the scale  $\mu^2 \sim m_t^2$ 
and ultimately these corrections reside in the
long-distance scales appearing in the observables, and PDFs
at the scale $\Lambda_{\rm{QCD}}$. 
Concretely, we can 
again write down a formal solution of the RGE for the 
matching coefficient below $m_t$ which follows from 
the anomalous dimension of the operator in 
\eqref{eq:lp-pdfc-op}, 
this time the natural reference scale is a low-energy
scale $\mu_s\sim \Lambda_{\rm IR}$
\begin{equation}
{\mathcal{C}}_{[i]}^{{\rm{LP}}}(s,t,\mu) = 
\exp\Big[2 S(\mu_s,\mu) -  a_{\gamma}(\mu_s,\mu) \Big]
\left(\frac{\Lambda^2_{\rm{IR}}}{\mu_s^2} \right)^{- a_\Gamma(\mu_s,\mu)}\!
{\mathcal{C}}_{[i]}^{{\rm{LP}}}(s,t,\mu_s) \,,
\label{eq:softRGE}
\end{equation}
and the strong coupling constant present in these anomalous 
dimensions of the functions defined in \eqref{eq:Sdef}
and \eqref{eq:adef}
is evaluated in the five-flavour scheme.
This solution can be used to run from the low scale to $m_t$
where it must be matched to the high-energy theory.
In the matching procedure we must
also take care of the massification 
logarithms $\propto C_A \ln \left(-{m_t^2}/{s}\right)$
starting at $\mathcal{O}(\alpha_s^3)$ in the amplitude which arise
due to soft massive top quarks appearing in loop corrections at
this scale. These corrections are universal and
have already been provided in Eq.~\eqref{eq:matchAtmt}.
They can be resummed to all orders using rapidity RGE. 
Last, and most pertinent to the problem at hand, 
are the logarithms originating in the top-quark mass
renormalisation procedure arriving with a $C_F$ colour prefactor.
Since there are no other sources of logarithms involving the 
top-quark mass at leading power, as we have shown in this article, 
these are now be predicted to all orders in $\alpha_s$ using 
\eqref{eq:hardRGE}.

\section{Controlling scheme uncertainties}
\label{sec:resultsplots}
In this section we discuss the impact of our MoR and EFT analysis on
the uncertainty budget of double Higgs production via gluon-fusion. 
To summarise, the structure of the leading power amplitude is remarkably simple: as observed previously at NLO in \cite{Baglio:2018lrj}, it turns out that the leading logarithms to all orders
arise only from the renormalisation of the top-quark mass, which itself
is well known in the literature. At next-to-leading logarithmic 
accuracy, there are additional contributions from 
universal IR matching (massification), which start contributing to the amplitude at NNLO. 
Our analysis provides a systematic understanding of the leading power leading logarithmic structure.

To quantify the numerical effect of the leading power leading mass logarithms, we study their impact on the one-loop amplitude and the finite remainder of the two-loop virtual amplitude. 
The finite remainder of the two-loop amplitude is obtained after UV renormalisation and IR subtraction.
We begin by defining the two-loop remainder and briefly recapping how it is affected by changes in the top-quark mass scheme.
As described in Section~\ref{sec:amplitudes}, we expand each of the bare form factors $A_i$ as a perturbative series in the bare strong coupling $\alpha_{s,0}$,
\begin{align}
A_i = \left( \frac{\alpha_{s,0}}{2 \pi} \right) A_i^{(0)}(m_{t,0}^2) + \left( \frac{\alpha_{s,0}}{2 \pi} \right)^2 A_i^{(1)}(m_{t,0}^2) + \mathcal{O}(\alpha_{s,0}^3),
\end{align}
where $m_{t,0}$ is the bare top-quark mass.
The UV renormalisation is then performed by re-expressing the bare quantities in terms of their renormalised counterparts according to the formulae,
\begin{align}\label{eq:uvren}
&\alpha_{s,0} \equiv \alpha_s\, Z_{\alpha_s} S_\epsilon^{-1} \left( \frac{\mu_R^2}{\mu_0^2} \right)^\epsilon,&
& m_{t,0}^2 \equiv m_t^2\, Z_m, &
\end{align}
with $S_\epsilon = (4 \pi)^\epsilon e^{-\gamma_E \epsilon}$, and multiplying the amplitude with $Z_G^{1/2}$ for each external gluon as dictated by the LSZ formula.
The renormalisation constants can be expanded in $\alpha_s$ as, $Z_{X} = 1 + \left( \frac{\alpha_s}{2\pi}\right) \delta Z_{X} + \mathcal{O}(\alpha_s^2)$, with $X=\alpha_s,G,m$.
Using the above prescription, the renormalised amplitude can then be written as,
\begin{align}
A_i^\mathrm{ren} = & S_\epsilon^{-1} \left( \frac{\mu_R^2}{\mu_0^2} \right)^\epsilon \left( \frac{\alpha_{s}}{2 \pi} \right) A_i^{(0)}(m_{t}^2) \nonumber \\
& + S_\epsilon^{-1} \left( \frac{\mu_R^2}{\mu_0^2} \right)^{\epsilon} \left( \frac{\alpha_{s}}{2 \pi} \right)^2 \left[ \left( \delta Z_G + \delta Z_{\alpha_s} \right)  A_i^{(0)}(m_{t}^2)
+ \delta Z_m m_t^2 \frac{\partial A_i^{(0)}(m_t^2)}{\partial m_t^2} \right ] \nonumber\\
& + S_\epsilon^{-2} \left( \frac{\mu_R^2}{\mu_0^2} \right)^{2\epsilon} \left( \frac{\alpha_{s}}{2 \pi} \right)^2 A_i^{(1)}(m_{t}^2) + \mathcal{O}(\alpha_s^3).
\label{eq:Aren}
\end{align}
where we have expanded in $\alpha_s$ and neglected terms of order $\alpha_s^3$.
The mass counterterm amplitude is defined as,
\begin{align}
\label{eq:Amct}
A_i^{(0),\mathrm{mct}}(m_t^2) \equiv m_t^2 \frac{\partial A_i^{(0)}(m_t^2)}{\partial m_t^2}.
\end{align}
It is often convenient to compute the derivative of the one-loop amplitude with respect to the top-quark mass using mass counterterm insertions. 

Explicit expressions for the strong coupling and gluon wave-function renormalisation constants are given in Appendix~\ref{app:renormalisationconstants}.
The most relevant renormalisation constant for the present work is that of the top-quark mass.
In the existing literature on Higgs pair production the top-quark mass has been renormalised in either the OS or $\overline{\mathrm{MS}}$ renormalisation schemes, the corresponding renormalisation constants are given by~\cite{Broadhurst:1991fy,Tarrach:1980up},
\begin{align}
&\delta Z_{m}^\mathrm{OS} = C_F \left( -\frac{3}{\epsilon} - 4 \right) \left( \frac{\mu_R^2}{m_t^2} \right)^\epsilon,&
&\delta Z_{m}^{\overline{\mathrm{MS}}} = C_F \left( -\frac{3}{\epsilon} \right) \left( \frac{\mu_R^2}{\mu_t^2} \right)^\epsilon,&
\end{align}
with the colour factor $C_F = (N^2_c-1)/(2N_c)$.

Collecting the terms of Eq.~\eqref{eq:Aren} according to the order in $\alpha_s$, we can write
\begin{align}
A_i^\mathrm{ren} = & \left( \frac{\alpha_{s}}{2 \pi} \right) A_i^{(0),\mathrm{ren}}(m_{t}^2) + \left( \frac{\alpha_{s}}{2 \pi} \right)^2 A_i^{(1),\mathrm{ren}}(m_{t}^2) + \mathcal{O}(\alpha_{s}^3), \\
A_i^{(0),\mathrm{ren}} (m_{t}^2) = & S_\epsilon^{-1} \left( \frac{\mu_R^2}{\mu_0^2} \right)^\epsilon A_i^{(0)}(m_{t}^2), \\
A_i^{(1),\mathrm{ren}}(m_{t}^2) = & S_\epsilon^{-2} \left( \frac{\mu_R^2}{\mu_0^2} \right)^{2\epsilon} A_i^{(1)}(m_{t}^2) \nonumber \\
& + S_\epsilon^{-1} \left( \frac{\mu_R^2}{\mu_0^2} \right)^{\epsilon} \left[ \left( \delta Z_G + \delta Z_{\alpha_s} \right)  A_i^{(0)}(m_{t}^2)
+ \delta Z_m A_i^{(0),\mathrm{mct}}(m_t^2) \right ],
\end{align}
The IR subtraction can be performed using the $I_1(\epsilon)$ operator of a given IR subtraction scheme (e.g. $q_T$, Catani, Catani-Seymour, SCET, $\ldots$). 
We write,
\begin{align}
\label{eq:Afin}
A_i^\mathrm{fin}(m_t^2) = & \left( \frac{\alpha_{s}}{2 \pi} \right) A_i^{(0),\mathrm{fin}}(m_{t}^2) + \left( \frac{\alpha_{s}}{2 \pi} \right)^2 A_i^{(1),\mathrm{fin}}(m_{t}^2) + \mathcal{O}(\alpha_{s}^3), \\
A_i^{(0),\mathrm{fin}}(m_t^2) = & A_i^{(0),\mathrm{ren}}(m_t^2), \\
A_i^{(1),\mathrm{fin}}(m_t^2)  = & A_i^{(1),\mathrm{ren}}(m_t^2) - I_1(\epsilon) A_i^{(0),\mathrm{ren}}(m_t^2).
\end{align}
Our ``full'' results are obtained using all available terms of the analytic high-energy/small-mass expansion of Ref.~\cite{Davies:2018qvx}, which was computed using the Catani IR subtraction scheme~\cite{Catani:1998bh} after subtracting a scale dependent logarithm (see Eq.~(13) and (14) of Ref.~\cite{Davies:2018qvx}).
For an expansion in $\alpha_s/(2\pi)$, their IR subtraction operator can be written to finite order in $\epsilon$ as,
\begin{align}
I_1^{\text{Catani}^\prime}(\epsilon) = -\frac{C_A}{\epsilon^2} -\frac{1}{\epsilon} \left[ \beta_0 + C_A \ln\left(\frac{\mu_R^2}{-s - i \delta} \right) \right] + C_A \left( \frac{\pi^2}{12} - \frac{1}{2} \ln^2 \left( \frac{\mu_R^2}{-s - i \delta} \right) \right) + \mathcal{O}(\epsilon). \label{eq:i1_catp}
\end{align}
with $\delta \rightarrow 0_+$ and $\beta_0 = 11 C_A/6 - 2/3 T_F n_f$.
The conversion to the SCET subtraction scheme of Ref.~\cite{Becher:2009qa} involves dropping the finite terms of Eq.~\eqref{eq:i1_catp}, explicitly,
\begin{align}
A_i^{(1),\text{SCET}} &= A_i^{(1),\text{Catani}^\prime} + \Delta I_\text{SCET} A_i^{(0),\text{ren}}, \\
\Delta I_\text{SCET} &= I_1^{\text{Catani}^\prime}-I_1^{\text{SCET}} = C_A \left( \frac{\pi^2}{12} - \frac{1}{2} \ln^2 \left( \frac{\mu_R^2}{-s- i \delta} \right) \right) + \mathcal{O}(\epsilon),
\end{align}
where $A_i^{(1),\text{Catani}^\prime}$ is the finite amplitude with $I_1(\epsilon)$ given by \eqref{eq:i1_catp}, and $A_i^{(1),\text{SCET}}$ is the finite amplitude obtained with $I_1(\epsilon)$ that contains only the pole parts of \eqref{eq:i1_catp}. 
For the results presented here, we use the SCET scheme with $\mu_R^2 = s$.
We also set the top-quark mass renormalisation scale $\mu_t^2 = s$.

We now turn our attention to the 
all-order structure of the small top-quark mass power-expanded $y_t^2$ box contribution to the $pp \rightarrow HH$ amplitude, which in the
$\overline{\mathrm{MS}}$ scheme
can be written as follows, 
\begin{align}\label{eq:schem-lo}
& \mathrm{LO}: & &\alpha_s y_t^2 ( \textcolor{blue}{\boldsymbol{c_0}} + m_t \, n_0 ), & \\
& \mathrm{NLO}: & &\alpha_s^2 y_t^2  ( \textcolor{darkgreen}{\boldsymbol{a_1 l_\mu}} + \textcolor{blue}{\boldsymbol{c_1}} + m_t \, n_1  ), & \\
& \mathrm{NNLO}: & &\alpha_s^3 y_t^2  ( \textcolor{darkgreen}{\boldsymbol{a_2 l_\mu^2}} + \textcolor{orange}{\boldsymbol{b_2} \boldsymbol{l_m}} + \textcolor{blue}{\boldsymbol{c_2}} + m_t \, n_2 ), & \\
& \mathrm{N}^3\mathrm{LO}: & &\alpha_s^4 y_t^2 ( \textcolor{darkgreen}{\boldsymbol{a_3 l_\mu^3}} + \textcolor{orange}{\boldsymbol{b_3} \boldsymbol{l_m^2}} + \textcolor{purple}{\boldsymbol{d_3 l_m}} + \textcolor{blue}{\boldsymbol{c_3}} + m_t \, n_3 ), & \\
& \mathrm{N}^i\mathrm{LO}: &
&\alpha_s^{i-1} y_t^2 ( \textcolor{darkgreen}{\boldsymbol{a_i l_\mu^i}} + \textcolor{orange}{\boldsymbol{b_4} \boldsymbol{l_m^{i-1}}} 
+   \textcolor{purple}{\boldsymbol{
 d_i l_m^{i-2}}} + \ldots + \textcolor{blue}{\boldsymbol{c_i}} + m_t n_i ).&\label{eq:schem-nklo}
\end{align}
with $l_\mu = \ln(\mu_t^2/s)$ and $l_m$ contains both $\ln(\mu_t^2/s)$ and $\ln(m_t^2/s)$ logarithms.
In our expressions, we have suppressed the dependence on the Higgs boson mass and consider only the leading (hard) term in the expansion around small Higgs boson mass. 
The green terms, $\textcolor{darkgreen}{\boldsymbol{a_i l_\mu^i}}$, are the small-mass leading power leading logarithms, they are known from the renormalisation group running of the top-quark mass.
The orange terms, $\textcolor{orange}{\boldsymbol{b_i} \boldsymbol{l_m^{i-1}}}$, are the leading power next-to-leading logarithms, they receive a contribution from the running of the top-quark mass and from massification.
The leading power constant coefficients $\textcolor{blue}{\boldsymbol{c_0}}$ and $\textcolor{blue}{\boldsymbol{c_1}}$, highlighted in blue, are known from the one- and two-loop fixed order calculation in the small mass limit.
The three-loop leading power constant coefficient, $\textcolor{blue}{\boldsymbol{c_2}}$, is currently unknown, but it can be obtained from a purely massless three-loop computation since the leading power amplitude receives contributions only from the hard region (i.e. a simple Taylor expansion around $m_t=0$), the relevant master integrals are known~\cite{Caola:2020dfu,Bargiela:2021wuy}.
The four-loop next-to-next-to-leading power logarithm coefficient, $\textcolor{purple}{\boldsymbol{d_3}}$, is currently unknown but can be obtained from knowledge of the three-loop constant $\textcolor{blue}{\boldsymbol{c_2}}$, once it has been computed.
The terms $n_0$ and $n_1$ contain the one-loop and two-loop beyond leading power (i.e. next-to-leading power, next-to-next-to-leading power, ...) results, the $n_0$ term can be obtained at any power by expanding the analytic one-loop result, the first 118 terms in the $m_t$ expansion of $n_1$ are known~\cite{Davies:2018qvx,Davies:2023vmj}.
The next-to-leading power and beyond coefficients starting from three-loops (i.e. $n_3, n_4, \ldots$) are currently unknown, understanding their structure would require the extension of the SCET factorisation theorem to NLP for $2 \rightarrow 2$ scattering of gluon and Higgs particles along the lines of the considerations presented in Section~\ref{sec:scet}.

\begin{figure}
     \centering
      \includegraphics[width=0.8\textwidth]{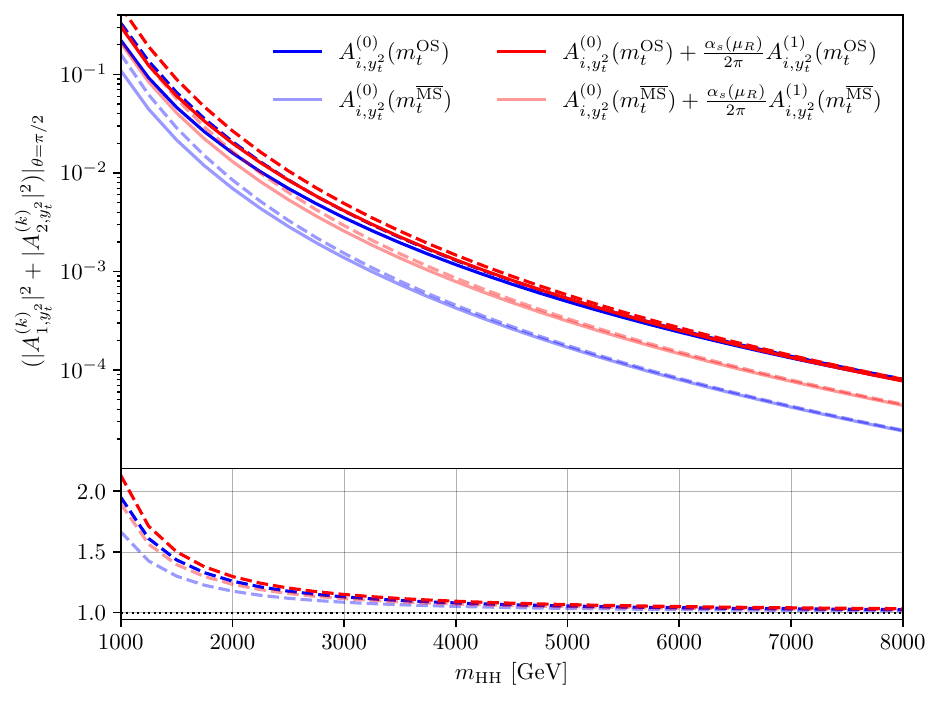}
      \caption{Plot demonstrating the validity of leading power expansion. Solid lines represent the full contribution to the LO and NLO squared amplitudes. The dark- and light-blue lines contain leading order one-loop amplitudes with top-quark mass in the $\mathrm{OS}$ and ${\overline{\mathrm{MS}}}$ scheme, respectively. The dark- and light-red lines show the same at two-loop order. The dashed lines contain the corresponding contributions at leading power in $m_t^2/s$. In the bottom panel we plot the ratio of the leading power to the corresponding unexpanded line. It is apparent that at high-energies the leading power version of each of the lines captures the behaviour of the corresponding full contribution.  }
      \label{fig:all_sq_full_vs_lp}
\end{figure}
As a first step in our phenomenological analysis, we
test the validity of the leading power expansion of 
the amplitudes at high-energies.
In Fig.~\ref{fig:all_sq_full_vs_lp}, we demonstrate that in the very high-energy regime the leading power ($y_t^2 m_t^0$) amplitude provides a good approximation of the full result.  
In this figure, we plot the sum of the squared amplitude for different cases. 
Namely, contributions
to $A_{1,y_t^2}$ and $A_{2,y_t^2}$
at one- and two-loops, both for the top-quark mass in the $\mathrm{OS}$ and ${\overline{\mathrm{MS}}}$ schemes.
For the conversion of the top-quark mass from the $\mathrm{OS}$ to the ${\overline{\mathrm{MS}}}$ scheme and the running of $\alpha_s$, we use the codes {\texttt{RunDec}} and {\texttt{CRunDec}} \cite{Chetyrkin:2000yt,Herren:2017osy}.
Note that when squaring the two-loop amplitude, we actually compute the interference between the one- and two-loop amplitudes, $ 2\, \mathrm{Re}[ A^{(0)}_{i,y_t^2} ( A^{(1)}_{i,y_t^2})^*]$.
Solid lines are
obtained using the expansion up to $\mathcal{O}(m_t^{16})$, including finite $m_H$ effects up to $\mathcal{O}(m_H^{2})$, as provided in Ref.~\cite{Davies:2018qvx}.
This expansion is known to reproduce the full NLO result with very high accuracy for the $\sqrt{s} \gtrsim 1$ TeV region considered here. 
The dashed lines contain the 
corresponding terms expanded to leading power in 
$m_t^2/s$. We see that at high-energies the leading 
power terms are responsible for the 
behaviour of the full amplitudes, i.e. the subleading 
power terms are negligible above the energy scale of
around $4$ TeV. 
At the lower end of the energy spectrum 
the leading power approximation breaks down and the 
subleading power terms become more important, as expected. 
We also note that the results plotted for the amplitudes 
at same order, but with top-quark mass in different
schemes, give rise to lines which do not overlap. The 
difference between these lines is taken as the 
top-quark mass scheme uncertainty,
as advocated in Ref.~\cite{Baglio:2020ini}.
In Appendix~\ref{app:full_vs_lp}, we additionally display the comparison between the full and leading power results for the real and imaginary of each form factor.
We remark that the uncertainty presented in~\cite{Baglio:2020ini}
is for the NLO $pp\to HH$ cross section, includes 
the virtual and real corrections. In our analysis we consider
solely the virtual corrections.

\begin{figure}
     \centering
     \begin{subfigure}[b]{0.49\textwidth}
         \centering
         \includegraphics[width=\textwidth]{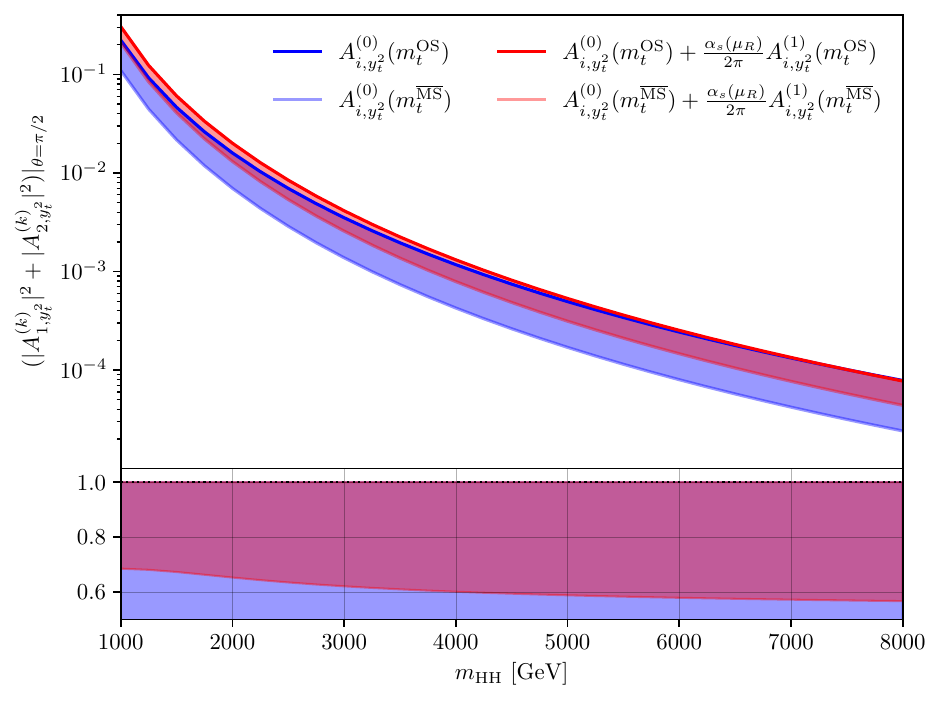}
         \caption{ }
         \label{sfig:all_sq_msbar_vs_os}
     \end{subfigure}
     \hfill
     \begin{subfigure}[b]{0.49\textwidth}
         \centering
         \includegraphics[width=\textwidth]{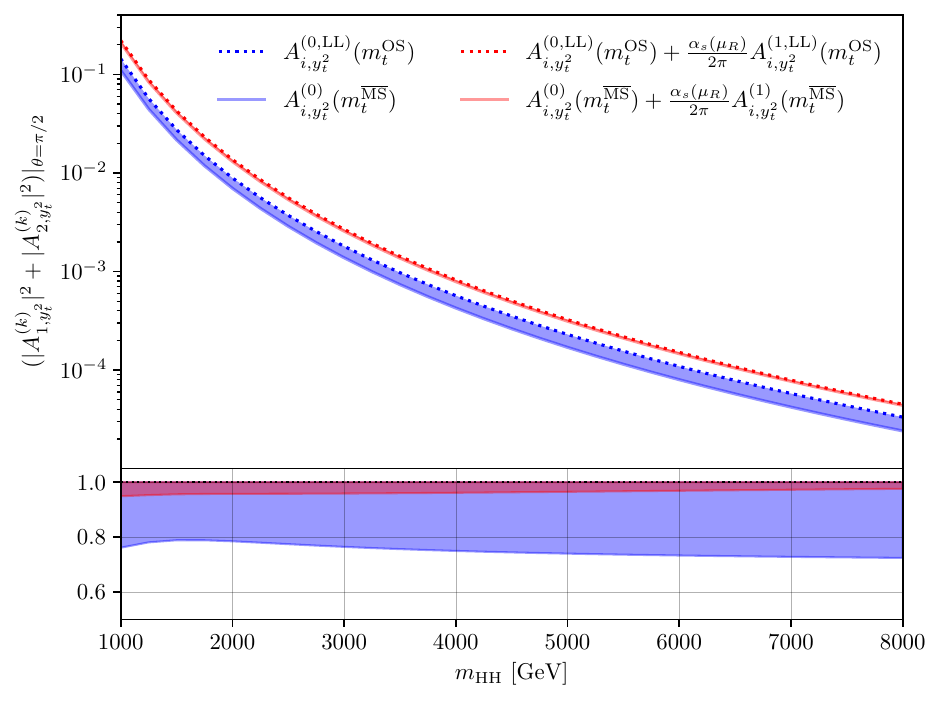}
         \caption{}
         \label{sfig:all_sq_msbar_vs_osll}
     \end{subfigure}
        \caption{Comparison of results for the sum of the squared form factors at one- and
        two-loop order
        where the top-quark mass is renormalised in the $\overline{\mathrm{MS}}$
        and ${\mathrm{OS}}$ (panel (a)) and the same quantities with the ${\mathrm{OS}}$
        result supplemented by the resummed tower of leading power leading logarithms
        (panel (b)). 
        Significant reduction in the size of the uncertainty due to the choice of the
        top-quark mass renormalisation scheme is observed. }
        \label{fig:all_sq_msbar_vs_osll}
\end{figure}

We plot the uncertainty bands due to the choice of the 
top-quark mass scheme at one- and two-loops in 
Fig.~\ref{sfig:all_sq_msbar_vs_os}. We see that at 
leading order the uncertainty on the virtual amplitude is $~60-70\%$ (blue band)
across the spectrum at high-energies,
where the small-$m_t^2$ expansion is valid.
With inclusion of the higher order corrections the uncertainty
is reduced to around $40\%$ (red band). 

At this point we
consider again the structure
of the amplitude at higher orders 
which, is laid out between
Eqs.~\eqref{eq:schem-lo}
and~\eqref{eq:schem-nklo}. 
In the $\overline{\mathrm{MS}}$
scheme, the logarithms 
$\textcolor{darkgreen}{\boldsymbol{a_i l_\mu^i}}$
depend on the scale $\mu_t$ which can be set to the order of $s$, rendering these explict logarithms small. The dependence on the
large ratio of scales, $m_t^2/s$, is captured to all orders in $\alpha_s$ 
implicitly through the running of top-quark mass. 
Since the bare top-quark mass can be renormalised in either scheme as in \eqref{eq:uvren}, 
the conversion factor between ${\mathrm{OS}}$ and $\overline{\mathrm{MS}}$ schemes 
is defied through the ratio of the corresponding $Z$-factors
\begin{eqnarray}
    \frac{m(\mu)}{M} = 
    \frac{Z^{ \mathrm{OS} }_m}{Z^{\overline{\mathrm{MS}}}_m}
    \equiv z_m(\mu) \,.
\end{eqnarray}
The quantity $z_m(\mu)$ 
has the following perturbative expansion
\begin{eqnarray}\label{eq:z-split}
    z_m(\mu) = \sum_{n\geq 0}\left(\frac{\alpha_s(\mu)}{2\pi}\right)^n \left(z_{m}^{n}(M) + z_{m}^{n,\log}(\mu)\right)\,.
\end{eqnarray}
In the above equation, $z_{m}^{n,\log}(\mu)$ contains only the $\mu$-dependent terms which vanish for $\mu=M$. 
The constant parts, $z_{m}^{n}(M)$,
have been computed up to four-loop order in Ref.~\cite{Marquard:2016dcn}, in our conventions they have an additional factor of 2 per loop order. Since the $gg\to HH$ amplitude is
known at NLO, the conversion
is typically truncated at the first order, as 
done in the plot in Fig.~\ref{fig:all_sq_full_vs_lp},
where we make use of,
\begin{eqnarray}\label{eq:nlo-conv}
    z_m(\mu) = 1 + \frac{\alpha_s(\mu)}{2\pi}
    \Big( -2C_F -\frac{3}{2}C_F \ln\frac{\mu^2}{M^2}\Big)
    \, + \mathcal{O}(\alpha_s^2),
\end{eqnarray}
which leads to a wide discrepancy between the
two schemes. Importantly, we can now also identify 
precisely why this is the case. Namely, the 
large logarithms which are captured implicitly
for the $\overline{\mathrm{MS}}$ scheme, have
only partially been restored in the corresponding
$\mathrm{OS}$ result through the truncated 
conversion factor. Indeed, only the first leading power leading 
logarithm is captured correctly using 
Eq.~\eqref{eq:nlo-conv}.
We argue that
the discrepancy between the two schemes obtained 
in this manner should not form part of the 
uncertainty budget for the amplitude, since 
the leading logarithms (and beyond) in the 
conversion factor are in fact {\emph{known}} to
all orders in perturbation theory
and can be reinstated using the renormalisation
group equations for $m(\mu)$ and $\alpha_s(\mu)$. 
In the previous sections we have found 
that this is the only source of leading power
leading logarithms, such that we can supplement the
$\mathrm{OS}$ result with a complete and consistent tower of these logarithms. 
In order to do this, we first resum the leading logarithms appearing in Eq.~\eqref{eq:nlo-conv} to all orders, this gives the scheme conversion formula,
\begin{align}\label{eq:schmeme-conv-all-order}
    &m^\text{LL}(\mu) = M \exp \left[ a_{\gamma_m}^\text{LL}(\mu)\right]\, z_m(M),&
    &a_{\gamma_m}^\text{LL}(\mu) = \frac{3 C_F}{2 \beta_0} \ln \left(1 - \frac{\alpha_s(\mu)}{2 \pi} \beta_0 \ln \left(\frac{\mu^2}{M^2} \right) \right).&
\end{align}
As a check, expanding the exponential in powers of $\alpha_s$ gives
the logarithms quoted in Appendix~C of \cite{Marquard:2016dcn}
up to the fourth loop order. 
Once these logarithms are included to all orders, either explicitly in the amplitude or implicitly in the definition of the running quark mass, the remaining mass scheme uncertainty is 
due only to subleading logarithms at leading power and subleading 
power terms.
In our numerical results, in addition to the leading logarithms generated by $a_{\gamma_m}^\text{LL}$, we take only the leading constant term in $z_m$, i.e. $z_m^0$.

We now define resummed amplitudes in the OS scheme supplemented by the complete tower of leading power leading mass logarithms, they are given by,
\begin{align}
A^{(j,\,\text{LL})}_{i,y_t^2}(m_t^\mathrm{OS}) = \left(\frac{m^\text{LL}(\mu_t)}{m_t^\mathrm{OS}} \right)^2 A^{(j)}_{i,y_t^2}(m_t^\mathrm{OS}). \label{eq:a_ll}
\end{align}
Note that since the power expansion of the amplitudes starts at order $y_t^2 m_t^0 \sim m_t^2$ the ratio $(m^\text{LL}(\mu_t)/m_t^\text{OS})^2$ effectively restores the (green) tower of leading logarithms in Eqs.~\eqref{eq:schem-lo}--\eqref{eq:schem-nklo}, leaving all other contributions in the original OS scheme.

We plot the comparison of the $A^{(j,\,\text{LL})}_{i,y_t^2}(m_t^\mathrm{OS})$ amplitudes to those in the $\overline{\mathrm{MS}}$ scheme
in Fig.~\ref{sfig:all_sq_msbar_vs_osll}.
We observe a very significant reduction in the size of the
uncertainty bands. The behaviour is expected, since as we argued the discrepancy between the results for the amplitudes obtained 
in different schemes is due to large logarithms taken into account implicitly in the 
$\overline{\mathrm{MS}}$ scheme but not in $\mathrm{OS}$, which we have now
explicitly reinstated via Eq.~\eqref{eq:a_ll}. The size of the mass scheme uncertainty, defined as the difference between the OS or $\text{OS}^\text{LL}$ result and the $\overline{\mathrm{MS}}$ result, is given for several different energies in Table~\ref{tab:virtual_mass_scheme_uncert}.

\begin{table}[]
\centering
\scalebox{0.905}{
\begin{tabular}{l|cc|cc}
              & \multicolumn{2}{c|}{LO}                                                                                                    & \multicolumn{2}{c}{NLO}                                                                                                   \\
$s$ {[}GeV{]} & \multicolumn{1}{c}{$\text{OS}$ [fb]} & \multicolumn{1}{c|}{$\text{OS}^\text{LL}$ [fb]} & \multicolumn{1}{c}{$\text{OS}$ [fb]} & \multicolumn{1}{c}{$\text{OS}^\text{LL}$ [fb]} \\ \hline
$3000.0$ &         $(3.52\times10^{-3})^{+0\%}_{-60.8\%}$  &         $(1.80\times10^{-3})^{+0\%}_{-23.6\%}$ &         $(4.14\times10^{-3})^{+0\%}_{-37.9\%}$  &         $(2.68\times10^{-3})^{+0\%}_{-4.2\%}$  \\
$4000.0$ &         $(1.17\times10^{-3})^{+0\%}_{-63.7\%}$  &         $(5.70\times10^{-4})^{+0\%}_{-25.1\%}$ &         $(1.32\times10^{-3})^{+0\%}_{-39.9\%}$  &         $(8.23\times10^{-4})^{+0\%}_{-3.9\%}$  \\
$5000.0$ &         $(4.96\times10^{-4})^{+0\%}_{-65.7\%}$  &         $(2.30\times10^{-4})^{+0\%}_{-26.0\%}$ &         $(5.34\times10^{-4})^{+0\%}_{-41.2\%}$  &         $(3.25\times10^{-4})^{+0\%}_{-3.5\%}$  \\
$6000.0$ &         $(2.44\times10^{-4})^{+0\%}_{-67.1\%}$  &         $(1.09\times10^{-4})^{+0\%}_{-26.7\%}$ &         $(2.54\times10^{-4})^{+0\%}_{-42.1\%}$  &         $(1.52\times10^{-4})^{+0\%}_{-3.2\%}$  \\
$7000.0$ &         $(1.33\times10^{-4})^{+0\%}_{-68.3\%}$  &         $(5.80\times10^{-5})^{+0\%}_{-27.2\%}$ &         $(1.35\times10^{-4})^{+0\%}_{-42.8\%}$  &         $(7.92\times10^{-5})^{+0\%}_{-2.8\%}$  \\
$8000.0$ &         $(7.86\times10^{-5})^{+0\%}_{-69.2\%}$  &         $(3.34\times10^{-5})^{+0\%}_{-27.6\%}$ &         $(7.76\times10^{-5})^{+0\%}_{-43.3\%}$  &         $(4.51\times10^{-5})^{+0\%}_{-2.4\%}$  \\
\end{tabular}
}
\caption{Comparison of the OS and OS$^\text{LL}$ scheme (see text) results and the remaining mass scheme uncertainty for the squared/interfered virtual amplitudes at one- and two-loop order as a function of $s$ at fixed $\theta=\frac{\pi}{2}$. 
Note that these numbers do not include the real radiation and subtraction term contributions.}
\label{tab:virtual_mass_scheme_uncert}
\end{table}

The uncertainty in the double Higgs production 
amplitudes due to the choice of the top-quark
mass scheme can be further reduced through
inclusion of terms constituting higher logarithmic
running of the top-quark mass in the conversion
factor, i.e. the $z(\mu)$ factor in Eq.~\eqref{eq:z-split}. 
This improvement amounts to extension of 
Eq.~\eqref{eq:schmeme-conv-all-order} to subleading 
logarithmic accuracy. 
However, in order to formally keep control over
next-to-leading logarithms we are also
required to perform resummation of the logarithms 
arising due to
universal IR matching (massification)
as discussed in Section~\ref{sec:attopmassscale}. 
This series starts at the three-loop order and we 
provide the first logarithm in Eq.~\eqref{eq:matchAtmt}.

The predictions can also be improved through resummation 
of large top-quark mass logarithms at subleading powers,
where their origin is more involved than at leading power 
as explored in our MoR analysis. Control over subleading power
terms would supplement the predictions at lower values
of the invariant mass of the double Higgs system where
the leading power approximation in seen to break down 
in Fig.~\ref{fig:all_sq_full_vs_lp}.
We leave both the resummation of universal IR matching
logarithms and subleading power terms for future studies.


\section{Summary and outlook}
\label{sec:summary}

One of the key objectives for the HL-LHC is to 
perform a measurement of the trilinear Higgs
self-coupling \cite{Cepeda:2019klc}. 
Precise theoretical predictions for the
relevant
cross sections 
are of critical importance in this endeavour.
However, as has been noted in the literature, 
theoretical predictions are currently dominated 
by the large uncertainty present
due to the dependence
of the corresponding amplitudes on the choice
of the top-quark mass scheme and scale~\cite{Baglio:2020ini}.
This uncertainty can be reduced by higher order
perturbative calculations. 
However, retaining the full mass dependence in
higher order calculations 
is extremely challenging. Curiously, 
it has also been observed that 
in the case of double Higgs production in gluon-gluon scattering,
at high-energies, 
the logarithmic dependence of the leading power contribution at NLO 
is fully determined by the choice of the top-quark
mass renormalisation scheme \cite{Baglio:2020ini}. In the language of 
the MoR, this corresponds to the fact that only the 
hard region contributes to the leading power amplitude
at this order. 

In this work, we have systematically studied the contribution of different regions to the amplitudes. 
Utilising an automated tool, we
found that at the level of {\emph{scalar integrals}}, it is 
possible that many regions give rise to leading, and in certain cases,
power-enhanced contributions. We also observed that at higher
orders new modes appear, as is expected for 
amplitudes with massive internal lines \cite{Ma:2023hrt}.
However, 
it remains true that at the level of the {\emph{amplitude}} the hard region captures the leading power
behaviour of the 
$gg\to HH$ process. 
This behaviour can be understood to all orders in perturbation 
theory once we cast the problem in the effective
field theory language and consider how power suppressed contributions 
arise. We use the SCET framework to build operators 
that capture the structure of the amplitude and find that regions other than the hard region can
indeed give contributions to the amplitude, but, these can only
arise at subleading powers due to helicity 
suppression. We explicitly demonstrate that there is no mixing of external gluons (or Higgs bosons) 
with leading power structures other than the hard region, and since helicity is conserved in the massless limit, this holds to 
all orders in perturbation theory. 
While we leave the derivation of an explicit factorisation
formula at subleading power for future work, the upshot 
of our analysis is a proof that the logarithmic behaviour of
the leading power amplitude can be predicted from the hard region and 
universal contributions, i.e. the choice of renormalisation scheme
for the top-quark mass and IR matching (massification). 

Leveraging our newfound understanding of the origin of the
large logarithms appearing in the leading power amplitudes,
we analyse the implications 
for the top-quark mass scheme choice uncertainty. First, we see that indeed the leading power approximation captures the behaviour 
of the full amplitude well at sufficiently high-energies in both the OS and $\overline{\mathrm{MS}}$ renormalisation schemes.
Ref.~\cite{Baglio:2020ini} advocates that the mass scheme uncertainty for the 
amplitude can be estimated by taking the envelope of the
results obtained in the $\mathrm{OS}$ 
and $\overline{\mathrm{MS}}$ schemes.
However, since the complete tower of leading mass logarithms is known at leading power, we argue that these logarithms should always be accounted for in the theoretical prediction and should {\emph{not}} form part of the uncertainty
budget. Following this reasoning, we include the complete set of leading logarithms into the OS result and find that for the squared/interfered virtual amplitude, the top-mass scheme uncertainty can be reduced from 65\% to 25\% at LO and from 40\% to 4\% at NLO.

Our handle on the mass
scheme uncertainties can be further improved
by including higher logarithmic evolution 
of the top-quark mass in the scheme conversion. Namely,
retaining more of the known logarithms in the 
conversion factor between the $\mathrm{OS}$ and
$\overline{\mathrm{MS}}$ schemes
which originate in the running of the top-quark mass. 
Moreover, the logarithms due to 
universal IR matching need to be
resummed and included in the predictions before formal NLL
accuracy for the result can be claimed at leading power. 
In our phenomenological study we also observe that the 
leading power approximation starts to break down 
at lower values for the invariant mass of the system where
power corrections become important. Subleading power 
contributions have a much richer factorisation structure 
as we have uncovered in our MoR analysis and it will be 
interesting from the theoretical development point of view 
to extend the framework to next-to-leading power.
To reduce the mass scheme uncertainty at and below the top-quark threshold, it may also prove useful to consider the all-order structure of mass corrections in a different expansion, for example, the heavy-top limit, the threshold expansion at $s \sim 4 m_t^2$, or the small-$p_T$ expansion.

Studies of the small mass expansion can be performed for
related processes such as Higgs production in association with a $Z$ boson, and $Z$ boson pair production. In these cases, the structure of the mass dependent
logarithms is more complicated already at leading power, 
which hints at a richer factorisation structure in the 
Therefore, a more involved EFT description and resummation is already needed at leading power.
Nonetheless, the results for these processes are important
from the phenomenological perspective of the HL-LHC 
theory precision targets. 


\subsubsection*{Acknowledgements} 

We want to thank M. Schnubel for providing clarification on \cite{Liu:2022ajh}, and R.S. would like to thank C. Savoini for very helpful discussions about massification.
S.E.J. and R.S. thank the ``Quantum Field Theory at the Frontiers of the Strong Interaction'' ESI workshop in Vienna
and the  
``EFT and multi-loop methods for advancing precision in collider and gravitational wave physics'' 
workshop at 
the Munich Institute for Astro-, Particle and BioPhysics (MIAPbP), which is funded by the Deutsche Forschungsgemeinschaft (DFG, German Research Foundation) under Germany's Excellence Strategy - EXC-2094 - 390783311, where parts of this work were completed. 
S.P.J. and S.E.J. were supported in part by STFC under grant ST/X003167/1 and the Royal Society University Research Fellowship (URF/R1/201268). R.S. is supported by the United States Department of Energy under Grant Contract DE-SC0012704.
Figures were drawn 
with \texttt{Jaxodraw}~\cite{Binosi:2008ig}.


\begin{appendix}

\section{{Renormalisation and decoupling constants}}
\label{app:renormalisationconstants}

In this appendix, we reproduce parameter renormalisation and decoupling constants as required in the main text.
The $\overline{\rm{MS}}$ renormalisation constant for the strong coupling 
up to the two-loop order is given by 
\begin{align}
    Z_{\alpha_s} = 1  - \left(\frac{ \alpha_s^{(n_f)} }{2 \pi }\right) \frac{ \beta_0}{  \epsilon }
    +  \left(\frac{\alpha_s^{(n_f)}}{2 \pi }\right)^2 \left(\frac{\beta_0^2}{ \epsilon^2}-\frac{\beta_1}{4 \epsilon}\right) \,,
\end{align}
where 
\begin{eqnarray}
   \beta_0 = \frac{11}{6} C_A - \frac{4}{6} T_F n_f, \qquad 
   \qquad \beta_1 = \frac{17}{ 3} C_A^2  - \frac{10}{3}C_A T_F n_f - 2 C_F T_F n_f\,.
\end{eqnarray}
The on-shell gluon wave-function renormalisation is given by,
\begin{equation}
 Z_g = 1 + \frac{\alpha_s^{(n_f)}}{2 \pi } T_F n_h 
    \left(-\frac{2}{3 \epsilon}-\frac{2}{3} \ln \left(\frac{\mu^2}{m_t^2}\right)
    -\frac{1}{3} \epsilon \ln ^2\left(\frac{\mu^2}{m_t^2}\right)-\frac{\pi ^2 \epsilon}{18}
    + \mathcal{O}(\epsilon^2)
    \right)+ \mathcal{O}(\alpha_s^2),
\end{equation}
where $n_f$ is the number of light quarks and $n_h$ is the number of heavy quarks. 

We also need the decoupling of the strong coupling constant, up to the first loop order it is given by~\cite{Bernreuther:1981sg},
\begin{eqnarray}\label{eq:decoupling}
    \zeta_{\alpha_s}&&=1+\frac{\alpha_s}{2 \pi }  T_F \bigg(\frac{2}{3} \ln \left(\frac{\mu^2}{m_t^2}\right)
   +\frac{1}{3} \epsilon \ln ^2\left(\frac{\mu^2}{m_t^2}\right)
    +\frac{\pi ^2 }{18}\epsilon
      +
    \frac{1}{9} \epsilon^2 \ln ^3\left(\frac{\mu^2}{m_t^2}\right)
\nonumber \\ && \hspace{2.5cm} 
 +\frac{\pi ^2 }{18} \epsilon^2 \ln \left(\frac{\mu^2}{m_t^2}\right)
 -\frac{2  \zeta_3}{9}\epsilon^2
      \bigg) + \mathcal{O}(\alpha_s^2).
\end{eqnarray}
for the case where we have only one heavy quark, the two-loop correction can be found in Eq.~(A.4) of \cite{Barnreuther:2013qvf}. 
Additionally, we provide the collinear spinor expansion needed in matching calculations performed in Section~\ref{sec:nlpmatching} 
\begin{align}
    u(q) = \left(1 + \frac{\left(\slashed{q}_{ \perp_1}+m_t\right)}{n_{1+}q}\frac{\slashed{n}_{1+}}{2} \right)    u_{c_1}(q) \;,  \nonumber\\
    v(q ) = \left(1 + \frac{\left(\slashed{q}_{  \perp_1}-m_t\right)}{n_{1+}q }\frac{\slashed{n}_{1+}}{2} \right)v_{c_1}(q) \;.
    \label{eq:spin_exp}
\end{align}

\section{Region expansion at two-loops}
\label{app:two_loop_regions}
In this appendix, we provide the regions for the 
remaining two top-level topologies at two-loops drawn in Fig.~\ref{fig:gghh_diag2l}. The regions for $\mathbf{P2}$
are given in Fig.~\ref{fig:regions2LP2} and
regions for $\mathbf{NP2}$
are given in Fig.~\ref{fig:regions2LNP2}. The purely hard regions 
are omitted. 

{
\begin{figure}[h!]
    \centering
    \scalebox{0.8}{\begin{tabular}{cccc}
    \begin{tikzpicture}[y=0.4cm,x=0.8cm]\diagPG{c}{c}{}{}{}{cc}{cc}{}{c}{}{}\end{tikzpicture}&
\begin{tikzpicture}[y=0.4cm,x=0.8cm]\diagPG{cc}{}{}{}{c}{c}{cc}{}{}{}{c}\end{tikzpicture}&
\begin{tikzpicture}[y=0.4cm,x=0.8cm]\diagPG{}{c}{c}{cc}{}{}{cc}{c}{}{}{}\end{tikzpicture}&
\begin{tikzpicture}[y=0.4cm,x=0.8cm]\diagPG{}{}{cc}{c}{c}{}{cc}{}{}{c}{}\end{tikzpicture}\\

$(-2,-2,0,0,0,-2,-2)$&
$(-2,0,0,0,-2,-2,-2)$&
$(0,-2,-2,-2,0,0,-2)$&
$(0,0,-2,-2,-2,0,-2)$\\

\begin{tikzpicture}[y=0.4cm,x=0.8cm]\diagPG{}{}{}{c}{c}{}{}{}{}{c}{}\end{tikzpicture}&
\begin{tikzpicture}[y=0.4cm,x=0.8cm]\diagPG{}{}{}{}{c}{c}{}{}{}{}{c}\end{tikzpicture}&
\begin{tikzpicture}[y=0.4cm,x=0.8cm]\diagPG{}{}{}{}{}{}{s}{}{}{}{}\end{tikzpicture}\\

$(0,0,0,-2,-2,0,0)$&
$(0,0,0,0,-2,-2,0)$&
$(0,0,0,0,0,0,-2)$
    \end{tabular}}
    \caption{Regions for the two-loop diagram $\mathbf{P2}$ in Fig.~\ref{fig:gghh_diag2l}.  Propagators and external lines are coloured
    \textbf{\textcolor{c}{orange}} for the first collinear mode, \textbf{\textcolor{cb}{blue}} for the second collinear mode, \textbf{\textcolor{s}{green}} for the soft modes, and \textbf{black} for the hard modes. Purely hard region is not depicted. \label{fig:regions2LP2}  }
\end{figure}

\begin{figure}[h!]
    \centering
    \scalebox{0.8}{\begin{tabular}{cccc}
    \begin{tikzpicture}[y=0.4cm,x=0.8cm]\diagNG{c}{c}{cc}{}{}{}{cc}{c}{}{}{}\end{tikzpicture}&
\begin{tikzpicture}[y=0.4cm,x=0.8cm]\diagNG{c}{c}{}{}{cc}{cc}{}{c}{}{}{}\end{tikzpicture}&
\begin{tikzpicture}[y=0.4cm,x=0.8cm]\diagNG{c}{c}{}{}{}{cb}{cb}{c}{cb}{}{}\end{tikzpicture}&
\begin{tikzpicture}[y=0.4cm,x=0.8cm]\diagNG{c}{c}{}{}{}{}{}{c}{}{}{}\end{tikzpicture}\\

$(-2,-2,-2,0,0,0,-2)$&
$(-2,-2,0,0,-2,-2,0)$&
$(-2,-2,0,0,0,-2,-2)$&
$(-2,-2,0,0,0,0,0)$\\

\begin{tikzpicture}[y=0.4cm,x=0.8cm]\diagNG{s}{}{}{}{s}{s}{}{}{}{}{}\end{tikzpicture}&
\begin{tikzpicture}[y=0.4cm,x=0.8cm]\diagNG{cc}{}{}{c}{c}{cc}{}{}{}{}{c}\end{tikzpicture}&
\begin{tikzpicture}[y=0.4cm,x=0.8cm]\diagNG{cc}{}{}{}{cc}{cc}{cc}{}{c}{}{}\end{tikzpicture}&
\begin{tikzpicture}[y=0.4cm,x=0.8cm]\diagNG{}{s}{s}{}{}{}{s}{}{}{}{}\end{tikzpicture}\\

$(-2,-1,0,-1,-2,-2,-1)$&
$(-2,0,0,-2,-2,-2,0)$&
$(-2,0,0,0,-2,-2,-2)$&
$(-1,-2,-2,-1,0,-1,-2)$\\

\begin{tikzpicture}[y=0.4cm,x=0.8cm]\diagNG{}{cc}{c}{c}{}{}{cc}{}{}{c}{}\end{tikzpicture}&
\begin{tikzpicture}[y=0.4cm,x=0.8cm]\diagNG{}{cc}{cc}{}{}{cc}{cc}{}{c}{}{}\end{tikzpicture}&
\begin{tikzpicture}[y=0.4cm,x=0.8cm]\diagNG{}{}{}{}{}{c}{c}{}{c}{}{}\end{tikzpicture}\\

$(0,-2,-2,-2,0,0,-2)$&
$(0,-2,-2,0,0,-2,-2)$&
$(0,0,0,0,0,-2,-2)$
    \end{tabular}}
    \caption{Regions for the two-loop diagram $\mathbf{NP2}$ in Fig.~\ref{fig:gghh_diag2l}.  Propagators and external lines are coloured
    \textbf{\textcolor{c}{orange}} for the first collinear mode, \textbf{\textcolor{cb}{blue}} for the second collinear mode, \textbf{\textcolor{s}{green}} for the soft modes, and \textbf{black} for the hard modes. Purely hard region is not depicted.
    \label{fig:regions2LNP2}}
\end{figure}
}

\clearpage
\section{Amplitudes: Full vs Leading Power}
\label{app:full_vs_lp}
\begin{figure}
     \centering
     \begin{subfigure}[b]{0.49\textwidth}
         \centering
         \includegraphics[width=\textwidth]{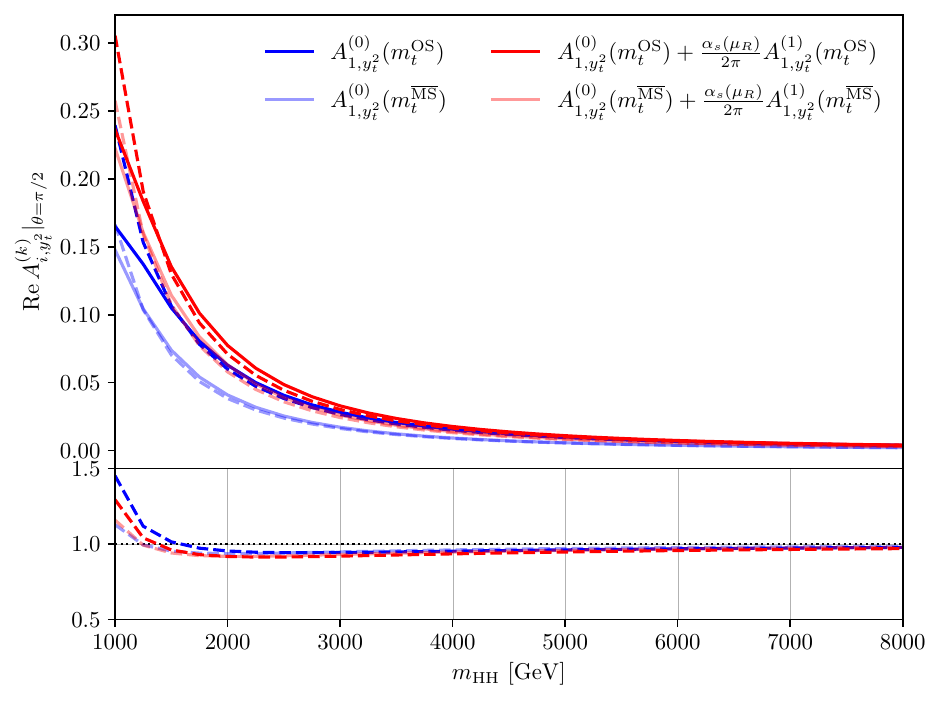}
         \caption{
         }
         \label{sfig:full_vs_lp_a1_re}
     \end{subfigure}
     \hfill
     \begin{subfigure}[b]{0.49\textwidth}
         \centering
         \includegraphics[width=\textwidth]{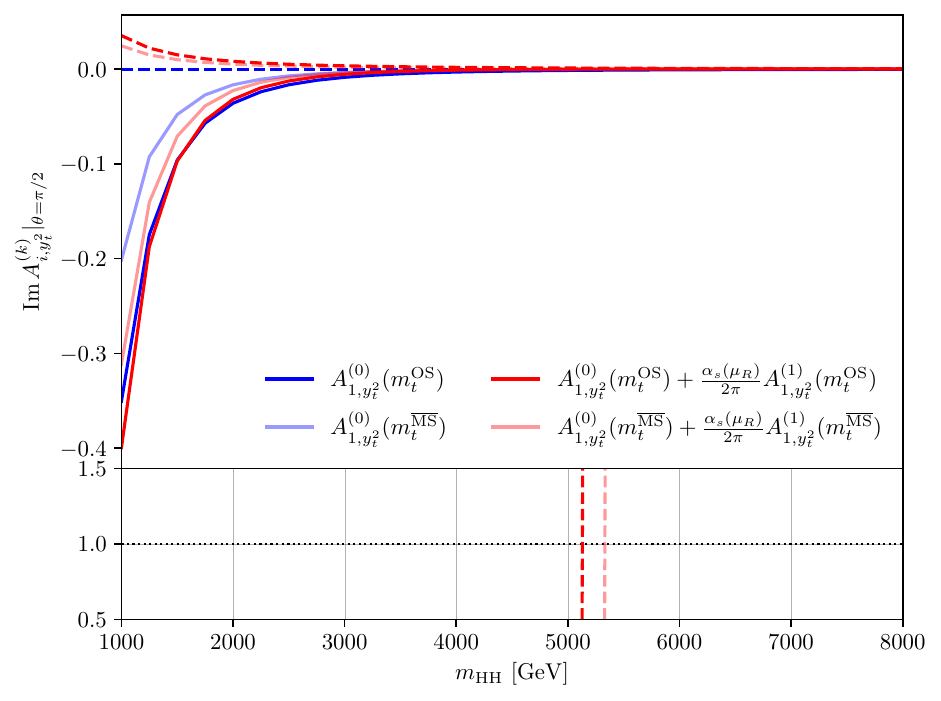}
         \caption{
         }
         \label{sfig:full_vs_lp_a1_im}
     \end{subfigure}
     \begin{subfigure}[b]{0.49\textwidth}
         \centering
         \includegraphics[width=\textwidth]{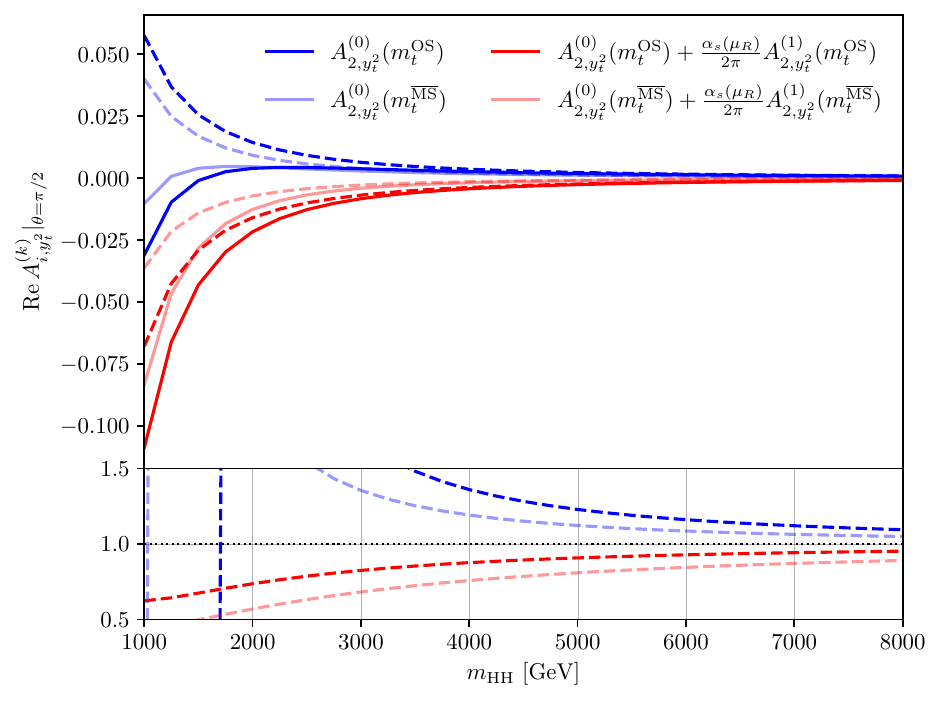}
         \caption{
         }
         \label{sfig:full_vs_lp_a2_re}
     \end{subfigure}
     \hfill
     \begin{subfigure}[b]{0.49\textwidth}
         \centering
         \includegraphics[width=\textwidth]{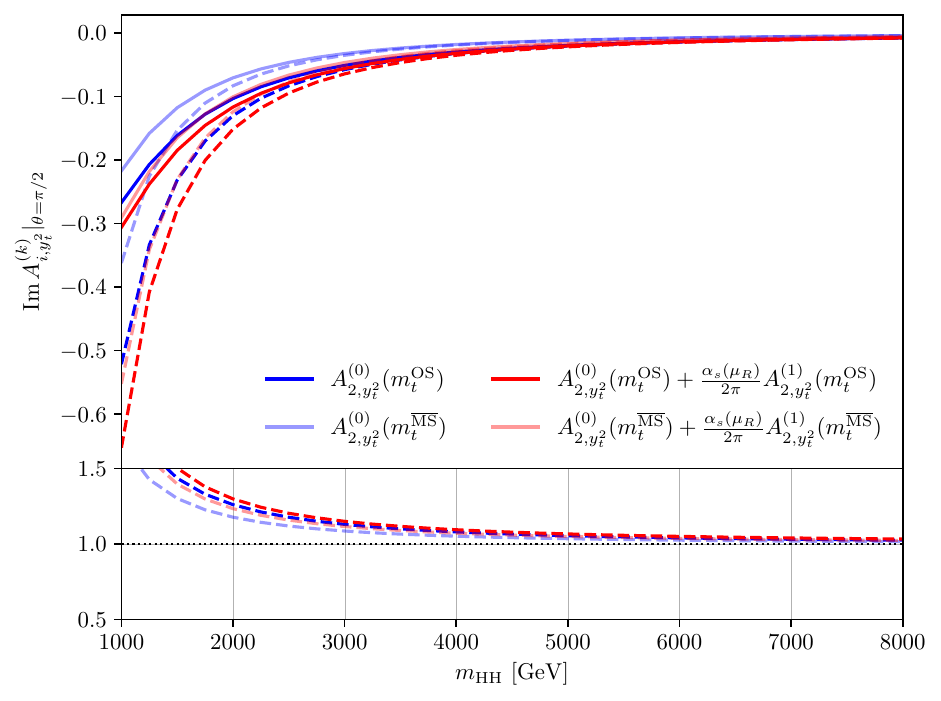}
         \caption{
         }
         \label{sfig:full_vs_lp_a2_im}
     \end{subfigure}
        \caption{Plots analogous to Fig.~\ref{fig:all_sq_full_vs_lp},
        testing the validity of the leading power expansion at one- and two-loop. Solid lines represent the full contributions and dotted lines are only the leading power terms. Here we show a breakdown of the different parts:
        In panels (a) and (b) we have the real and imaginary contributions
        to $A_{1,y_t^2}$, respectively. In panels (c) and (d)
        we show the same information for the $A_{2,y_t^2}$ form factor.}
        \label{fig:a1_a2_full_vs_lp}
\end{figure}

\begin{figure}
     \centering
     \begin{subfigure}[b]{0.49\textwidth}
         \centering
         \includegraphics[width=\textwidth]{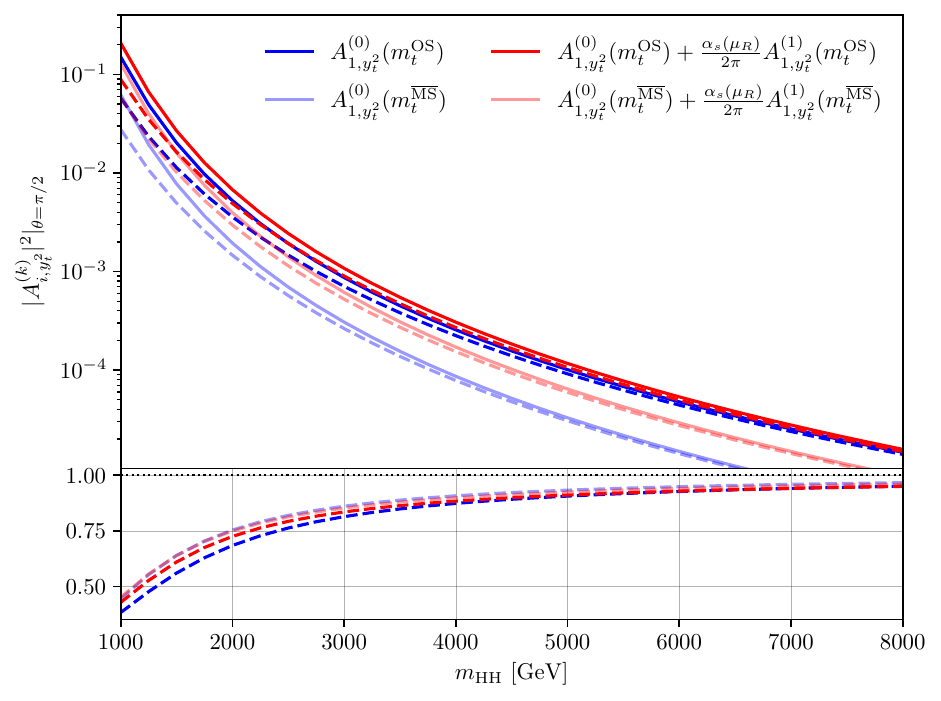}
         \caption{
         }
         \label{sfig:full_vs_lp_a1_sq}
     \end{subfigure}
     \hfill
     \begin{subfigure}[b]{0.49\textwidth}
         \centering
         \includegraphics[width=\textwidth]{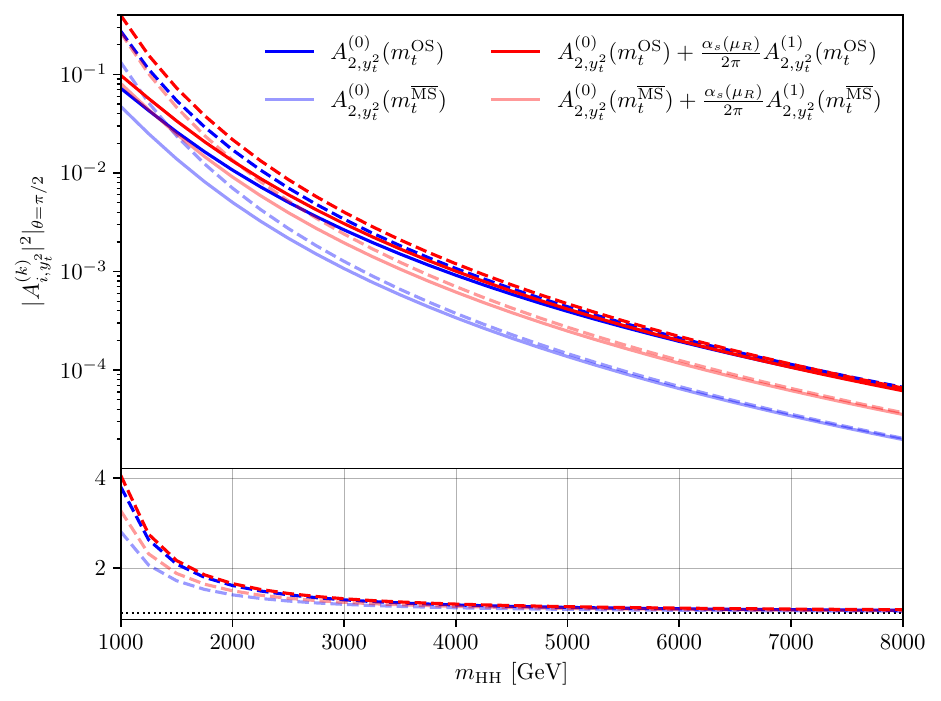}
         \caption{
         }
         \label{sfig:full_vs_lp_a2_sq}
     \end{subfigure}
        \caption{Comparison of full versus the leading-power result at one- and two-loop for square of each form factor. Solid lines represent the full result and dotted lines the leading power contributions.  }
        \label{fig:a1_a2_sq_full_vs_lp}
\end{figure}

Here, we present a breakdown of the contributions of each individual form factor, $A_{1,y_t^2}$ and $A_{2,y_t^2}$, to the total amplitude. 
In Fig.~\ref{fig:a1_a2_full_vs_lp}, we 
plot the full contributions
to the real and imaginary 
parts of $A_{1,y_t^2}$ and $A_{2,y_t^2}$, and their leading power 
counterparts. For the case of the real parts, we see that the
leading power terms for $A_{1,y_t^2}$
exhibit a better agreement over wide range of the invariant mass of the system, $m_{{\rm{HH}}}$, than the $A_{2,y_t^2}$ amplitudes.
For the real parts of $A_{2,y_t^2}$, including higher order
effects extends the range of validity of the leading power approximation 
down approximately $2.5\mathrm{TeV}$, here the power suppressed effects will have a larger impact than in the case of $A_{1,y_t^2}$. 
In Fig.~\ref{fig:a1_a2_sq_full_vs_lp}, we show the individual contributions from $A_{1,y_t^2}$ and $A_{2,y_t^2}$ at the amplitude-squared level, both show good agreement between full and leading power contributions.

\begin{figure}
     \centering
     \begin{subfigure}[b]{0.49\textwidth}
         \centering
         \includegraphics[width=\textwidth]{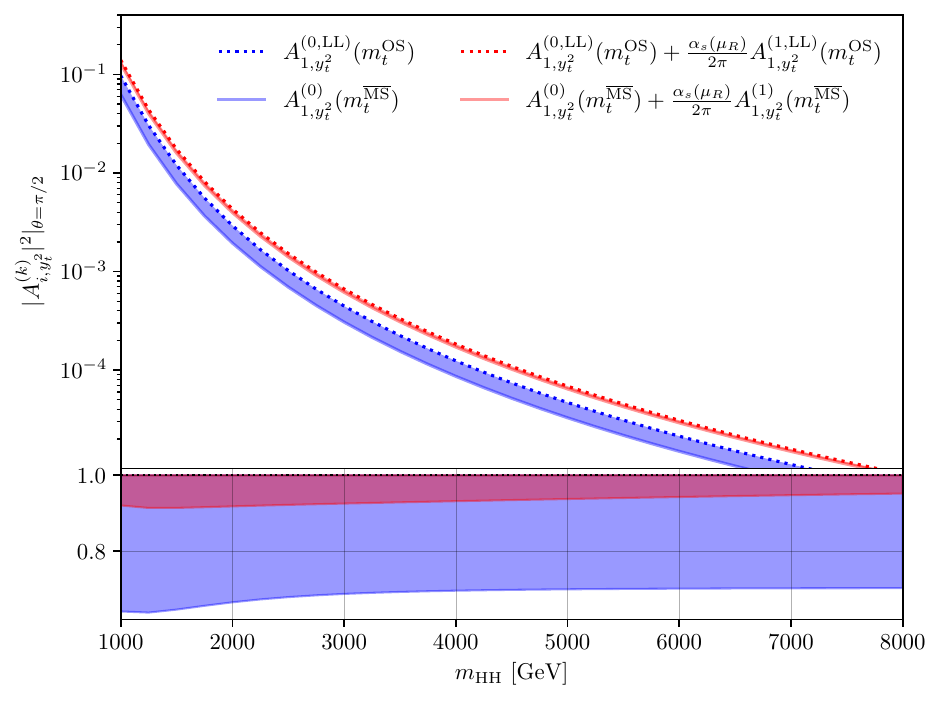}
         \caption{
         }
         \label{sfig:msbar_vs_osll_a1_sq}
     \end{subfigure}
     \hfill
     \begin{subfigure}[b]{0.49\textwidth}
         \centering
         \includegraphics[width=\textwidth]{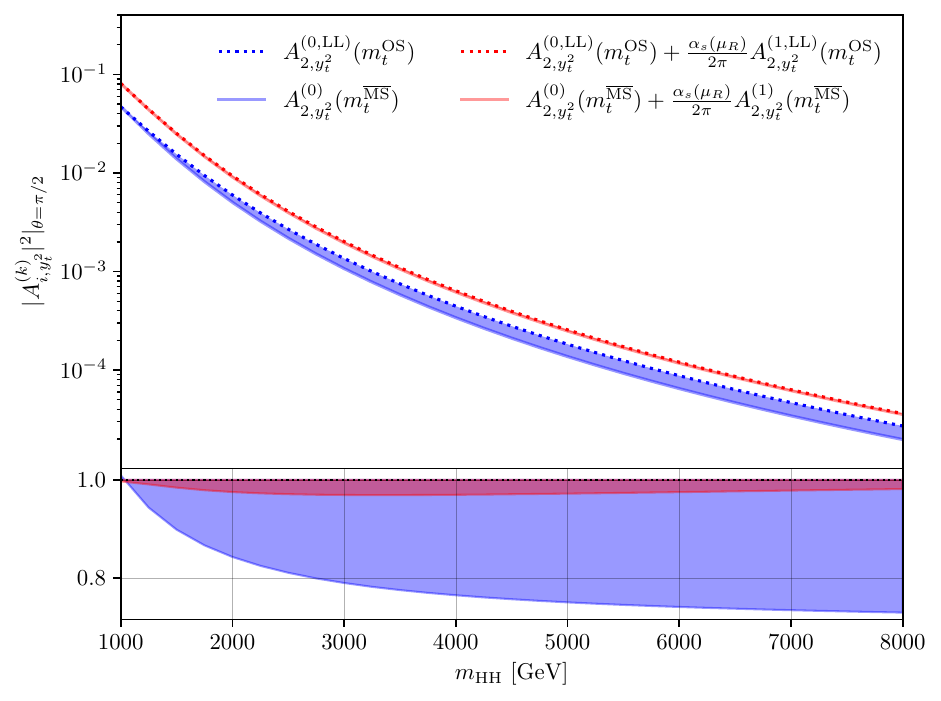}
         \caption{
         }
         \label{sfig:msbar_vs_osll_a2_sq}
     \end{subfigure}
        \caption{Plots showing the uncertainty bands due to the choice of the top-quark mass renormalisation scheme after supplementing the $\mathrm{OS}$ result with leading large logarithms according to the prescription described in the main text.
        Plot is analogous to Fig.~\ref{sfig:all_sq_msbar_vs_osll},
        but instead of the sum of the square of the form factors, here we show the  $A_{1,y_t^2}$  and  $A_{2,y_t^2}$ squared form factors individually in panels (a) and (b), respectively. }
        \label{fig:a1_a2_sq_msbar_vs_osll}
\end{figure}

In Fig.~\ref{fig:a1_a2_sq_msbar_vs_osll}, we present the reduction in uncertainty due to the choice of the top-quark mass scheme, as described in the main text, for the individual form factors. This plot presents the breakdown of Fig.~\ref{fig:all_sq_msbar_vs_osll} into the separate form factors, $A_{1,y_t^2}$ and $A_{2,y_t^2}$. 

\clearpage
\end{appendix}

\clearpage
\bibliography{gghh}

\end{document}